\documentclass[11pt, final]{article}
\pdfoutput=1
\usepackage{JVnotes}
\usepackage{braket}
\usepackage[dvipsnames]{xcolor} 
\usetikzlibrary{arrows.meta}
\definecolor{ochre}{rgb}{0.6, 0.0, 0.0}

 \newcommand{\rH}{r_{+}}
 \newcommand{\tR}{t_{\text{R}}}
 \newcommand{\tL}{t_{\text{L}}}
 \newcommand{\rR}{r_{\text{R}}}
 \newcommand{\rL}{r_{\text{L}}}
 \newcommand{\vphi}{\varphi}
 
 \newcommand{\TSK}{\mathcal{T}_{\text{SK}}}
 \newcommand{\TSKI}[1]{\mathcal{T}_{\text{SK}_{#1}}}
 \newcommand{\CSK}{\mathcal{C}_{\text{SK}}}
 \newcommand{\grSKI}[1]{\text{grSK}_{#1}}
 \newcommand{\Op}{\mathcal{O}}
 \newcommand{\OpR}{\mathcal{O}_{\text{R}}}
  \newcommand{\OpRI}[1]{\mathcal{O}_{\text{R}_{#1}}}
  \newcommand{\OpLI}[1]{\mathcal{O}_{\text{L}_{#1}}}
  \newcommand{\OpdI}[1]{\mathcal{O}_{\text{d}_{#1}}}
  \newcommand{\JRI}[1]{J_{\text{R}_{#1}}}
  \newcommand{\JLI}[1]{J_{\text{L}_{#1}}}
  \newcommand{\JdI}[1]{J_{\text{d}_{#1}}}
 \newcommand{\OpL}{\mathcal{O}_{\text{L}}}
 \newcommand{\JR}{J_{\text{R}}}
 \newcommand{\JL}{J_{\text{L}}}
 
 \newcommand{\Opd}{\mathcal{O}_{\text{d}}}
 \newcommand{\Opa}{\mathcal{O}_{\text{av.}}}

 \newcommand{\OpKMS}{\mathcal{O}_{\text{KMS}}}
 
 \newcommand{\JKMS}{J_{\text{KMS}}}
 \newcommand{\Kin}{K^{\text{in}}}
 \newcommand{\Gin}{G_{\text{in}}}

 \newcommand{\Gout}{G_{\text{out}}}

  \newcommand{\Gbb}{\mathcal{G}}
  \newcommand{\GbbM}{\hat{\mathcal{G}}}
  \newcommand{\Mbb}{\hat{\mathcal{M}}}

  \newcommand{\Gbdy}[1]{\mathbb{G}^{#1}}
 
 \newcommand{\DPlus}{\mathbb{D}_{+}}
 \newcommand{\DMinus}{\mathbb{D}_{-}}
 \newcommand{\vx}{\vec{x}}
 \newcommand{\vk}{\vec{k}}
 \newcommand{\vw}{\omega}
 \newcommand{\nB}{n_{\text{B}}}
 \newcommand{\mP}{\mathfrak{p}}
 \newcommand{\mG}{\mathfrak{G}}
 \newcommand{\mg}{\mathfrak{g}}
 
 \newcommand{\mW}{\mathbb{W}}
 \newcommand{\Jm}{\hat{\mathcal{J}}}

 \newcommand{\mF}{\mathfrak{F}}

\catcode`,\active

\catcode`\,12

\newcommand*\pFqRegskip{8mu}
\catcode`,\active
\newcommand*\pFqReg{\begingroup
        \catcode`\,\active
        \def ,{\mskip\pFqRegskip\relax}%
        \dopFqReg
}
\catcode`\,12
\def\dopFqReg#1#2#3#4#5{%
        {}_{#1}\textbf{F}_{#2}\biggl(\genfrac..{0pt}{}{#3}{#4};#5\biggr)%
        \endgroup 
        }

\catcode`,\active

\catcode`\,12

 \newcommand{\mw}{\mathfrak{w}}
 \newcommand{\mq}{\mathfrak{q}}
 
 \newcommand{\tDelta}{\tilde{\Delta}}

\makeindex

\graphicspath{{./Images/}}
\title{A Holographic prescription for generalized Schwinger-Keldysh contours}
\author[a]{Martin Ammon,}
\author[a]{Jette Germerodt,}
\author[a]{Christoph Sieling,}
\author[a]{and Julio Virrueta}

\affiliation[a]{
	Theoretisch-Physikalisches Institut, Friedrich-Schiller-Universit\"at Jena, Max-Wien-Platz 1, D-07743 Jena, Germany
}

\emailAdd{martin.ammon@uni-jena.de}
\emailAdd{jette.germerodt@uni-jena.de}
\emailAdd{christoph.sieling@uni-jena.de}
\emailAdd{julio.virrueta@uni-jena.de}

\abstract{
We provide a holographic prescription to compute real-time thermal correlators with arbitrary operator ordering. In field theory, these correlation functions are captured by a multi-fold Schwinger-Keldysh time contour.  We propose a holographic dual for these contours, which generalizes the gravitational Schwinger-Keldysh geometry previously advocated in the literature.  Our geometry consists of multiple AdS-black holes glued together at the future and past horizons, with matching conditions determined by unitarity and the KMS condition. As a proof of concept, we solve for a probe scalar field in this geometry and compute bulk-bulk and bulk-boundary propagators, in terms of which we evaluate the 4-point functions at tree-level. We show that in perturbation theory, the lowest-order diagrams that contribute non-trivially to the out-of-time order four-point function are exchange diagrams which explore the full four-fold geometry. Furthermore, these diagrams reduce to a simple factorized expression. We propose a conjecture on the structure of higher
order observables and provide a partial proof by studying a subset of the contributing diagrams. 

}

\begin{document}
\maketitle

\section{Introduction}\label{sec:Intro}

The dynamics of quantum fields at finite temperature play a central role across many areas of theoretical physics, from condensed matter to cosmology. In the context of the holographic correspondence, thermal states offer a controlled setting for probing the quantum aspects of black holes, especially in asymptotically AdS spacetimes \cite{Maldacena:1997re, Witten:1998qj}. 

While the physics of systems at thermal equilibrium can be captured by well-understood Euclidean field theory techniques, the study of dynamics out of equilibrium requires a Lorentzian setting. This is commonly captured by an analytic continuation, where the appropriate $i\epsilon$ prescription accounts for time ordering, as shown by the reconstruction theorem \cite{Osterwalder:1973dx}. The analytic continuation can be avoided by employing a purely Lorentzian setting, provided by the Schwinger-Keldysh (SK) formalism \cite{Schwinger:1960qe,Keldysh:1964ud}. This can be further extended to account for open quantum systems \cite{Feynman:1963fq}. For systems with a holographic dual description, understanding out-of-equilibrium dynamics provides an avenue to study the gravitational path integral for Lorentzian spacetimes.

Several prescriptions have been developed to compute real-time observables in holography \cite{Herzog:2002pc,Son:2002sd,vanRees:2009rw,Skenderis:2008dg,deBoer:2018qqm}. Among them, a successful approach identifies the SK generating functional with a bulk path integral over a complexified AdS black hole geometry \cite{Glorioso:2018mmw}. This method has been effective in the analysis of open systems \cite{Jana:2020vyx,Loganayagam:2022zmq} and effective hydrodynamic actions \cite{Ghosh:2020lel,He:2021jna,He:2022jnc,He:2022deg}.

Despite these successes, certain 4-point functions lie beyond the reach of the conventional Schwinger-Keldysh framework; these are denoted as out-of-time-order correlators (OTOCs). A particular example is the correlator
\begin{equation}
C(t) = -\expval{[W(t),V(0)]^2}_{\beta}\,,
\end{equation}
which has a close relationship with the phenomenon of quantum chaos \cite{Larkin:1959}, since its early-time exponential growth of $C(t)  \sim \frac{1}{N^2}e^{\lambda t}$ can be related to the classical notion of a Lyapunov exponent. This behavior is of key importance for black hole physics, which can be shown to be maximally chaotic systems \cite{Shenker:2013pqa,Maldacena:2015waa}.
Existing holographic treatments of the OTOC typically rely on shockwave geometries or eikonal approximations \cite{Shenker:2014cwa, Shenker:2013pqa,Roberts:2014isa, Caceres:2023zft, Kawamoto:2025hnw, Chua:2025vig, Roberts:2016wdl, Roberts:2014ifa, Blake:2016wvh}, which are tailored to specific kinematic regimes. A more general, contour-based prescription would provide a more comprehensive handle on real-time gravitational dynamics.

Capturing the full structure of an OTOC requires a generalization of the Schwinger-Keldysh contour. A 2-fold time contour suffices for standard time-ordered or nested correlators, but computing genuine OTOCs requires at least a 4-fold contour \cite{Aleiner:2016eni,Haehl:2017qfl}. This naturally raises the question of how to construct holographic duals for such multi-fold contours, and how to evaluate real-time Witten diagrams in these more elaborate backgrounds.

In this work, we propose a holographic construction of a bulk geometry dual to general time-folded SK contours. We focus on the 4-fold case relevant for the OTOC, but our construction naturally generalizes to higher folds. Our solution consists of multiple Lorentzian AdS black hole segments glued along their horizons in a pattern dictated by the boundary contour. This complexified geometry provides a non-trivial saddle point of the real-time gravitational path integral, and it allows for the systematic computation of Witten diagrams. In particular, we show that contributions to the OTOC arising from contact interactions vanish identically, while exchange diagrams yield the non-trivial structure of the OTOC.

This paper is organized as follows. In Section \ref{sec:TimeF}, we review real-time contour techniques and the constraints of unitarity and KMS symmetry. Section \ref{sec:HoloTimeContours} revisits the gravitational Schwinger-Keldysh (grSK) geometry and generalizes it to 4-fold contours. In Sections \ref{sec:probe_scalars}, we analyze the dynamics of probe scalars in this extended geometry. We show that contact diagrams vanish, while exchange diagrams yield non-trivial contributions to OTOCs. We conclude in Section \ref{sec:discussion} with a discussion of possible generalizations and future directions. In Appendix \ref{app:grSK2}, we discuss some further details of the construction as well as how to recover the well-known 2-fold dual. The details on the Witten diagrams calculations can be found in Appendix \ref{app:higherCommutators}, and we evaluate the four-point function for a specific example in Appendix \ref{app:toy_model}. Finally, in Appendix \ref{app:VRees} we outline an alternative construction of the multi-fold geometry.

\section{Real-time correlators and time-folds}\label{sec:TimeF}

In this section, we review several aspects of the Schwinger-Keldysh (SK) formalism in quantum field theory, as well as its generalizations for multiple time-folds. Our main goal is to set up the notation, as well as the main physical properties that will guide our holographic construction. For very comprehensive reviews on the subject we recommend \cite{Chou:1984es,Haehl:2016pec,Haehl:2017eob,Haehl:2017qfl,Liu:2018crr}.

\begin{figure}[h!]
\centering 
\includegraphics[scale=0.75]{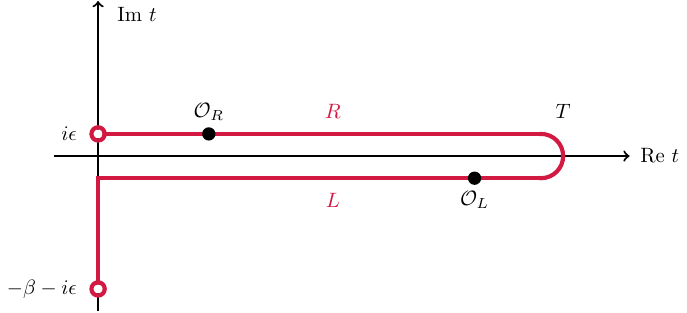}
\caption{The Schwinger-Keldysh contour for a thermal state $\rho_{\beta}$. We include a segment of Euclidean time evolution, preparing the state, and the endpoints are identified as $\tau \sim \tau+\beta$. }
\label{fig:SKContourBasic}
\end{figure}

The SK approach differs from the common understanding of perturbative QFT, where we compute the S-matrix between an initial and a final state. These S-matrix elements are obtained from the time-ordered vacuum correlators, which in turn can be obtained from the Feynman path integral. In contrast, the SK formalism directly targets real-time correlation functions, without requiring asymptotic in/out states, making it particularly well-suited for studying non-equilibrium or strongly coupled systems. These correlators are computed using a path integral along a contour in a complexified time coordinate, with the necessary operator insertions along the contour (see fig.\,\ref{fig:SKContourBasic}).

Although the SK formalism is applicable to very general states (see, for instance, \cite{Sivakumar:2024iqs,Botta-Cantcheff:2019apr,Martinez:2021uqo}), here we will focus mainly on a system prepared in a thermal state, with inverse temperature $\beta$, characterized by a Gibbs density matrix $\rho_{\beta}$. This state is prepared using a Euclidean path integral, with the Euclidean time compactified on a circle $S^{1}_{\beta}$, of circumference $\beta$.

The correlators are captured by a generating functional of the form
\begin{equation}\label{eq:SKGenFun1}
Z_{\text{SK}}[J] = \expval{\TSK\, \exp \left(i\int_{\CSK}dt\, \Op J\right)}_{\beta} = \text{Tr}\left[\rho_{\beta}\,\TSK\, \exp \left(i\int_{\CSK}dt\, \Op J\right)\right]\,,
\end{equation}
where we suppressed integration over the spatial directions. The expectation value is evaluated following a contour ordering, as denoted by the symbol $\TSK$.

An equivalent way to understand this path integral is to consider a double theory. Instead of a single collection of operators $\{\Op^{(i)}(t)\}$ where $t\in\CSK$, we take two sets of operators $\left\{\Op^{(i)}_{\text{R}}(t),\Op^{(i)}_{\text{L}}(t)\right\}$; now $t$ is a real variable, but the operators are defined only in the forward or backward segments of $\CSK$. The corresponding generating functional is
\begin{equation}\label{eq:SKGenFun2}
Z_{\text{SK}}[J_R,J_L] = \expval{\TSK\, \exp \left(i\int dt\, \OpR \,\JR - i\int dt\, \OpL\, \JL\right)}_{\beta}\,,
\end{equation}
where the sign accounts for the orientation of the SK contour\footnote{We do not consider operator insertions along the thermal cycle, which correspond to correlators with respect to some excited state \cite{Botta-Cantcheff:2019apr,Christodoulou:2016nej,Martinez:2021uqo}.}.

Using the Schwinger-Keldysh formalism, we can compute all correlators of the form
\begin{align}
\label{eq:notOTOC}
	\expval{ \mathcal{T}\, [\Op(x_1) \dots \Op(x_K)]\, \overline{\mathcal{T}}\,[\Op(x_{K+1})\dots \Op(x_N)]}\,,
\end{align}
where now $\mathcal{T}$ acts on operators in the first branch of the contour and implements time ordering, while $\overline{\mathcal{T}}$ acts on the second branch and enforces anti-time ordering.

This is already a rich class of correlators but, as outlined in the introduction, it does not capture all possible cases. For example, the correlator $\expval{ \Op(t_4)\Op(t_2)\Op(t_3)\Op(t_1)}$, with $t_1 < t_2 < t_3< t_4$, cannot be computed using the traditional Schwinger-Keldysh contour. This is an example of an out-of-time-order correlator. 

\begin{figure}[h!]
\centering 
\includegraphics[scale=0.75]{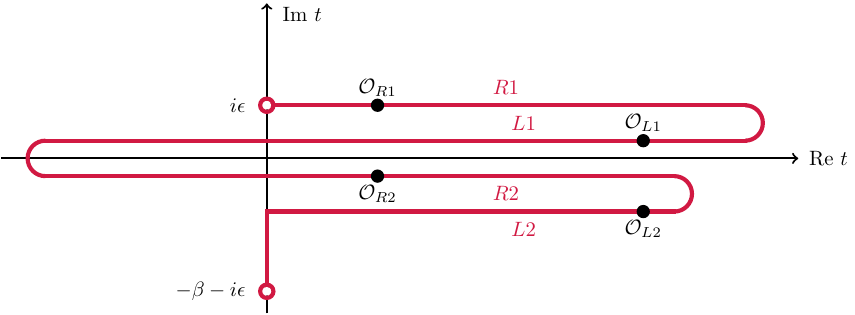}
\caption{$4-$fold contour for the thermal state. }
\label{fig:4FoldThermal}
\end{figure}

It is clear that, to capture the OTOCs within a generating functional, we must generalize the SK contour to include multiple time n-folds (see fig.\,\ref{fig:4FoldThermal} for the case $n=4$) or we must consider more than two copies of the operator spectrum.

The corresponding generating functional is
\begin{equation}
Z_{\text{SK}_n}[\JR^{(i)},\JL^{(i)}] = \expval{\exp \left[i\sum_{i=1}^{n/2}\int dt\left(\JR^{(i)}\OpR^{(i)} - \JL^{(i)}\OpL^{(i)}\right)\right]}_{\beta}\,,
\end{equation}
which now captures also the OTOCs.

\subsection{Unitarity and KMS constraints}
The doubling of degrees of freedom in the SK contour leads to a large degree of redundancies in the formulation, which in turn capture the constraints imposed on the thermal correlators by the KMS condition and microscopic unitarity \cite{Haehl:2016pec,Haehl:2016uah}.

\begin{figure}[h!]
\centering 
\includegraphics[scale=0.75]{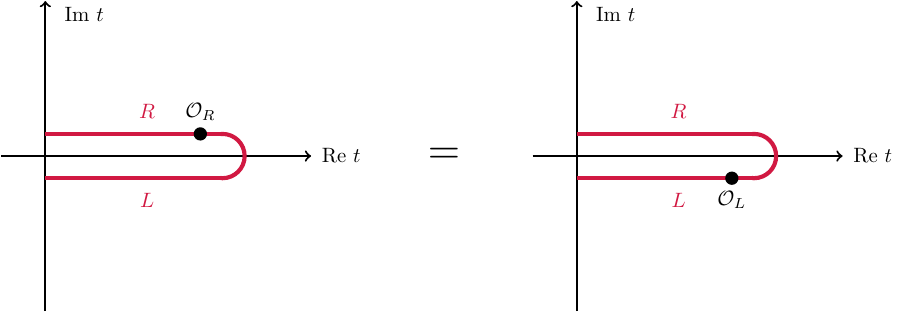}
\caption{Two equivalent SK correlators, related by unitarity.}
\label{fig:EquivSKUnitarity}
\end{figure}

From the point of view of the complex contour, the condition of microscopic unitarity can be seen as the possibility of collapsing the contour in the absence of operator insertions.\footnote{The evolution along the SK contour involves a unitary operator $U$ and its inverse $U^{-1}$; in the absence of insertions, their product collapses to the identity, $UU^{-1}=\mathbb{I}$.} This contour collapsing can also be understood as the fact that we can represent the same single-copy correlator as two equivalent SK ones (see fig.\,\ref{fig:EquivSKUnitarity}).

This redundancy motivates the introduction of a different basis for the operators $\{\OpR,\OpL\}$, known as the average/difference basis:
\begin{equation}\label{eq:AvDifBasis}
\Opa = \frac{1}{2}\left(\OpR + \OpL\right)\,, \qquad \Opd = \OpR- \OpL\,.
\end{equation}
In this basis, the unitarity condition is expressed as
\begin{equation}\label{eq:UnitConstStrong}
\expval{\TSK \, \Opd(t)\cdots} = 0\,,
\end{equation}
where the ellipsis stands for operator insertions at times $t_i<t$. 

It is clear that this condition implies a weaker version
\begin{equation}\label{eq:UnitConstWeak}
\expval{\TSK\,\prod_{i=1}^{k}\Opd(t_i)} = 0\,,
\end{equation}
for arbitrary times $t_i$. In our future analysis, we will mostly use this weaker version of the unitarity condition.

Moreover, thermal correlators satisfy the KMS condition \cite{Kubo:1957mj,Martin:1959jp}, which results from the cyclic property of the trace and the specific form of the thermal density matrix. In the case of two-point functions, the KMS condition is
\begin{equation}\label{eq:KMS2Point}
\expval{\Op_1(t)\Op_2(0)}_{\beta} = \expval{\Op_2(0)\Op_1{(t-i\beta)}}_{\beta}\,,
\end{equation}
and this can be generalized for arbitrary $n$-point correlators \cite{Haehl:2017eob}.

Just as in the case of the unitarity constraint, we can implement the KMS condition by introducing a new basis for the operators. This is known as the advanced/retarded basis\footnote{In the literature, the advanced/retarded operators are often referred to as $\Op_{\text{F}}$ and $\Op_{\text{P}}$. Here we dispensed with this notation for the sake of clarity.}:
\begin{equation}\label{eq:diffKMS}
\Opd(t) = \OpR(t) - \OpL(t)\,, \qquad \OpKMS(t) = \OpR(t) - \OpL(t-i\beta)\,,
\end{equation}
in terms of which, the KMS condition implies
\begin{equation}\label{eq:KMSConstWeak}
\expval{\TSK\,\prod_{i=1}^{k}\OpKMS(t_i)} = 0\,,
\end{equation}
which also has a stronger version when only the earliest operator insertion is a KMS operator.

The KMS constraints are easier to implement in frequency space. To this end, we introduce the following shorthand notation: $x^{\mu}=(t,\vx)$ and $k^{\mu}=(\vw,\vk)$, together with
\begin{equation}\label{eq:MomSpace}
\int_{k} = \int \frac{d\vw}{2\pi}\int\frac{d^{d-1}\vk}{(2\pi)^{d-1}}\,,
\end{equation}
and, for future applications, we also define the time-reversed momentum $\bar{k}^{\mu} = (-\vw,\vk)$. 

The momentum-space correlators are then
\begin{equation}\label{eq:MomSpaceCorr}
\expval{\TSK\,\Op_1(x_1)\cdots \Op_n(x_n)} = \prod_{i=1}^{n}\int_{k_i} e^{ik_i\cdot x_i}\expval{\TSK\,\Op_1(k_1)\cdots \Op_n(k_n)}\,,
\end{equation}
where the labels in the operators may refer to the SK species or a genuinely different kind of operators. 

The weaker versions of both the unitarity and KMS conditions are easy to implement in momentum space:
\begin{equation}\label{eq:MomSpaceUnitKMS}
\expval{\TSK\, \prod_{i=1}^{n}\Opd(k_i)}_{\beta} = 0\,, \qquad \expval{\TSK\, \prod_{i=1}^{n}\OpKMS(k_i)}_{\beta} = 0\,,
\end{equation}
where 
\begin{equation}\label{eq:diffKMSMom}
\Opd(k) = \OpR(k) - \OpL(k)\,, \qquad \OpKMS(k) = \OpR(k) - e^{-\beta\vw}\OpL(k)\,.
\end{equation}

On the other hand, the stronger version of the constraints is not evident in momentum space since it is captured within the analytic structure of the correlators.

The unitarity and KMS conditions also constrain correlators in the multi-time-fold contours, and the collection of independent correlators has been classified in \cite{Haehl:2017qfl}. Following \cite{Chaudhuri:2018ymp}, we introduce a generalized advanced/retarded basis for the 4-fold geometry:
\begin{equation}\label{eq:FPBasis4Fold}
\begin{split}
\OpKMS(\vw,\vk) &= \OpRI{1}(\vw,\vk) - e^{-\beta\vw}\OpLI{2}(\vw,\vk)\,, \\ 
\OpdI{1}(\vw,\vk) &= \OpRI{1}(\vw,\vk) - \OpLI{1}(\vw,\vk)\,, \\ 
\OpdI{2}(\vw,\vk) &= \OpLI{1}(\vw,\vk) - \OpRI{2}(\vw,\vk)\,, \\
\OpdI{3}(\vw,\vk) &= \OpRI{2}(\vw,\vk) - \OpLI{2}(\vw,\vk)\,,
\end{split}
\end{equation}
where we refer to fig.\,\ref{fig:4FoldThermal} for the definition of operators in the L/R basis. The last three of these combinations enforce the unitarity constraint, which can be understood as the possibility of displacing operators along the contour, leading to the vanishing of the corresponding correlators. Similarly, the first of these combinations enforces the KMS condition, which can be seen as moving an operator from the first to the last branch of the contour along the thermal circle. 

The notion of the advanced/retarded basis can be similarly extended to an  $n-$fold contour as
\begin{equation}\label{eq:FPBasisnFold}
\begin{split}
\OpKMS(\vw,\vk) &= \OpRI{1}(\vw,\vk) - e^{-\beta\vw}\OpLI{n/2}(\vw,\vk)\,, \\ 
\OpdI{1}(\vw,\vk) &= \OpRI{1}(\vw,\vk) - \OpLI{1}(\vw,\vk)\,, \\ 
\OpdI{2}(\vw,\vk) &= \OpLI{1}(\vw,\vk) - \OpRI{2}(\vw,\vk)\,, \\
&\;\;\vdots \\
\OpdI{n-2}(\vw,\vk) &= \OpLI{n/2-1}(\vw,\vk) - \OpRI{n/2}(\vw,\vk)\,, \\ 
\OpdI{n-1}(\vw,\vk) &= \OpRI{n/2}(\vw,\vk) - \OpLI{n/2}(\vw,\vk)\,.
\end{split}
\end{equation}

For the purpose of writing a generating functional of correlators, we also introduce the corresponding advanced/retarded basis for the sources. In the case of the 4-fold contour, this is
\begin{equation}\label{eq:FPBasisSources4}
\begin{split}
\JKMS(\vw,\vk) &= \nB(\vw)\left[-\JRI{1}(\vw,\vk) + \JLI{1}(\vw,\vk) - \JRI{2}(\vw,\vk) + \JLI{2}(\vw,\vk)\right]\,,\\
\JdI{1}(\vw,\vk) &= \nB(\vw)\left[-e^{\beta\vw}\JRI{1}(\vw,\vk) + \JLI{1}(\vw,\vk) - \JRI{2}(\vw,\vk) + \JLI{2}(\vw,\vk)\right]\,,\\
\JdI{2}(\vw,\vk) &= \nB(\vw)\left[e^{\beta\vw}\left(\JRI{1}(\vw,\vk) - \JLI{1}(\vw,\vk)\right) + \JRI{2}(\vw,\vk) - \JLI{2}(\vw,\vk)\right]\,,\\
\JdI{3}(\vw,\vk) &= \nB(\vw)\left[e^{\beta\vw}\left(-\JRI{1}(\vw,\vk) + \JLI{1}(\vw,\vk) - \JRI{2}(\vw,\vk)\right) + \JLI{2}(\vw,\vk)\right]\,,\\
\end{split}
\end{equation}
where
\begin{equation}\label{eq:BoltzFact}
\nB(\vw) = \frac{1}{e^{\beta\vw}-1}
\end{equation}
is the usual Bose-Einstein distribution. This basis for the sources can be easily extended for the n-fold contour along the same lines as in \eqref{eq:FPBasisnFold}.

In terms of these sources, we can write
\small
\begin{equation}
\int_k \sum_{i=1}^{n/2}\left(\JRI{i}(-k)\OpRI{i}(k) - \JLI{i}(-k)\OpLI{i}(k)\right) = \int_k \left(\JKMS(-k)\OpKMS(k) + \sum_{i=1}^{n-1}(-1)^{i}\JdI{i}(-k)\OpdI{i}(k)\right)\,,
\end{equation}
\normalsize
so that the generating functional is
\begin{equation}
Z_{\text{SK}_n}[\JKMS,\JdI{i}] = \expval{\TSKI{n}\,\exp\,   i\int_k \left(\JKMS(-k)\OpKMS(k) + \sum_{i=1}^{n-1}(-1)^{i}\JdI{i}(-k)\OpdI{i}(k)\right)}_{\beta}.
\end{equation}

In the next section, we construct a gravitational dual for the generating functional of connected correlators $W_{\text{SK}_n}[\JKMS,\JdI{i}] = \log Z_{\text{SK}_n}[\JKMS,\JdI{i}]$, which we compute as a saddle-point of the bulk path integral.

\section{Holographic time-fold contours}\label{sec:HoloTimeContours}
\subsection{A review of the grSK geometry}
As noticed in the introduction, the Schwinger-Keldysh contour has a well-known holographic dual given by the gravitational Schwinger-Keldysh geometry (grSK) \cite{Glorioso:2018mmw,Jana:2020vyx,Loganayagam:2022zmq}. In this section, we will extend this construction to include the 4-fold time contour (fig.\,\ref{fig:4FoldThermal}) and argue how it can be further extended for arbitrary time-folds. To this end, we first review the construction of the original grSK geometry.

The main idea behind the construction is to consider a complexified geometry, which is a solution to Einstein equations, i.e., a saddle of the gravitational path integral, and whose asymptotic boundary is equal to $\CSK\times \mathbb{R}^{d-1}$. 

Drawing inspiration from the SK contour (fig.\,\ref{fig:SKContourBasic}), we can decompose the geometry into three segments. First, we have an Euclidean region:
\begin{equation}\label{eq:EucCigar}
ds^2 = r^2f(r)d\tau^2 + \frac{dr^2}{r^2f(r)} + r^2 d\vx^2_{d-1}\,, \qquad f(r) = 1- \frac{\rH^{d}}{r^{d}}\,,
\end{equation}
where $\tau\sim\tau+\beta$. This prepares the system in a thermal state, with inverse temperature
\begin{equation}\label{eq:HawT}
\beta = \frac{4\pi}{|f'(\rH)|}\,.
\end{equation}

The other two segments are Lorentzian solutions and consist of two identical copies of the AdS-Schwarzschild black hole:
\begin{equation}\label{eq:BHGeo}
ds^2 = - r^2 f(r)dt^2 + \frac{dr^2}{r^2f(r)} + r^2 d\vx^2_{d-1}\,,
\end{equation}
where we consider only the domains of outer communication ($r\geq\rH$ and $t\geq0$). In order to distinguish between these two solutions, we use coordinates $(\tL,\rL)$ and $(\tR,\rR)$, respectively.

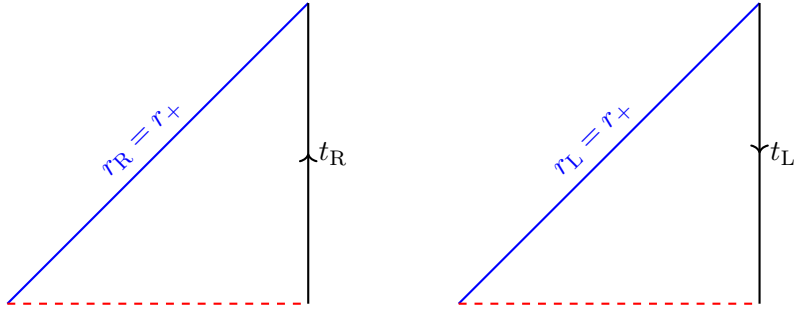
\begin{figure}[H]
\centering
\begin{tikzpicture}
\draw[blue,thick,-] (0,0)  -- (4,4) node[midway, above, sloped] {$\rR=r_+$};
\draw[red,thick,dashed] (0,0) -- (4,0);
\draw[black,thick, ->-] (4,0) -- (4,4) node[midway, right] {$\tR$};
\draw[blue,thick,-] (6,0)  -- (10,4) node[midway, above, sloped] {$\rL=r_+$};
\draw[red,thick,dashed] (6,0) -- (10,0);
\draw[black,thick, ->-] (10,4) --  (10,0) node[midway, right] {$\tL$};
\end{tikzpicture}
\caption{Penrose diagrams for two copies of the domain of outer communication of the AdS black hole. The gravitational Schwinger-Keldysh geometry is obtained by identifying the horizons (blue) and joining the regions $t=0$ (dashed, red) to an Euclidean black hole geometry.}
\label{Fig:grSK1}
\end{figure}

The combined geometry is built by joining the Euclidean and Lorentzian segments at the time-reflection symmetric slice: $\tR=0+i\epsilon$ and $\tL= 0 -i(\beta+\epsilon)$, and the black holes across their future horizon (see fig.\,\ref{Fig:grSK1}). This second joining condition accounts for the turning point of the SK contour, taken to be at $t\rightarrow\infty$.\footnote{In general, the turning point of the SK contour occurs at an arbitrary time $T$, larger than the time at which any operator is inserted. The extension to $T\rightarrow\infty$ is allowed by unitarity.}

\begin{figure}[H]
\begin{center}
\begin{tikzpicture}[scale=0.8]
\draw[thick,color=black,fill=black] (-5,0) circle (0.45ex);
\draw[thick,color=ochre,fill=ochre] (5,1) circle (0.45ex) node [right] {$\zeta(r+i\epsilon)=0$};
\draw[thick,color=ochre,fill=ochre] (5,-1) circle (0.45ex) node [right] {$\zeta(r-i\epsilon)=1$};
\draw[very thick, color=black] (-5,0) node [below] {$\scriptstyle{\rH}$} -- (5,0);
\draw[thick,color=ochre, -] (5,1) -- (-4,1) node [midway, above] {$r+i\epsilon$};
\draw[thick,color=ochre,-] (-4,-1) -- (5,-1) node [midway, below] {$r-i\epsilon$};
\draw[thick,color=ochre,-] (-4,1) arc (45:315:1.414);
\end{tikzpicture}
\end{center}
\caption{Hankel contour describing the complexified radial coordinate.}
\label{Fig:HankelContour}
\end{figure}
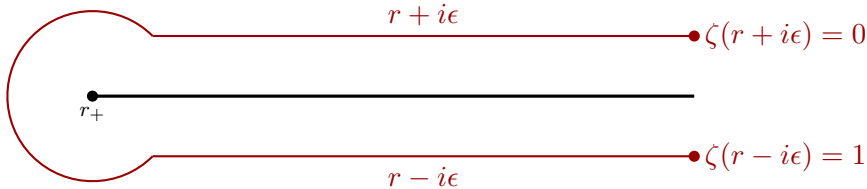

This last procedure requires some special attention. In order to perform the joining at the horizon, we require a coordinate system that extends across the horizon between both Lorentzian segments. This is easily achieved by considering complexified coordinates, where we take $\tR=t+i\epsilon$ and $\tL= t-i\epsilon$ (as suggested by the SK contour), as well as $\rR = r-i\epsilon$, $\rL = r+i\epsilon$. These coordinates are valid away from the horizon, while in a neighborhood of the horizon we introduce $r=\rH - i\epsilon e^{i\theta}$, $\theta\in[0,\pi]$, which provides a coordinate chart valid across the two Lorentzian geometries. Of course, instead of thinking about coordinate patches, we may consider a complexified radial coordinate constrained to the contour depicted in fig.\,\ref{Fig:HankelContour}. This second approach was used in the original proposal \cite{Glorioso:2018mmw,Jana:2020vyx}, and it allows an elegant presentation of the solution. However, in the generalization to higher number of time-folds, it will be better to think in terms of coordinate patches.

It will be convenient to introduce ingoing coordinates:
\begin{equation}\label{eq:BHGeoIng}
ds^2 = -r^2f(r)dv^2 + 2dvdr + r^2d\vx^2_{d-1}\,, \qquad dv = dt + \frac{dr}{r^2f(r)}\,,
\end{equation}
so that the metric remains smooth when changing from one Lorentzian segment to the next, although the radial coordinate gains a constant contribution (the total manifold remains smooth since $r$ follows the Hankel contour).

We may introduce a mocked tortoise coordinate, $\zeta(r)$, of the form
\begin{equation}\label{eq:zetaDef}
\zeta(r) = \frac{2}{i\beta}\int^{r} \frac{dr'}{r'^2 f(r')} + \zeta_0\,.
\end{equation}

In the present case, with just two Lorentzian segments, and taking the view of having a single complexified radial coordinate, we choose the normalization such that
\begin{equation}\label{eq:zetaMono}
\lim_{r\rightarrow\infty}\left(\zeta(r+i\epsilon) - \zeta(r-i\epsilon)\right) = 1\,,
\end{equation}
while the integration constant, $\zeta_0$, can be chosen such that
\begin{equation}\label{eq:zetagrSK}
\lim_{r\rightarrow \infty}\zeta(r+i\epsilon) = 0\,.
\end{equation}
This allows us to cover the whole Lorentzian geometry with a single coordinate patch $(v,\zeta,\vx)$. For future reference, we note that the requirement of having only two Lorentzian segments is needed only in the definition of the analytic continuation of $\zeta(r)$ from one segment to the next, while the definition \eqref{eq:zetaDef} holds even in the case of multiple segments. 

This procedure for the analytic continuation of the radial coordinate is the fundamental ingredient of the grSK prescription, as first proposed in \cite{Glorioso:2018mmw}. The monodromy gained when moving from one of the Lorentzian segments to the other one, is such that it agrees with the periodicity of the Euclidean coordinate $\tau$, this guarantees a smooth extension of the origin of the Euclidean geometry to the horizon of the two Lorentzian segments. Furthermore, it was shown in \cite{Loganayagam:2022zmq,Loganayagam:2022teq} that this contour leads to well-defined higher-order correlators and, in particular, it regularizes any potential singular vertices of the form $\frac{1}{f(r)}$ in the calculation of Witten diagrams.
\subsection{The 4-fold geometry}
Now we proceed to generalize this construction for the multi-fold geometry. We shall use the case of the 4-fold case as a working example. We denote as $\grSKI{n}$ the geometry corresponding to the $n$-fold time contour, with $\grSKI{2}$ being the 2-fold geometry we just discussed.

As before, we consider an Euclidean geometry preparing the thermal state, but now we take four copies of the AdS-Schwarzschild black hole. For two of them, we still restrict ourselves to the domain of outer communication and join them to the Euclidean segment at their point of time-reflection symmetry. 

\begin{figure}[H]
\centering
\begin{tikzpicture}[scale=0.6]
\draw[blue,thick,-] (0,0)  -- (4,4) node[midway, above, sloped] {$r=r_+$};
\draw[red,thick,dashed] (0,0) -- (4,0);
\draw[black,thick, ->-] (4,0) -- (4,4) node[midway, right] {$t_1$};
\draw[blue,thick,-] (5,0)  -- (9,4) node[midway, above, sloped] {$r=r_+$};
\draw[ForestGreen,thick,-] (5,0) -- (9,-4) node[midway, below, sloped] {$r=r_+$};;
\draw[black,thick, ->-] (9,4) --  (9,-4) node[midway, right] {$t_2$};
\draw[cyan,thick,-] (10,0)  -- (14,4) node[midway, above, sloped] {$r=r_+$};
\draw[ForestGreen,thick,-] (10,0) -- (14,-4) node[midway, below, sloped] {$r=r_+$};;
\draw[black,thick, ->-] (14,-4) -- (14,4)  node[midway, right] {$t_3$};
\draw[cyan,thick,-] (15,0)  -- (19,4) node[midway, above, sloped] {$r=r_+$};
\draw[red,thick,dashed] (15,0) -- (19,0);
\draw[black,thick, ->-]  (19,4) -- (19,0) node[midway, right] {$t_4$};
\end{tikzpicture}
\begin{tikzpicture}[scale=0.7]
\draw[blue,thick,-] (0,0)  -- (4,4) node[midway, above, sloped] {$r=r_+$};
\draw[red,thick,dashed] (0,0) -- (3.5,-1);
\draw[black,thick, ->-] (3.5,-1) -- (4,4) node[midway, right] {$t_1$};
\draw[black,thick, ->-] (4,4) --  (5,-4) node[midway, left] {$t_2$};
\draw[ForestGreen,thick,-] (0,0) -- (5,-4) node[midway, below, sloped] {$r=r_+$};
\draw[black,thick, ->-] (5,-4) -- (6,4)  node[midway, right] {$t_3$};
\draw[cyan,thick,-] (54/13,36/13) -- (6,4) node[midway, above, sloped] {$r=r_+$};
\draw[cyan, thick, dotted,-] (0,0)  -- (54/13,36/13) ;
\draw[black,thick, ->-]  (6,4) -- (6.5, -0.5) node[midway, right] {$t_4$};
\draw[red,thick,dotted] (0,0) -- (5.448,-0.419);
\draw[red,thick,dashed] (5.448,-0.419) -- (6.5,-0.5);
\fill[orange!60,opacity=0.5] (0,0) -- (4,4) -- (3.5,-1) -- cycle;
\fill[orange!60, opacity=0.3] (0,0) -- (4,4) -- (5,-4) -- cycle;
\fill[orange!60,opacity=0.5] (54/13,36/13) -- (6,4) -- (5,-4) -- cycle;
\fill[orange!60, opacity=0.3] (5.448,-0.419) -- (6,4) -- (6.5, -0.5) -- cycle;
\end{tikzpicture}
\caption{Bulk geometry dual to the 4-fold time contour. The horizons are identified according to their color coding. The red-dashed lines are then connected to an Euclidean black hole geometry (not depicted). We use unitarity to extend the time contour to $t\rightarrow\pm \infty$, as needed in order to perform the matches on the future and past horizons.}
\label{Fig: grSK4}
\end{figure}
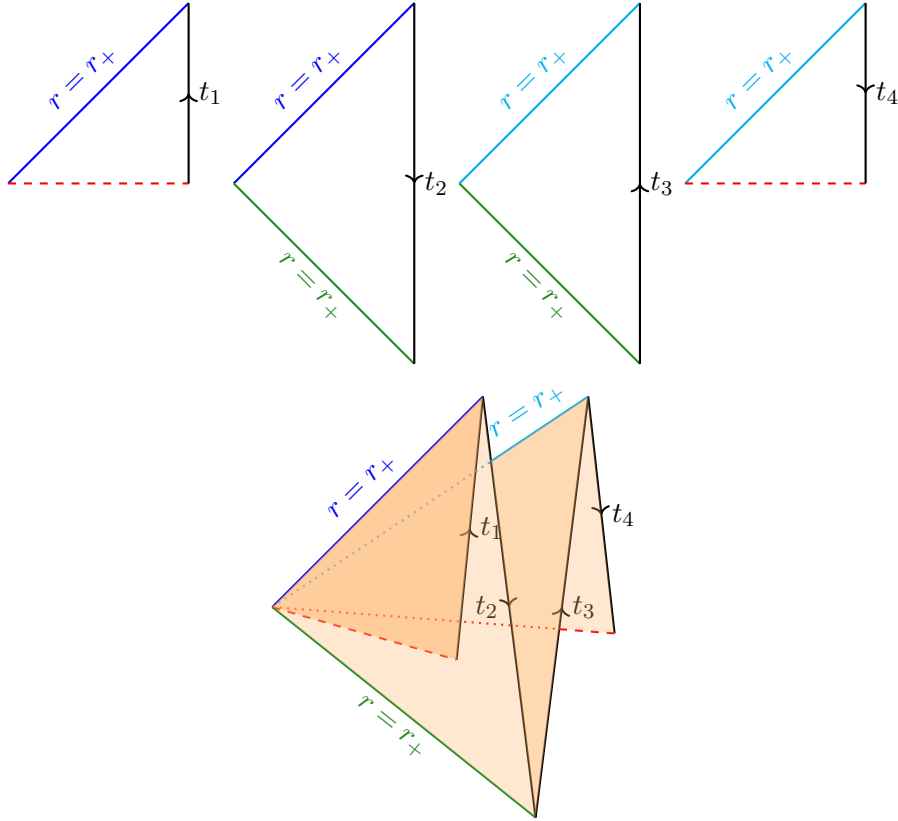

For the other two Lorentzian segments, we will consider the full exterior of the black hole. The reason for this is that, unlike the SK contour which has a single turning point at $t\rightarrow\infty$, the 4-fold contour includes two future turning points and one past turning point (see fig.\,\ref{fig:4FoldThermal}). As in the case of the $\grSKI{2}$ geometry, we then extend these turning points into the bulk by joining adjacent Lorentzian segments across either their future or past horizons (see fig.\,\ref{Fig: grSK4}).

Once again, the joining across the horizons requires the introduction of coordinate patches covering any two segments of interest. To this end, it will be convenient to introduce the outgoing coordinate:
\begin{equation}
ds^2 = -r^2 f(r)du^2 - 2dudr + r^2 d\vx_{d-1}^2\,, \qquad du=dt-\frac{dr}{r^2f(r)}\,,
\end{equation}
which we use to implement the matching at the past horizons. 

The main idea now will be to extend the procedure we employed in the case of $\grSKI{2}$ locally for every pair of Loretzian segments we wish to match. The first Lorentzian segment is then described by coordinates $(v,r_1,\vx)$, and similarly for the second Lorentzian segment but with a different radial coordinate $r_2$. Across their common future horizons we introduce a coordinate patch consisting of $(v,\rH - i\epsilon e^{i\theta},\vx)$, with $\theta\in[0,\pi]$, i.e., following the Hankel contour fig.\,\ref{Fig:HankelContour}.  For the next matching, we first change coordinates from ingoing to outgoing (this can be done at any region where the two patches overlap), and once again we follow a Hankel contour to introduce a coordinate patch which covers the region between the past horizons. This procedure is then performed once more, changing back into ingoing coordinates and introducing the necessary Hankel contours at the second pair of future horizons. The final result is the $\grSKI{4}$ geometry depicted in fig.\,\ref{Fig: grSK4}. This procedure can be naturally extended for an arbitrary number of time-folds.

A clear issue with this prescription is that we cannot introduce a single coordinate patch that covers the entirety of the Lorentzian manifold. The two-sheeted Riemann structure could be globally defined for $\grSKI{2}$, but for $\grSKI{n}$ with $n>2$ this is no longer possible. Hence the analytic structure of the geometry remains somewhat obscure due to the piece-wise nature of the construction. It may be possible that a single coordinate system with a complexified radial coordinate in a multi-sheeted Riemann surface exists, but while such a construction would be highly convenient, we shall not further explore this in the present work.

\section{Probe scalars in 4-fold geometry}\label{sec:probe_scalars}
Even though we have now constructed the bulk geometry dual to the $n$-fold time contour, the discussion has remained somewhat formal. In order to be more concrete, it will be convenient to study the dynamics of fields in the $\grSKI{n}$ manifold. This will allow us to study the multi-sheeted structure of the geometry, which was previously obscured by the absence of a unified coordinate system.

For simplicity, we consider a minimally coupled massive scalar field, whose action takes the form
\begin{equation}\label{eq:SAction}
S[\vphi] = \int d^{d+1}x\sqrt{-g}\left[-\frac{1}{2}\left(\nabla_{A}\vphi\nabla^{A}\vphi + \Delta(\Delta-d)\vphi^2\right) + V(\vphi)\right] + S_{\text{ct}}\,,
\end{equation}
where $S_{\text{ct}}$ stands for the usual counterterms required by holographic renormalization to render the on-shell action finite. We impose Dirichlet boundary conditions for $\vphi$ at the asymptotic boundary \footnote{As it is well-known, for $\Delta\in (\frac{d}{2}-1,\frac{d}{2})$, it is possible to impose Neumann conditions \cite{Witten:2001ua}, in which case the scaling dimension of the dual operators is not $\Delta$ but $d-\Delta$. Our construction works the same way in this case, but we must include additional boundary terms, as required by the variational principle. The same holds for the non-Markovian scalars introduced in \cite{Ghosh:2020lel}.}.

The potential $V(\vphi)$ is assumed to be polynomial in the fields and to be controlled by a coupling constant $\lambda\ll1$. We will then study the boundary correlations functions in a perturbative expansion, which are computed in terms of Witten diagrams on the $\grSKI{n}$ geometry.

While this example will suffice to illustrate how the $\grSKI{n}$ geometry captures the generating functional of correlations for the $n$-fold time contour, it can be generalized to more complicated situations, such as having multiple interacting fields as well as non-minimally coupled scalars. In particular, the latter can be used to capture the dynamics of vector and tensor fields using the \textit{designer scalar} formalism explored in \cite{Ghosh:2020lel,He:2021jna,He:2022jnc,He:2022deg}.

\subsection{Gaussian dynamics}\label{subsec:gaussian}
While the traditional $\grSKI{2}$ geometry suffices for computing any two-point function, we still consider the Gaussian dynamics of the scalar field in $\grSKI{n}$, as a preparation for the study of the Witten diagrams computing higher order correlators.

On any of the Lorentzian sections, we may write the equation of motion for the scalar field in the Fourier domain. In the ingoing coordinates, we perform an expansion
\begin{equation}\label{eq:InfModes}
\vphi(v,r,\vx) = \int_k\, e^{-i\vw v+ i \vk\cdot\vx}\,\phi(r,k)\,,
\end{equation}
where we employ the shorthand notation introduced in \eqref{eq:MomSpace}. Using this decomposition, the equation of motion is
\begin{equation}\label{eq:EoMIn}
\frac{1}{r^{d-1}}\DPlus\left(r^{d-1}\DPlus\,\phi(r,k)\right) +\left(\vw^2-\left(\vk^2+\Delta(\Delta-d)r^2\right)f(r)\right)\phi(r,k)=0\,,
\end{equation}
where we introduced the differential operator:
\begin{equation}\label{eq:DPlus}
\mathbb{D}_{\pm} = r^2f(r)\frac{d}{dr}\mp i\vw\,.
\end{equation}
This way of writing the equation of motion makes the action of the time-reversal transformation, $v\rightarrow -v + i\beta\zeta$, manifest. In particular, we notice that 
\begin{equation}\label{eq:DPConj}
e^{\beta\vw\zeta}\DPlus e^{-\beta\vw\zeta} = \DMinus\,,
\end{equation}
and from this property, it is easy to see that given a solution $\phi(r,k)$ of \eqref{eq:EoMIn}, we can find its time-reversal conjugate as $\phi^{\text{rev.}}(r,k) = e^{-\beta\vw \zeta}\phi(r,\bar{k})$, which is again a solution of the equation of motion. 

In writing these expressions, we employ the function $\zeta(r)$ for convenience, as defined by the equation \eqref{eq:zetaDef}. 
However, we emphasize that, unlike the case of $\grSKI{2}$, it is not globally defined: its shift as one moves from one Lorentzian section to the next is not necessarily $\zeta\rightarrow\zeta-1$. 

The behavior of the fields near the asymptotic boundary is
\begin{equation}
\phi(r\rightarrow\infty,k) = c_1\, r^{\Delta-d}\left(1+\cdots\right) + c_2\, r^{-\Delta}\left(1+\cdots\right)\,,
\end{equation}
where, for simplicity, we assume that $2\Delta-d\notin\mathbb{Z}$ in order to avoid logarithmic branches. The first term above corresponds to non-normalizable modes and hence $c_1$ is fixed by the Dirichlet conditions:
\begin{equation}
\label{eq:DirichletBoundaryConditions}
\lim_{r\rightarrow\infty}r^{d-\Delta}\phi(r,k) = J(k)\,,
\end{equation}
where $J(k)$ is identified with the source of the dual field on the given segment of the time-fold contour.

Near the horizon, the fields admit an expansion of the form
\begin{equation}\label{eq:NearHorizon}
\phi(r,k) = b_1(\vw,\vk)(1+\cdots) + b_2(\vw,\vk)(r-\rH)^{\frac{i\vw\beta}{2\pi}}\left(1+\cdots\right)\,,
\end{equation}
where the ellipsis stands for terms that vanish polynomially as $r\rightarrow\rH$. We see that the first of these terms is analytic at the horizon, while the second is not. As such, the first term has a trivial extension from one Lorentzian segment to the next, while the second term picks up a non-trivial monodromy under analytic continuation across the horizon, reflecting the multi-sheeted structure of the radial coordinate.

All these features may be captured by introducing an ingoing bulk-to-boundary propagator, $\Gin(r,k)$, which is a solution of the equation of motion satisfying the condition:
\begin{equation}\label{eq:BdyCondIng}
\lim_{r\rightarrow \infty}r^{d-\Delta}\Gin(r,k) = 1\,, \quad \lim_{r\rightarrow\rH}\Gin(r,k) = \text{regular}\,,
\end{equation}
and a corresponding outgoing propagator
\begin{equation}
\Gout(r,k) = e^{-\beta\vw \zeta}\,\Gin(r,\bar{k})\,,
\end{equation}
whose non-analytic behavior at the horizon is entirely captured by the function $\zeta(r)$. We recall that $\bar{k}^{\mu}=(-\vw,\vk)$.

The most general solution is then
\begin{equation}\label{eq:genSol}
\begin{split}
\vphi(v,r,\vx) &= 
\int_k e^{-i\vw v+ i\vk\cdot\vx}\left(a_1(\vw,\vk)\,\Gin(r,k) + a_2(\vw,\vk)\,\Gout(r,k)\right)\,, 
\end{split}
\end{equation}
where the coefficients are constraint by the asymptotic boundary condition to be $a_1+a_2=J$.

This analysis can be performed in exactly the same way in outgoing coordinates, with
\begin{equation}
\vphi(u,r,\vx) = \int_k\, e^{-i\vw u+ i \vk\cdot\vx}\,\tilde{\phi}(k,r)\,,
\end{equation}
and the equation of motion
\begin{equation}\label{eq:EoMOut}
\frac{1}{r^{d-1}}\DMinus\left(r^{d-1}\DMinus\,\tilde{\phi}(k,r)\right) +\left(\vw^2-\left(\vk^2+\Delta(\Delta-d)r^2\right)f(r)\right)\tilde{\phi}(k,r)=0\,.
\end{equation}

By performing the change between ingoing and outgoing coordinates in \eqref{eq:genSol}, it is easy to see that we can write
\begin{equation}\label{eq:InOutRel}
\begin{split}
\vphi(u,r,\vx) &= 
\int_k e^{-i\vw u+ i\vk\cdot\vx}\left(a_1(\vw,\vk)\, \Gout(r,\bar{k}) + a_2(\vw,\vk)\, \Gin(r,\bar{k})\right)\,.
\end{split}
\end{equation}
The coefficients $a_1$ and $a_2$ must be the same as in the ingoing case, since the two solution must match in a region where the ingoing and outgoing coordinates overlap. However, the analytic properties of the ingoing and outgoing bulk-to-boundary propagators are reversed, with the ingoing propagator being non-analytic at the past horizon while the outgoing propagator is fully regular.

In this way, we can construct the solution to the equation of motion on any of the Lorentzian sections, with corresponding asymptotic boundary conditions. Next, we consider how these solutions are related to each other by the matching conditions across the future and past horizons.

Let us recall how this is done in the simple situation of the $\grSKI{2}$ geometry. As noticed in the previous section, in this case we can define $\zeta(r)$ such that is gains a shift $\zeta\rightarrow\zeta-1$ as we move from one Lorentzian segment to the next, and, as we just discussed, this captures all the possible monodromies of the solution. If we then require the scalar field to be continuous across the horizon, we must equate the coefficients in the expansion so that
\begin{equation}
\begin{split}
\phi_1(r,k) &= c_1 \,\Gin(r,k) + c_2 \,e^{-\beta \vw \zeta} \,\Gin(r,\bar{k})\,,\\ 
\phi_2(r,k) &= c_1 \,\Gin(r,k) + c_2  \,e^{\beta \vw (1-\zeta)}\, \Gin(r,\bar{k})\,,
\end{split}
\end{equation}
and the two integration constants are then determined by the asymptotic boundary conditions. The result is
\begin{equation}\label{eq:grSKsol1}
\begin{split}
\phi_1(r,k) &= \left((1+\nB(\vw))\JR - \nB(\vw)\JL\right) \,\Gin(r,k) + \nB(\vw)(\JL-\JR) \, e^{-\beta \vw \zeta}\,\Gin(r,\bar{k})\,,\\ 
\phi_2(r,k) &= \left((1+\nB(\vw))\JR - \nB(\vw)\JL\right) \,\Gin(r,k) + \nB(\vw)(\JL-\JR)  \,e^{\beta \vw (1-\zeta)}\, \Gin(r,\bar{k})\,.
\end{split}
\end{equation}
Of course, we can write this solution in a more compact manner as
\begin{equation}\label{eq:grSKsol2}
\begin{split}
\phi(\zeta,k) &= \left((1+\nB(\vw))\JR - \nB(\vw)\JL\right) \,\Gin(r,k) + \nB(\vw)(\JL-\JR) \,e^{-\beta \vw \zeta} \,\Gin(r,\bar{k})\,,
\end{split}
\end{equation}
where the field is defined along all of the Hankel contour, instead of in a piecewise fashion. \footnote{We use the convention where the solution, as written in \eqref{eq:grSKsol2}, lives in the first branch of the Hankel contour (the R branch), and to get the second branch we do the shift $\zeta\rightarrow\zeta-1$. This is the reverse of the convention employed in \cite{Jana:2020vyx}.}

With this solution at hand, it is possible to evaluate the Gaussian contribution to the generating functional in terms of the renormalized on-shell action; this generating functional has been shown to satisfy the KMS and unitarity constraints \cite{Glorioso:2018mmw,Jana:2020vyx}. 

We now proceed to consider the case of a scalar in the $\grSKI{n}$ geometry, taking $n=4$ as a working example. An immediate problem we encounter is that, at first glance, the system appears to be overdetermined. On any of the Lorentzian segments, the scalar field obeys a second order ordinary differential equation in the Fourier domain, and as such it seems entirely determined by a pair of boundary conditions, while we now would like to impose distinct conditions in four or more asymptotic boundaries. The answer to this puzzle lies on the analytic continuation of $\zeta(r)$ across the multiple horizons.

As we have emphasized several times now, in the case of $\grSKI{2}$, we were able to define $\zeta(r)$ such that $\zeta\rightarrow\zeta-1$ as we cross the future horizon. We will now relax this condition, and consider several possible monodromies for the non-analytic solution $\Gout(r,k)$. We can write this in terms of $\zeta(r)$, with $\zeta_{(\alpha)}(r) \rightarrow \zeta_{(\alpha)}(r) + \alpha$, where $\alpha$ is some constant. We remark that the integration constant in the definition of $\zeta_{(\alpha)}(r)$ is taken so that $\zeta_{(\alpha)}(r)\rightarrow 0$ in the asymptotic boundary. 

The more general solution to the equation of motion is then a linear combination of the analytic solution $\Gin(r,k)$ and a collection of $\Gout^{(\alpha)}(r,k)$. In practice, we will show that it is sufficient to consider two distinct analytic continuations:
\begin{equation}\label{eq:4FoldAnsatzIn}
\begin{split}
\phi_1(r,k) &= a_1 \,\Gin(r,k) + a_2 \,e^{-\beta \vw \zeta_+} \,\Gin(r,\bar{k})+ a_3 \,e^{-\beta \vw \zeta_-} \,\Gin(r,\bar{k})\,,\\ 
\phi_2(r,k) &= b_1 \,\Gin(r,k) +b_2 \,\Gin(r,k) + b_3 \,e^{-\beta \vw \zeta_+} \,\Gin(r,\bar{k})+ b_4 \,e^{-\beta \vw \zeta_-} \,\Gin(r,\bar{k})\,,\\ 
\phi_3(r,k) &= c_1 \,\Gin(r,k) +c_2 \,\Gin(r,k) + c_3 \,e^{-\beta \vw \zeta_+} \,\Gin(r,\bar{k})+ c_4 \,e^{-\beta \vw \zeta_-} \,\Gin(r,\bar{k})\,,\\ 
\phi_4(r,k) &= d_1 \,\Gin(r,k) + d_2 \,e^{-\beta \vw \zeta_+} \,\Gin(r,\bar{k})+ d_3 \,e^{-\beta \vw \zeta_-} \,\Gin(r,\bar{k})\,,\\ 
\end{split}
\end{equation}
where $\zeta_{\pm}$ is defined by the condition
\begin{equation}\label{eq:defzetapm}
\zeta_{\pm}\rightarrow \zeta_{\pm} \pm 1\,,
\end{equation}
across any of the horizons. This can be seen as an ambiguity on the choice of orientation of the Hankel contour. We remark that this choice is made to simplify the analysis below, but we could also work with more general monodromies, without changing the final result for the dynamics, as discussed in Appendix \ref{app:grSK2}.

In the expression above, it may seem unnecessary to separate the contributions from $\Gin(r,k)$ into two pieces on the solutions for the second and third Lorentzian segments. However, we must remember that we also need to impose matching conditions at the past horizon, where these functions become non-analytic. Concretely, we may use \eqref{eq:InOutRel} to write the solutions in outgoing coordinates as
\begin{equation}\label{eq:4FoldAnsatzOut}
\begin{split}
\tilde{\phi}_2(r,k) &= b_1 \,e^{\beta \vw \zeta_+} \,\Gin(r,k) +b_2 \,e^{\beta \vw \zeta_-}\,\Gin(r,k) + b_3 \,\Gin(r,\bar{k})+ b_4  \,\Gin(r,\bar{k})\,,\\ 
\tilde{\phi}_3(r,k) &= c_1 \,e^{\beta \vw \zeta_+} \,\Gin(r,k) +c_2\,e^{\beta \vw \zeta_-} \,\Gin(r,k) + c_3 \,\Gin(r,\bar{k})+ c_4  \,\Gin(r,\bar{k})\,,\\ 
\end{split}
\end{equation}
where the tilde refers to the Fourier mode in outgoing coordinates.

This construction can be extended for $n>4$ by including more fields $\phi_i(r,k)$, $i=1,2,\cdots n$. However, we still require only two distinct analytic continuations of $\zeta(r)$. This follows from the fact that each horizon connects only two distinct Lorentzian segments. It is easy to count the degrees of freedom, boundary conditions, and matching conditions. 

We now proceed to impose the matching conditions across the horizons, following fig.\,\ref{Fig: grSK4}. At the first future horizon, we match the first and second Lorentzian segments
\begin{equation}
a_1 = b_1 +b_2\,, \qquad a_2 e^{-\beta\vw}  = b_3 \,, \qquad a_3 e^{\beta\vw} = b_4\,.
\end{equation}
Notice we match the two non-analytic contributions separately, as they are distinguished by their analytic continuation. This is so that the solution remains continuous in the coordinate patch covering the two Lorentzian segments.

The next matching, between the second and third segments, is imposed at the past horizon; we then need to use \eqref{eq:4FoldAnsatzOut}, leading to
\begin{equation}
b_1 e^{\beta\vw}  = c_1\,, \qquad  b_2 e^{-\beta\vw} = c_2\,, \qquad b_3 + b_4 = c_3 + c_4\,.
\end{equation}
Finally, the third and fourth segments are again matched across a future horizon, hence:
\begin{equation}
c_1 + c_2 = d_1 \,, \qquad  c_3 e^{-\beta\vw}  = d_2\,, \qquad c_4 e^{\beta\vw} = d_3\,.
\end{equation}

In addition to the matching at the future and past horizons, we also require the first and last solution to relate to each other at the point of time reflection symmetry, as required by their matching across the Euclidean region (see fig.\,\ref{fig:4FoldThermal}) \footnote{Here we assume no sources are present in the Euclidean boundary, and the solutions may differ by a normalizable contribution. A more general setting with Euclidean sources can be found in \cite{Christodoulou:2016nej}.}
\begin{equation}
\begin{split}
\phi_1(v=i\frac{\beta}{2}\zeta,\vk) &= \int \frac{d\vw}{2\pi}e^{\frac{1}{2}\beta\vw\zeta}\left(a_1 \,\Gin(r,k) + (a_2+a_3) \,e^{-\beta \vw \zeta} \,\Gin(r,\bar{k})\right)\,, \\  
\phi_4(v=i\frac{\beta}{2}\zeta - i\beta,\vk) &= \int \frac{d\vw}{2\pi}e^{\frac{1}{2}\beta\vw\zeta}e^{-\beta\vw}\left(d_1 \,\Gin(r,k) + (d_2+d_3) \,e^{-\beta \vw \zeta} \,\Gin(r,\bar{k})\right)\,, 
\end{split}
\end{equation} 
where we recall that, in ingoing coordinates, the points of time-reflection symmetry are located at $v=i\frac{\beta}{2}\zeta$ and $v=i\frac{\beta}{2}\zeta - i\beta$.
In writing these expressions, we combined the two non-analytic solutions since the distinction between them is irrelevant for this matching, and notice that $\phi_4$ is displaced by $-i\beta$ along the thermal circle. 

Since we have $\text{Im}(\zeta)>0$, the integral over frequency is performed by closing the contour in the upper half-plane. The function $\Gin(r,k)$ has no poles in the upper half-plane, and so only the terms proportional to $\Gin(r,\bar{k})$ lead to a non-trivial condition:
\begin{equation}
a_2+a_3 = e^{-\beta\vw}(d_2+d_3)\,.
\end{equation}
 
 In total, the matching conditions at the horizons and at the Euclidean region result in the following conditions: 
 \begin{equation}\label{eq:matchcond4}
\begin{split}
a_1 = b_1 +b_2\,, \qquad a_2 e^{-\beta\vw}  &= b_3 \,, \qquad a_3 e^{\beta\vw} = b_4\,,\\ 
b_1 e^{\beta\vw}  = c_1\,, \qquad  b_2 e^{-\beta\vw} &= c_2\,, \qquad b_3 + b_4 = c_3 + c_4\,, \\ 
c_1 + c_2 = d_1 \,, \qquad  c_3 e^{-\beta\vw}  &= d_2\,, \qquad c_4 e^{\beta\vw} = d_3\,,\\ 
a_2+a_3 &= e^{-\beta\vw}(d_2+d_3)\,.
\end{split}
 \end{equation}

Finally, we impose the asymptotic boundary conditions:
\begin{equation}
\lim_{r\rightarrow \infty} r^{\Delta-d}\phi_i(r,k) = J_i(k)\,, \qquad J_i = \{\JRI{1}, \JLI{1}, \JRI{2}, \JLI{2}\}\,,
\end{equation}
which leads to
\begin{equation}\label{eq:AsympCond4}
\begin{split}
\JRI{1}(k) &= a_1 + a_2+a_3\,,\\ 
\JLI{1}(k) &= b_1 + b_2 +b_3 + b_4\,,\\ 
\JRI{2}(k) &= c_1+c_2+c_3 + c_4\,,\\ 
\JLI{2}(k) &= d_1 + d_2 + d_3 \,.
\end{split}
\end{equation}

We can then solve the linear system given by \eqref{eq:matchcond4} and \eqref{eq:AsympCond4}. The result can be conveniently expressed as a matrix
\small
\begin{equation}\label{eq:solMatrixLR}
\begin{pmatrix}
\phi_1 \\ 
\phi_2 \\ 
\phi_3 \\ 
\phi_4
\end{pmatrix}
=\nB(\vw)
\begin{pmatrix}
e^{\beta\vw}\Gin -\Gout && -\left(\Gin - \Gout\right) && \Gin -\Gout && -\left(\Gin - \Gout\right) \\ 
e^{\beta\vw}\left(\Gin - \Gout\right) && -\Gin + e^{\beta\vw}\Gout && \Gin -\Gout && -\left(\Gin - \Gout\right) \\ 
e^{\beta\vw}\left(\Gin - \Gout\right) && -e^{\beta\vw}\left(\Gin - \Gout\right) && e^{\beta\vw}\Gin -\Gout && -\left(\Gin - \Gout\right)\\ 
e^{\beta\vw}\left(\Gin - \Gout\right) && -e^{\beta\vw}\left(\Gin - \Gout\right) && e^{\beta\vw}\left(\Gin - \Gout\right) && -\Gin +e^{\beta\vw}\Gout 
\end{pmatrix}
\begin{pmatrix}
\JRI{1}\\ 
\JLI{1}\\ 
\JRI{2}\\ 
\JLI{2}
\end{pmatrix}\,.
\end{equation}
\normalsize
In writing these expressions, we have not differentiated between the possible analytic continuations of $\zeta(r)$. This makes the expressions shorter, but obscures their continuity property across the black hole horizons\footnote{In the language of coordinate patches, the solution \eqref{eq:solMatrixLR} has been written in terms of ingoing patches away from the horizon crossings.}.

We see that the structure of the solution is clear. In this basis, the diagonal components correspond to non-normalizable modes, while the off-diagonal components are fully normalizable. Furthermore, if we write also the $\grSKI{2}$ solution as a matrix, we see that the solution above indeed collapses to the simpler $2$-fold case as we turn off specific sources (see equation \eqref{eq:grSK2BBdryMatrix}). 

The different factors of $e^{\beta\vw}$ in the solution \eqref{eq:solMatrixLR} are the result of the continuity condition across the horizons and, in turn implement the KMS condition. This can be made manifest by changing the basis for the sources into the generalized advanced/retarded basis \eqref{eq:FPBasisSources4}, in terms of which
\begin{equation}\label{eq:solMatrixFP}
\begin{pmatrix}
\phi_1 \\ 
\phi_2 \\ 
\phi_3 \\ 
\phi_4
\end{pmatrix}
=
\begin{pmatrix}
\Gout && -\Gin && 0 && 0 \\ 
0 && -\Gin  &&  -\Gout && 0 \\ 
0 && 0 && -\Gout && -\Gin\\ 
e^{\beta\vw}\Gout && 0 && 0 && -\Gin  
\end{pmatrix}
\begin{pmatrix}
\JKMS\\ 
\JdI{1}\\ 
\JdI{2}\\ 
\JdI{3}
\end{pmatrix}\,.
\end{equation}

As advertised earlier, we can now perform a simple counting argument to see that the generalization for arbitrary $n$ holds. In this case, the most general solution will depend on $6$ parameters from the first and last Lorentzian segments, and $4(n-2)$ parameters for all the other sections, for a total of $2(2n-1)$ undetermined coefficients. There are a total of $n-1$ future and past horizons, and at each we have three conditions corresponding to the matching of the analytic and the two non-analytic solutions; in addition to this, we have $n$ conditions in the asymptotic boundaries and one last constraint arising from the relation to the Euclidean segment. In total, we have $3(n-1)+n+1 = 2(2n-1)$ independent linear equations, which exactly match the number of undetermined parameters. This counting argument confirms that our procedure generalizes naturally to $n>4$, and that in all cases it suffices to consider only two analytic branches of $\zeta(r)$. We notice that, in the case of $\grSKI{2}$, we could also consider the two possible analytic continuations, $\zeta_{\pm}(r)$, but the matching to the Euclidean region reduces this to the simpler solution \eqref{eq:grSKsol2} (see Appendix \ref{app:grSK2}).

We can then evaluate the Gaussian on-shell action, which localizes at the boundary by virtue of the equation of motion. The general form of the action is then
\begin{equation}
S_{\text{on-shell}}^{(2)} = \lim_{r\rightarrow\infty}\frac{1}{2}\sum_{i=1}^{4}(-1)^{i+1}\int_k r^{d-\Delta}\phi_{i}(r,-k)\pi_{i}(r,k)\,,
\end{equation}
where $\pi_i(r,k)$ is the renormalized canonical conjugate momentum in the radial direction, given by
\begin{equation}
\pi_i(r,k)=-\left(r^{\Delta-1}\DPlus \phi_i(r,k) + \text{counterterms}\right)\,.
\end{equation}

 We then evaluate this action using the solution \eqref{eq:solMatrixFP}, the result is
\begin{equation}\label{eq:OnShellActFP}
S_{\text{on-shell}}^{(2)} = -\frac{1}{2}\int_k \Kin(\vw,\vk)\left[\JKMS\left(\JdI{1}^{\dagger} - e^{\beta\vw}\JdI{3}^{\dagger}\right) +\JdI{1}^{\dagger}\JdI{2} - \JdI{2}\JdI{3}^{\dagger}\right] + \text{c.c.}\,,
\end{equation}
where $J^{\dagger}(k) = J(-k)$ and we define
\begin{equation}\label{eq:KinDef}
\Kin(\vw,\vk) = -\lim_{r\rightarrow\infty} \left(r^{\Delta-1}\DPlus\Gin(r,k) + \text{counterterms}\right)\,,
\end{equation}
which corresponds to the retarded two-point function on the boundary.

The fact that the on-shell action has no diagonal terms, such as $\JKMS^{\dagger}\JKMS$ or $\JdI{i}^{\dagger}\JdI{i}$ implies via functional differentiation that $\expval{\OpKMS(-k)\OpKMS(k)} = \expval{\OpdI{i}(-k)\OpdI{i}(k)}=0$, in agreement with the weak forms of the unitarity and KMS conditions (see \cite{Haehl:2017eob,Chaudhuri:2018ymp}).

As anticipated, the expression for the on-shell action contains no new dynamical information — it could be obtained entirely from the $\grSKI{2}$ geometry by a trivial extension of the sources. In order to truly obtain non-trivial results, we must consider higher order correlators.

\subsection{Contact Witten diagrams}\label{subsec:witten_contact}
Now that we understand the Gaussian dynamics of a scalar field in the $\grSKI{n}$ geometry, we turn our attention to interactions, generated by polynomial potentials of the form $V(\vphi)=\lambda\vphi^n$.

We first consider the contact Witten diagrams. For the case of cubic interactions and the $\grSKI{4}$ geometry, these diagrams produce a contribution to the on-shell action of the form:
\begin{equation}\label{eq:OnShell3}
\begin{split}
S^{(3)}_{\text{contact}} &= \lambda\int_{\grSKI{4}} dr \sqrt{-g} \int_{k_j} \phi(r,k_1)\phi(r,k_2)\phi(r,k_3) \,\delta^{(d)}\left(\sum_{i=1}^3 k_i\right)\\ 
&=
\lambda\sum_{a=1}^{4}(-1)^{a+1}\int_{\rH}^{\infty} dr \sqrt{-g} \int_{k_j} \phi_i(r,k_1)\phi_i(r,k_2)\phi_i(r,k_3) \,\delta^{(d)}\left(\sum_{i=1}^3 k_i\right)\,.
\end{split}
\end{equation}
In the first line, the integral over the radial coordinate formally runs over the entirety of the generalized gravitational Schwinger-Keldysh geometry. We can decompose the expression in the second line into simpler integrals over a single black hole geometry. Notice that we must be mindful of the orientation of the contour. The fields $\phi_i(r,k)$ ($i=1,\cdots 4$) can be read directly from \eqref{eq:solMatrixLR}, which we can use to write the contribution to the cubic on-shell action in terms of the sources and the bulk-to-boundary propagators $\Gin$ and $\Gout$.

In order to easily compute the different contributions, it will be convenient to introduce some additional notation. We notice that the solution \eqref{eq:solMatrixLR} is written in terms of three basic linear combinations of the $\Gin$ and $\Gout$ solutions:
\begin{equation}\label{eq:gCombo}
\begin{split}
\mg_1(r,k) = e^{\beta\vw}\Gin(r,k) - \Gout(r,k)\,, \quad \mg_2 &= \Gin(r,k) - e^{\beta\vw} \Gout(r,k)\,, \\
 \mg_3(r,k) = \Gin(r,k)&-\Gout(r,k)\,,
\end{split}
\end{equation}
where the first two are non-normalizable, while the last is fully normalizable.

We then define the left/right bulk-to-boundary propagator as diagonal matrices whose entries correspond to sections of the solution supported by the corresponding sources
\begin{equation}\label{eq:BtoBdyVecLR}
\begin{split}
\Gbdy{\text{R}_1}(r,k) &= \nB(\vw)\,\text{diag} \left(\mg_1(r,k)\,, e^{\beta\vw}\mg_3(r,k)\,, e^{\beta\vw}\mg_3(r,k)\,, e^{\beta\vw}\mg_3(r,k)\right)\,, \\
\Gbdy{\text{L}_1}(r,k) &= -\nB(\vw)\,\text{diag}\left(\mg_3(r,k)\,, \mg_2(r,k)\,, e^{\beta\vw} \mg_3(r,k)\,, e^{\beta\vw} \mg_3(r,k)\right)\,, \\ 
\Gbdy{\text{R}_2}(r,k) &=\nB(\vw)\,\text{diag}\left(\mg_3(r,k)\,, \mg_3(r,k)\,, \mg_1(r,k)\,, e^{\beta\vw}\mg_3(r,k)\right)\,, \\
\Gbdy{\text{L}_2}(r,k) &= -\nB(\vw)\,\text{diag}\left(\mg_3(r,k)\,, \mg_3(r,k)\,, \mg_3(r,k)\,, \mg_2(r,k)\right)\,,
\end{split}
\end{equation}
and corresponding generalization for $\grSKI{n}$. Of course, the diagonal entries are just the columns of the matrix in \eqref{eq:solMatrixLR}, the rows corresponding to the expression of the propagator on the $i$-th Lorentzian segment. The convenience of this matrix notation will become more apparent once we discuss exchange diagrams in the following sub-section.

In terms of these propagators, we can read off the contributions to \eqref{eq:OnShell3} coming from the different sources. For instance, a contribution with sources only on the first segment reads
\begin{equation}
\begin{split}
S^{(3)}_{\text{contact}} &\supset \lambda \int_{k_j} \int_{\rH}^{\infty} dr \sqrt{-g} \, \Tr \left(\hat{S}\,\Gbdy{\text{R}_1}(r,k_1) \Gbdy{\text{R}_1}(r,k_2) \Gbdy{\text{R}_1}(r,k_3)\right)\JRI{1}(k_1)\JRI{1}(k_2)\JRI{1}(k_3)\,,\\ 
\end{split}
\end{equation}
where we omitted the momentum-preserving delta-function, and defined
\begin{equation}\label{eq:sMatrix}
\hat{S} = \text{diag}\left(1,-1,1,-1\right) 
\end{equation}
in order to account for the orientation of the contour.

We can also introduce corresponding bulk-to-boundary propagators in the generalized F/P basis, by simply reading the columns of the matrix in \eqref{eq:solMatrixFP}:
\begin{equation}\label{eq:BtoBdyVecFP}
\begin{split}
\Gbdy{\text{KMS}}(r,k) &= \text{diag}\left(\Gout(r,k)\,, 0\,, 0\,, e^{\beta\vw} \Gout(r,k)\right)\,, \\ 
\Gbdy{\text{d}_1}(r,k) &= -\text{diag}\left(\Gin(r,k)\,, \Gin(r,k)\,, 0\,, 0\right)\,, \\ 
\Gbdy{\text{d}_2}(r,k) &=-\text{diag}\left(0\,, \Gout(r,k)\,, \Gout(r,k)\,, 0\right)\,, \\
\Gbdy{\text{d}_3}(r,k) &= -\text{diag}\left(0\,, 0\,, \Gin(r,k)\,, \Gin(r,k)\right)\,.
\end{split}
\end{equation}
On this basis, the generalization for arbitrary $n$ is especially clear since most of the entries of the vectors are zero, and we simply alternate between pairs of $\Gin$ and $\Gout$:
\begin{equation}\label{eq:BtoBdyFPGeneraln}
\begin{split}
\Gbdy{\text{KMS}}(r,k) &= \text{diag}\left(\Gout(r,k)\,, 0\,, \cdots\,, 0\,, e^{\beta\vw} \Gout(r,k)\right)\,, \\ 
\Gbdy{\text{d}_1}(r,k) &= -\text{diag}\left(\Gin(r,k)\,, \Gin(r,k)\,, 0\,, \cdots\,, 0\right)\,, \\ 
\Gbdy{\text{d}_2}(r,k) &=-\text{diag}\left(0\,, \Gout(r,k)\,, \Gout(r,k)\,, 0\,, \cdots\,, 0\right)\,, \\
\Gbdy{\text{d}_3}(r,k) &= -\text{diag}\left(0\,, 0\,, \Gin(r,k)\,, \Gin(r,k)\,, 0\,, \cdots\,, 0\right)\,, \\ 
&\vdots \\ 
\Gbdy{\text{d}_{n-1}}(r,k) &= - \text{diag}\left(0\,, 0\,, \cdots\,, 0\,, \Gin(r,k)\,, \Gin(r,k)\right)\,.
\end{split}
\end{equation}

Using the propagators \eqref{eq:BtoBdyVecFP}, we see that
\small
\begin{equation}
\begin{split}
S^{(3)}_{\text{contact}} &\supset \lambda \int_{k_j} \int_{\rH}^{\infty} dr \sqrt{-g}\,\Tr\left(\hat{S}\, \Gbdy{\text{d}_1}_a(r,k_1) \Gbdy{\text{d}_1}_a(r,k_2) \Gbdy{\text{d}_1}_a(r,k_3)\right)\JdI{1}(k_1)\JdI{1}(k_2)\JdI{1}(k_3)\,,\\ 
&=\lambda \int_{k_j}\int_{\rH}^{\infty} dr \sqrt{-g}\left[-\Gin(k_1)\Gin(k_2)\Gin(k_3)  + \Gin(k_1)\Gin(k_2)\Gin(k_3)\right]\JdI{1}(k_1)\JdI{1}(k_2)\JdI{1}(k_3) \\ 
&= 0\,.
\end{split}
\end{equation}
\normalsize
A similar cancellation occurs for all diagonal source structures, such as $\JKMS^3$ and $\JdI{i}^3$, due to the structure of the bulk-to-boundary propagators \eqref{eq:BtoBdyVecFP}\footnote{For the case of $\JKMS^3$, one also needs to use frequency conservation, implemented as $\vw_1+\vw_2+\vw_3=0$.}.

As with the Gaussian action, the vanishing of diagonal contact contributions reflects the constraints imposed by unitarity and KMS symmetry. This serves as a non-trivial consistency check of the holographic prescription. Notice this calculation is immediately generalized for the $\grSKI{n}$ geometry, for arbitrary $n$.

It is now easy to see then, that the only non-zero contributions to the cubic on-shell action are off-diagonal in the sources, such as $\JKMS\JdI{1}\JdI{1}$, and furthermore these contributions are written in terms of just two types of integrals:
\begin{equation}
\begin{split}
\int dr \sqrt{-g}\, e^{-\beta\vw_3\zeta}\,\Gin(r,k_1)\,\Gin(r,k_2)\,\Gin(r,\bar{k}_3)\,, \\ 
\int dr \sqrt{-g}\, e^{\beta\vw_1\zeta}\,\Gin(r,k_1)\,\Gin(r,\bar{k}_2)\,\Gin(r,\bar{k}_3)\,.
\end{split}
\end{equation}

Similarly, all non-vanishing contributions contact diagram contributions to n-point correlators are of the form:
\begin{equation}
\begin{split}
&\int dr \sqrt{-g}\, e^{-\beta\vw_n\zeta}\,\Gin(r,k_1)\,\Gin(r,k_2)\,\cdots\,\Gin(k_{n-1},r)\,\Gin(\bar{k}_n,r)\,, \\ 
&\int dr \sqrt{-g}\, e^{-\beta(\vw_n+\vw_{n-1})\zeta}\,\Gin(r,k_1)\,\Gin(r,k_2)\,\cdots\,\Gin(\bar{k}_{n-1},r)\,\Gin(\bar{k}_n,r)\,, \\ 
&\qquad \hspace{5cm} \vdots\\ 
&\int dr \sqrt{-g}\, e^{\beta\vw_1\zeta}\,\Gin(r,k_1)\,\Gin(r,\bar{k}_2)\,\cdots\,\Gin(\bar{k}_n,r)\,.
\end{split}
\end{equation}
This structure for the contact diagrams follows from essentially the same argument employed in \cite{Jana:2020vyx,Loganayagam:2022zmq} for the standard $\grSKI{2}$ geometry.  

Furthermore, we notice that, as we increase the order of the correlator, many more diagrams vanish than we would expect from the unitarity and KMS conditions. Concretely, diagrams of not \textit{neighboring} sources, such as $\JKMS\JdI{1}\JdI{2}$ always vanish. These two features reflect the fact that contact diagrams cannot capture all of the higher order thermal correlators. As we shall argue shortly, contact Witten diagrams provide no additional contributions to the on-shell action that could not be obtained by adequate generalization of the sources, just as we observed for the Gaussian action.

Unlike vacuum correlators, thermal correlators are constrained by the KMS condition, which ensures that all thermal three-point functions can be captured by the standard Schwinger-Keldysh contour. Additional time-folds — and the full structure of $\grSKI{n}$— are only needed for four-point functions and higher. The collection of independent correlators for a given $n-$fold can be captured using the language of spectral functions, as defined in \cite{Chaudhuri:2018ymp}.

For two- and three-point functions, all the thermal correlators can be written in terms of a single spectral function: $\expval{[\Op_1,\Op_2]}_{\beta}$, and $\expval{[[\Op_1,\Op_2],\Op_3]}_{\beta}$, respectively. However, for the case of four-point functions we have not one but two spectral functions: the nested commutator, $\expval{[[[\Op_1,\Op_2],\Op_3],\Op_4]}_{\beta}$, and the commutator-squared $\expval{[\Op_1,\Op_2][\Op_3,\Op_4]}_{\beta}$. 

Guided by this structure, let us compute the contact diagram contributions to $\expval{[\Op(t),\Op(0)]^2}_{\beta}$, which contains contribution from genuine OTOCs. In general, this correlator can be computed using the 4-fold contour as a specific permutation of the basic contour-ordered correlator.
\begin{equation}\label{eq:Comm2}
\begin{split}
\braket{\left[\Op(t,\vx),\Op(0)\right]^2}_{\beta} &= \int_{k_i} e^{-i(\vw_1+\vw_3)t+i (\vk_1+\vk_3)\cdot\vx}\left[\braket{\TSKI{4}\,\OpRI{1}(k_1)\OpLI{1}(k_2)\OpRI{2}(k_3)\OpLI{2}(k_4)}_{\beta}  \right.\\ 
&\quad 
\left.
- \braket{\TSKI{4}\,\OpRI{1}(k_1)\OpLI{1}(k_2)\OpRI{2}(k_4)\OpLI{2}(k_3)}_{\beta} \right.\\
&\quad 
\left. 
-\braket{\TSKI{4}\,\OpRI{1}(k_2)\OpLI{1}(k_1)\OpRI{2}(k_3)\OpLI{2}(k_4)}_{\beta} \right. \\ 
&\quad
\left.
+ \braket{\TSKI{4}\,\OpRI{1}(k_2)\OpLI{1}(k_1)\OpRI{2}(k_4)\OpLI{2}(k_3)}_{\beta}\right]\,,
\end{split}
\end{equation}
where we find it more convenient to employ the left/right basis. With our choice of frequency and momenta, the first and last contributions correspond to OTOCs, while the second and third are (anti-) time-ordered contributions.

For a $\frac{\lambda}{4!}\vphi^4$ interaction, all four contributions are obtained from the same diagram, just with different permutations of the momentum labels. As argued earlier, all the non-trivial contributions to the correlator can be written  in terms of the following radial integrals
\begin{equation}
\begin{split}
\mF_1(k_1;k_2,k_3,k_4)&=
\lambda\int dr\,\sqrt{-g} e^{-\beta\vw_1\zeta}\Gin(r,\bar{k}_1)\Gin(r,k_2)\Gin(r,k_3)\Gin(r,k_4)\,,\\
\mF_2(k_1,k_2;k_3,k_4)&=
\lambda\int dr\,\sqrt{-g} e^{-\beta(\vw_1+\vw_2)}\Gin(r,\bar{k}_1)\Gin(r,\bar{k}_2)\Gin(r,k_3)\Gin(r,k_4)\,,\\
\mF_3(k_1,k_2,k_3;k_4)&=
\lambda\int dr\,\sqrt{-g} e^{\beta\vw_4}\Gin(r,\bar{k}_1)\Gin(r,\bar{k}_2)\Gin(r,\bar{k}_3)\Gin(r,k_4)\,.\\
\end{split}
\end{equation}
These functions have clear properties under permutations of the momentum labels, which we emphasize by the use  of a semi-colon in the arguments. This kind of integrals have been studied for the case of the BTZ black hole in \cite{Loganayagam:2022zmq} or for higher dimensional black holes in a gradient expansion \cite{Loganayagam:2022teq,Rangamani:2023mok}. 

The contact contributions to the 4-point correlator are then
\small
\begin{equation}
\begin{split}
\braket{\TSKI{4}\,\Op(k_1)\Op(k_2)\Op(k_3)\Op(k_4)}_{\beta} &=
    \left(e^{\beta\vw_1}-1\right)\left(\mF_1(k_1;k_2,k_3,k_4)-\mF_3(k_2,k_3,k_4;k_1)\right) \\ 
&\quad
    + e^{\beta\vw_1}\left(e^{\beta\vw_2}-1\right)\left(\mF_1(k_2;k_1,k_3,k_4)- \mF_3(k_1,k_3,k_4;k_2)\right)\\ 
&\quad
    +e^{\beta(\vw_1+\vw_2)}\left(e^{\beta\vw_3}-1\right)\left(\mF_1(k_3;k_1,k_2,k_4)-\mF_3(k_1,k_2,k_4;k_3)\right) \\ 
&\quad
    +\left(1-e^{-\beta\vw_4}\right)\left(\mF_1(k_4;k_1,k_2,k_3)-\mF_3(k_1,k_2,k_3;k_4)\right)\\
&\quad 
    +\left(1-e^{\beta(\vw_1+\vw_2)}\right)\left(\mF_2(k_1,k_2;k_3,k_4)-\mF_2(k_3,k_4;k_1,k_2)\right) \\ 
&\quad
    +\left(e^{-\beta\vw_4}-e^{\beta\vw_1}\right)\left(\mF_2(k_1,k_4;k_2,k_3) - \mF_2(k_2,k_3;k_1,k_4)\right)\\ 
&\quad 
    +\left(1+e^{\beta\vw_1}(e^{\beta\vw_2}-1)-e^{-\beta\vw_4}\right)\left(\mF_2(k_1,k_3;k_2,k_4) - \mF_2(k_2,k_4;k_1,k_3)\right)\,,
\end{split}
\end{equation}
\normalsize
and from this expression we may compute $\braket{\left[\Op(t,\vx),\Op(0)\right]^2}_{\beta}$ according to \eqref{eq:Comm2}. 

While the full expression is somewhat intricate, the symmetry properties of the functions $\mF_i$ under permutations lead to a cancellation of all terms
\begin{equation}
\braket{\left[\Op(t,\vx),\Op(0)\right]^2}_{\beta}|_{\text{contact}} =0\,.
\end{equation}

We conclude that contact diagrams yield no contribution to $\braket{\left[\Op(t,\vx),\Op(0)\right]^2}_{\beta}$. They fail to probe the second spectral function and thus cannot capture the genuinely out-of-time-ordered nature of higher-point thermal correlators. 
More generally, all contact diagrams contributions to $\braket{\left[\Op(t,\vx),\Op(0)\right]^n}_{\beta}$ vanish due to the properties of the diagrams under permutations of the momentum labels.\footnote{The technical reasoning can be found in Appendix \ref{app:higherCommutators}, especially equation \eqref{eq:commutatorOnSameLeg}. }

We can provide a heuristic geometric argument as to why contact diagrams do not capture all the information contained in the $\grSKI{n}$ geometry and, in particular, the OTOC, which is part of the commutator-squared. If we consider equation \eqref{eq:OnShell3}, we see that as we go from the first to the second line, the interactions resulting from a contact diagram occur piecewise on the geometric contour, i.e., we have no interactions relating two separate Lorentzian segments of the full geometry. Thus, contact diagrams reproduce the same contributions already captured by the simpler $\grSKI{2}$ geometry, as we saw directly in the calculation before; this is further supported by the observation before that diagrams of non-neighboring sources vanish. In order to obtain a non-trivial contribution to the commutator-squared, we must turn our attention to interactions that relate different Lorentzian segments, that is, we consider exchange diagrams. 

It is important to remark that the statement of no contact diagram contributions to the OTOC is constrained only to our proposed bulk geometry. This geometry is certainly an extrema of the gravitational action, and it is consistent with the constraints of unitarity and KMS, as we have showed before. However, we do not claim it to be the dominant saddle. It may be possible that there exist a different bulk solution where the contact diagrams do contribute to the OTOC.

\subsection{The bulk-to-bulk propagator and exchange Witten diagrams}\label{subsec:witten_exchange}

The new ingredient we require to compute exchange Witten diagrams is the bulk-to-bulk propagator. For the case of $\grSKI{2}$, the bulk-to-bulk propagator was computed in \cite{Loganayagam:2022zmq} in the language of a single, complexified, radial coordinate. Here we will generalize this result for the $\grSKI{n}$ geometry in a piece-wise fashion using the same matrix notation implemented to describe the bulk-to-boundary propagators, \eqref{eq:BtoBdyVecLR} and \eqref{eq:BtoBdyVecFP}. 

On any of the Lorentzian segments, the bulk-to-bulk propagator for minimally coupled scalars is defined as a solution to the inhomogeneous equation
\begin{equation}\label{eq:BtoBDef}
\left(-\Box + m^2\right)\Gbb(X,X') = \frac{1}{\sqrt{-g}}\delta^{(d+1)}(X-X')\,,
\end{equation}
where the differential operator acts on the unprimed coordinates. For future applications, we recall that the propagator is not symmetric under $X\leftrightarrow X'$, due to the lack of translational symmetry in the radial direction. As in the case of the bulk-to-boundary propagator, we will solve this equation in the Fourier domain.

It is well-known that the inhomogeneous equation \eqref{eq:BtoBDef} can be solved in terms of functionally independent solutions to the homogeneous equation \eqref{eq:EoMIn} using the so-called \textit{Wronskian trick}. The bulk-to-bulk propagator is then given by
\begin{equation}\label{eq:BtoBpreSol}
\Gbb(r,r';k) = \frac{1}{W(r')}\left[\theta(r-r')u_1(r)u_2(r') + \theta(r'-r)u_1(r')u_2(r)\right]\,,
\end{equation}
where  $\theta(r)$ is the Heaviside step function and $u_{1,2}(r)$ are functionally independent solutions to the homogeneous equation, whose Wronskian is $W(r') = \{u_1,u_2\}(r')$.

We can easily show by direct calculation that the Wronskian obeys the first order differential equation
\begin{equation}
r^2 f(r)\frac{d\,W}{dr} - 2i\vw W(r) = 0\,,
\end{equation}
with solution $W(r) = \mathcal{N}^{-1}(k)e^{-\beta\vw\zeta(r)}$\footnote{The normalization constant $\mathcal{N}(k)$ is chosen so that the solution of the differential equation agrees with the choice of homogeneous solutions, see \cite{Loganayagam:2022zmq}.}. In this analysis, we only need the fact that $\zeta(r)$ is a solution to the equation \eqref{eq:zetaDef}, and we do not need to specify its analytic continuation across the horizon; we shall indicate this later.

At this point, the solutions $u_{1,2}(r)$ could be some linear combination of $\Gin$ and $\Gout$. In order to determine the specific expressions, we require the bulk-to-bulk propagator to be fully normalizable in both $r$ and $r'$. Due to the step function, this implies the function $u_{1}(r)$ must be proportional to the normalizable linear combination $\mg_3 = \Gin-\Gout$, while $u_2(r)$ remains unconstrained. In order to guarantee functional independence, we may take $u_2$ to be a non-normalizable combination such as $\mg_1$ or $\mg_2$, defined in \eqref{eq:gCombo}. 

Since the analysis above holds whenever both points lie within the same Lorentzian segment, we can write
\begin{equation}\label{eq:BtoBDiag}
\begin{split}
\Gbb_{aa}(r,r',k) = \mathcal{N}(k)e^{\beta\vw\zeta(r')}\nB(\vw)^2&\left[\theta(r-r')\left(\Gin(r,k)-\Gout(r,k)\right)u_a(r') \right. \\ 
&
\left.
+ \theta(r'-r)\left(\Gin(r',k)-\Gout(r',k)\right)u_a(r)\right]\,,
\end{split}
\end{equation}
with $a=1,2,\dots, n$. The functions $u_a(r)$ are some choice of non-normalizable combination of $\Gin$ and $\Gout$, while the prefactor of $\nB(\vw)^2$ is introduced for convenience.

Before we determine the precise combinations $u_a(r)$, let us consider the connecting points of the bulk-to-bulk propagator, which lie in different Lorentzian segments. In this case, the right-hand side of \eqref{eq:BtoBDef} is always zero and, requiring the propagator to be fully normalizable, we have
\begin{equation}\label{eq:BtoBNonDiag}
\Gbb_{ab}(r,r';k) = \mathcal{N}(k)e^{\beta\vw\zeta(r')}\nB(\vw)^2 c_{ab}(\vw)\left(\Gin(r,k)-\Gout(r,k)\right)\left(\Gin(r',k)-\Gout(r',k)\right)\,,
\end{equation}
for $a\neq b$. The $c_{ab}(\vw)$ are at this point undetermined normalization coefficients, to be fixed by consistency with unitarity and KMS.

We can combine the results in \eqref{eq:BtoBDiag} and \eqref{eq:BtoBNonDiag} into a single matrix expression
\begin{equation}\label{eq:BtoBMatrix}
\begin{split}
\GbbM(r,r';k) =
\mathcal{N}(k)e^{\beta\vw\zeta(r')}\nB(\vw)^2&\left[\theta(r-r')\left(\left(\Gin(r,k)-\Gout(r,k)\right)\Mbb(r',k)\right) \right. \\ 
&
\left.
 + \theta(r'-r)\left(\left(\Gin(r',k)-\Gout(r',k)\right)\Mbb(r,k)\right)
\right]\,,
\end{split} 
\end{equation}
where $\Mbb(r,k)$ is an $n\times n$ matrix whose diagonal components are the functions $u_a(r,k)$ while its off-diagonal entries are the the coefficients $c_{ab}(\vw)\mg_{3}(r,k)$\,.

In order to determine the matrix $\Mbb(r,k)$, we need additional conditions. Following the derivation of the bulk-to-bulk propagator in \cite{Loganayagam:2022zmq}, we require that the contributions of exchange diagrams to the on-shell action satisfy the unitarity and KMS conditions. As we have seen, these conditions are easily implemented on the generalized F/P basis, where they imply that there are no contributions of the form $\JKMS^m$ or $\JdI{i}^m$. 

Consider, for instance, the $\JdI{1}^4$ contribution to the on-shell action in a theory with cubic interactions, arising from an exchange Witten diagram. This is given by
\small
\begin{equation}
\begin{split}
S^{(4)}_{\JdI{1}^4} &= \lambda^2 \sum_{a,b}\int_{\rH}^{\infty} dr \sqrt{-g} \int_{\rH}^{\infty} dr' \sqrt{-g}\, \Gbdy{\text{d}_1}_{aa}(r,k_1) \Gbdy{\text{d}_1}_{aa}(r,k_2) 
 \GbbM_{ab}(r,r';k_1+k_2)
\Gbdy{\text{d}_1}_{bb}(r',k_3)\Gbdy{\text{d}_1}_{bb}(r',k_4)\,,
\end{split}
\end{equation}
\normalsize
where we omitted both the momentum integrals and the momentum-preserving delta-function.
The bulk-to-boundary propagators $\Gbdy{}$ were defined in \eqref{eq:BtoBdyVecLR}. 
Similar expressions can be written for all other $\JdI{i}^4$ and $\JKMS^{4}$. 

Once again, it will be convenient to write this expression as a trace over a matrix product, just as we did with the contact diagrams. To this end, we introduce a matrix $\Jm$, which is unity on all its entries
\begin{equation}\label{eq:JMatrix}
\Jm_{ab} = 1\,.
\end{equation}
In terms of this expression, we may write
\small
\begin{equation}\label{eq:4DiagramM}
\begin{split}
S^{(4)}_{\JdI{1}^4} &= \lambda^2 \int_{\rH}^{\infty} dr \sqrt{-g} \int_{\rH}^{\infty} dr' \sqrt{-g}\,\Tr\left(\Jm\, \Gbdy{\text{d}_1}(r,k_1) \Gbdy{\text{d}_1}(r,k_2) 
 \GbbM(r,r';k_1+k_2)
\Gbdy{\text{d}_1}(r',k_3)\Gbdy{\text{d}_1}(r',k_4)\right)\,.
\end{split}
\end{equation}
\normalsize
It is then a somewhat tedious but straightforward exercise to demand that all the exchange diagrams satisfy the constraints of unitarity and the KMS condition. This allows us to determine the matrix $\Mbb$ to be
\begin{equation}\label{eq:BtoB4FoldAux}
\Mbb(r,k) = 
\begin{pmatrix}
-\mg_1 & e^{\beta\vw}\mg_3 & -e^{\beta\vw}\mg_3 & e^{\beta\vw}\mg_3\\ 
\mg_3 & -\mg_2 & e^{\beta\vw}\mg_3 & - e^{\beta\vw}\mg_3\\ 
-\mg_3 & \mg_3 & -\mg_1& e^{\beta\vw}\mg_3\\ 
 \mg_3 & -\mg_3 & \mg_3 & -\mg_2
\end{pmatrix}\,,
\end{equation}
where the $\mg_i$ were defined in \eqref{eq:gCombo}.

Notice that we accounted for the orientation sign within the definition of this matrix instead of writing an explicit dependence on the matrix $\hat{S}$, as we did for the contact diagrams.   

We now have all the ingredients required to compute the exchange diagram contribution to the commutator-squared \eqref{eq:Comm2}. In terms of our notation, the basic contour ordered correlators are
\small
\begin{equation}\label{eq:4PointExchange}
\begin{split}
\braket{\TSKI{4}\OpRI{1}(k_1)\OpLI{1}(k_2)\OpRI{2}(k_3)\OpLI{2}(k_4)} &= \\
&\hspace{-4.5cm}
\lambda^2\left(1+e^{2\beta(\vw_1+\vw_2)}\right)\int_{r,r'}\Tr\left(\Jm \Gbdy{\text{R}_1}(r,k_1)\Gbdy{\text{L}_1}(r,k_2)\GbbM(r,r',k_1+k_2) \Gbdy{\text{R}_2}(r',k_3) \Gbdy{\text{L}_2}(r',k_4)\right) \\ 
&\hspace{-4.5cm} 
+
\lambda^2\left(1+e^{2\beta(\vw_1+\vw_3)}\right) \int_{r,r'}\Tr\left(\Jm \Gbdy{\text{R}_1}(r,k_1)\Gbdy{\text{L}_1}(r',k_2)\GbbM(r,r',k_1+k_3)\Gbdy{\text{R}_2}(r,k_3)\Gbdy{\text{L}_2}(r',k_4)\right)\\ 
&\hspace{-4.5cm} 
+
\lambda^2\left(1+e^{2\beta(\vw_1+\vw_4)}\right) \int_{r,r'}\Tr\left(\Jm \Gbdy{\text{R}_1}(r,k_1)\Gbdy{\text{L}_1}(r',k_2)\GbbM(r,r',k_1+k_4)\Gbdy{\text{R}_2}(r',k_3)\Gbdy{\text{L}_2}(r,k_4)\right)\,.
\end{split}
\end{equation}
\normalsize

In this expression, we have the three familiar contributions from the s, t, and u channels. However, as we mentioned earlier, the bulk-to-bulk propagator is not symmetric in $r\leftrightarrow r'$ due to a lack of translational symmetry and, in principle, this produces three additional contributions. We account for these additional channels by employing the identity
\begin{equation}\label{eq:BtoBPropId}
\GbbM_{ab}(r,r',k) = e^{-2\beta\vw}\GbbM_{ba}(r',r,-k)\,.
\end{equation}

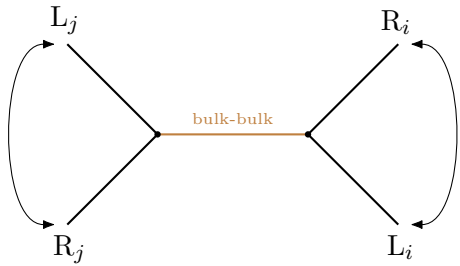
\begin{figure}[H]
\centering
\def\tkzscl{0.4}
\begin{tikzpicture}[baseline={([yshift=-.5ex]current bounding box.center)},vertex/.style={anchor=base,
    circle,fill=black!25,minimum size=18pt,inner sep=2pt},scale=\tkzscl]
    \coordinate[label=above:$\text{L}_j$\,] (u_m2) at (-23,3);
     \coordinate[label=above:$\text{R}_i$\,] (u_m1) at (-12,3);
        \coordinate[label=below:\,$\text{R}_j$] (d_m2) at (-23,-3);
          \coordinate[label=below:\,$\text{L}_i$] (d_m1) at (-12,-3);

        \draw[thick] (u_m2) -- (-20,0);
        \draw[thick] (u_m1) -- (-15,0);
      
        \draw[thick] (d_m2) -- (-20,0);
        \draw[thick] (d_m1) -- (-15,0);

        \draw[thick,  brown] (-15,0) -- node[above] {\tiny bulk-bulk} (-20, 0);
        \fill (-20, 0) circle (3pt); 
        \fill (-15, 0) circle (3pt); 
       
        \draw[{Latex[round]}-{Latex[round]},solid, shorten >= 5pt,shorten <= 5pt] (u_m2) to[out=180,in=180, looseness=1] (d_m2);
        \draw[{Latex[round]}-{Latex[round]},solid, shorten >= 5pt,shorten <= 5pt] (u_m1) to[out=0,in=0, looseness=1] (d_m1); 
    \end{tikzpicture}\:
    \caption{Witten diagram for the s-channel contribution to the correlator. The arrows indicate the necessary permutations to compute the spectral function, as in \eqref{eq:Comm2}.}
    \label{fig:sChannel}
\end{figure}

Just as in the case of the contact diagrams, the explicit expression of \eqref{eq:4PointExchange} is rather complicated and not very illustrative, but we observe a large amount of simplifications when we evaluate not \eqref{eq:4PointExchange} but the linear combination \eqref{eq:Comm2}. Concretely, with our choice of frequency and momenta, the only contributing diagram is the s-channel and its permutations (see fig.\,\ref{fig:sChannel}).

Furthermore, the contribution from this channel has a special feature: the radial integrals factorize\footnote{Of course, we do not observe a full factorization of the diagram, which would lead to a trivial result. The two factors are still connected by their dependence on the momenta.}:
\begin{equation}\label{eq:ExcC2}
\begin{split}
\braket{\left[\Op(t,\vx),\Op(0)\right]^2}_{\beta} &= \int_{k_i} e^{-i(\vw_1+\vw_3)t+i(\vk_1+\vk_3)\cdot\vx}\,
\nB^2(\vw_1+\vw_2) e^{\beta(\vw_1+\vw_2)}\mathcal{N}(k_1+k_2) \\ 
&\quad 
\times
 \left(1+e^{2\beta(\vw_1+\vw_2)}\right)\mW(k_1,k_2)\mW(k_3,k_4)\,,
\end{split}
\end{equation}
where we define 
\begin{equation}\label{eq:WInt}
\begin{split}
\mW(k_1,k_2) &= \int_{r}\left(\Gin(r,k_1)\Gout(r,k_2)-\Gin(r,k_2)\Gout(r,k_1)\right)\\ 
&\quad
\times \left(\Gin(k_1+k_2,r)-\Gout(k_1+k_2,r)\right)\,.
\end{split}
\end{equation}

This factorization is not common in the calculation of exchange Witten diagrams, due to the presence of the step functions $\theta(r-r')$. In the case the commutator-squared, the specific linear combination \eqref{eq:Comm2} is such that only the off-diagonal contributions in the bulk-to-bulk propagator contribute leading to the factorization. This also explains why the contact diagrams do not contribute as we may think of them as exchanges with a purely diagonal bulk-to-bulk propagator of the form $\hat{S}\,\delta(r-r')$. Relatedly, in the heuristic geometric argument presented early, we need an object that relates distinct branches of the $\grSKI{4}$ geometry, and this is precisely the role of the off-diagonal components in the bulk-to-bulk propagator.

While we used the example of a minimally coupled scalar for concreteness, the results obtained so far are very general. In particular, the structure of the bulk-to-bulk propagator follows from basic principles, such as unitarity and KMS, and it holds for more general fields. The information regarding which specific system we study is contained entirely in the form of the functions $\Gin(r,k)$. The integrals \eqref{eq:WInt} are of a similar form as those studied in \cite{Loganayagam:2022zmq}, and may be directly evaluated for the toy model studied there (see Appendix \ref{app:toy_model}).

We notice that a similar factorization channel has been argued for in \cite{Meltzer:2020qbr}. There the argument was carried out within the context of the 2-fold contour for the vacuum state, by evaluating the so-called causal commutator, which could then be analytically continued into the spectral function we studied here.

To see that a correlation function of the form \eqref{eq:ExcC2} cannot be obtained from the grSK$_2$ geometry, we can take the full antisymmetrization of that object in frequency space and observe that it does not vanish. As there are only two different Schwinger-Keldysh labels in the two-fold geometry, the full anti-symmetrization of any 4-point function vanishes. 

\subsection{Higher order spectral functions: a conjecture}\label{subsec:higher_order_spectral_functions}
The discussion above may be extended to higher-order thermal correlators. In particular, we may consider $\expval{[\Op(t),\Op(0)]^n}_{\beta}$. In the Fourier domain, this higher-order spectral function may be computed along the same lines as \eqref{eq:Comm2}, as a permutation of a basic $2n$-point function, captured holographically by $\grSKI{2n}$:
\begin{equation}\label{eq:commn}
\expval{\TSKI{2n}\OpRI{1}(k_1)\OpLI{1}(k_2)\cdots\OpRI{n}(k_{2n-1})\OpLI{n}(k_{2n})} \pm \left(k_{2j +1} \leftrightarrow k_{2j+2}\right)\,,
\end{equation}
where $\pm \left(k_{2j +1} \leftrightarrow k_{2j+2}\right)$ denote all permutations generated by $k_{2j +1} \leftrightarrow k_{2j+2}$ for $j \in \mathds{N}_0$ and the sign is given by the parity of the permutation.

In agreement with our previous heuristic argument, it is easy to see that no contact diagrams contribute to this correlator. Furthermore, since we expect such a correlator must \textit{explore} the full geometry, then we conjecture that
\begin{align}
\label{eq:TheConjectureMain}
	\braket{\left[\Op(t,\vx),\Op(0)\right]^n}_{\beta}|_{\text{Diagrams with at most $n-2$ internal lines}} &= 0\qquad \text{for all }n\, ,  
\end{align}
for tree-level diagrams.

\begin{figure}[H]
\centering
\def\tkzscl{0.4}
\begin{tikzpicture}[baseline={([yshift=-.5ex]current bounding box.center)},vertex/.style={anchor=base,
    circle,fill=black!25,minimum size=18pt,inner sep=2pt},scale=\tkzscl]
    \coordinate[label=above:$\text{L}_j$\,] (u_m2) at (-20,3);
     \coordinate[label=above:$\text{R}_i$\,] (u_m1) at (-15,3);
           \coordinate[label=above:$\text{R}_k$\,] (u_0) at (-10,3);
        \coordinate[label=above:$\text{L}_m$\,] (u_1) at (-5,3);
        \coordinate[label=above:$\text{L}_q$\,] (u_2) at (0,3);
        \coordinate[label=below:\,$\text{R}_j$] (d_m2) at (-20,-3);
          \coordinate[label=below:\,$\text{L}_i$] (d_m1) at (-15,-3);
       \coordinate[label=below:\,$\text{L}_k$] (d_0) at (-10,-3);
        \coordinate[label=below:\,$\text{R}_m$] (d_1) at (-5,-3);
        \coordinate[label=below:\,$\text{R}_q$] (d_2) at (0,-3);

        \draw[thick] (u_m2) -- (-20,0);
        \draw[thick] (u_m1) -- (-15,0);
        \draw[thick] (u_0) -- (-10,0);
        \draw[thick] (u_1) -- (-5,0);
        \draw[thick] (u_2) -- (0,0);
        \draw[thick] (d_m2) -- (-20,0);
        \draw[thick] (d_m1) -- (-15,0);
        \draw[thick] (d_0) -- (-10,0);
        \draw[thick] (d_1) -- (-5,0);
        \draw[thick] (d_2) -- (0,0);
        
        \draw[thick,  brown] (-17,0) -- node[above] {\tiny bulk-bulk} (-20, 0);
        \draw[thick,  brown] (-12,0) -- node[above] {\tiny bulk-bulk} (-15, 0);
        \draw[thick,  brown] (-7,0) -- node[above] {\tiny bulk-bulk} (-10, 0);
        \draw[thick,  brown] (-2,0) -- node[above] {\tiny bulk-bulk} (-5, 0);
\draw[thick,  brown, dotted] (0,0) -- (1.5, 0);
 \draw[thick,  brown] (-17,0) --   (-15, 0);
 \draw[thick,  brown] (-12,0) --  (-10, 0);
 \draw[thick,  brown] (-7,0) --   (-5, 0);
 \draw[thick,  brown] (-2,0) --   (-0, 0);        
        \fill (-20, 0) circle (3pt); 
        \fill (-15, 0) circle (3pt); 
        \fill (-10, 0) circle (3pt); 
        \fill (-5, 0) circle (3pt); 
        \fill (0, 0) circle (3pt);
        
        \draw[{Latex[round]}-{Latex[round]},solid, shorten >= 5pt,shorten <= 5pt] (u_m2) to[out=180,in=180, looseness=1] (d_m2);
        \draw[{Latex[round]}-{Latex[round]},solid, shorten >= 5pt,shorten <= 5pt] (u_m1) to[out=180,in=180, looseness=1] (d_m1);
        \draw[{Latex[round]}-{Latex[round]},solid, shorten >= 5pt,shorten <= 5pt] (u_0) to[out=180,in=180, looseness=1] (d_0);
        \draw[{Latex[round]}-{Latex[round]},solid, shorten >= 5pt,shorten <= 5pt] (u_1) to[out=180,in=180, looseness=1] (d_1);
        \draw[{Latex[round]}-{Latex[round]},solid, shorten >= 5pt,shorten <= 5pt] (u_2) to[out=180,in=180, looseness=1] (d_2);
    \end{tikzpicture}\:
\caption{Witten diagram for non-vanishing tree-level contribution to \eqref{eq:commn}. The arrows indicate the permutations of the external SK labels.}
    \label{fig:nonVanishing}
\end{figure}
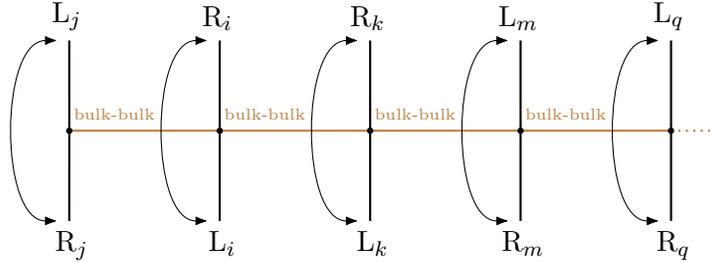

In Appendix \ref{app:higherCommutators} we provide a partial proof of this conjecture by considering a subset of the possible diagrams, those where each vertex has at most two internal propagators. 
An example of a diagram of that form with $n-1$ internal lines contributing to the higher order commutator is the one depicted in fig.\,\ref{fig:nonVanishing}. According to the conjecture \eqref{eq:TheConjectureMain}, this is the lowest order for which a diagram can contribute. 

This diagram generalizes the s-channel contribution we studied for the commutator-squared, and remarkably, it exhibits the same factorization property. We postpone the study of more general tree-level diagrams (those with three or more internal propagators per vertex) for future work.


\section{Discussion}\label{sec:discussion}

We have proposed a holographic dual for a generalized Schwinger-Keldysh contour, in the form of a multi-sheeted geometry which we denote as $\grSKI{n}$. We described the general procedure for computing higher-order real-time correlators in terms of Witten diagrams in this geometry. Our construction generalizes the previous contour prescriptions applied to capture the holographic dual of the standard Schwinger-Keldysh contour \cite{Glorioso:2018mmw,Jana:2020vyx} and, in particular, it allows for a direct calculation of fully out-of-time order correlators. While we focus much of our attention on the case of the 4-fold geometry, our procedure applies to an arbitrary number of time-folds. Similarly, for the sake of simplicity, we considered only the dynamics of probe scalars, but more general observables, such as the correlation functions of conserved currents, can be easily incorporated along the lines studied in \cite{Loganayagam:2022zmq}. The Gaussian sector of these probe scalars in $\grSKI{n}$ reproduces the familiar results of $\grSKI{2}$.

We would like to highlight some specific features of our constructions and the main results we are able to deduce from them.

\textbf{Piecewise geometry and analytic continuation.} 
The $\grSKI{n}$ geometry consists of a collection of Lorentzian geometries, as well as a Euclidean section preparing the initial state of the system. A remarkable feature of our construction is its simplicity, with the only required building blocks being the ingoing bulk-to-boundary propagator and the corresponding bulk-to-bulk propagator; the latter being required to compute exchange Witten diagrams. From these ingredients, we are able to set up the calculation of any $n$-point correlator. The non-trivial nature of the geometry emerges only at the level of the analytic continuation across the different Lorentzian geometries and can be seen as the core feature of our prescription. We show that the analytic continuation we propose is highly constrained by the physical requirements of microscopic unitarity and the KMS condition, which translate into specific monodromies for the non-analytic bulk-to-boundary propagators. 

\textbf{Higher order spectral functions.} 
Despite the few building blocks we require in our prescription, we show that new independent observables emerge as we consider higher-order correlators and the corresponding multitude of inequivalent time orderings. Concretely, we show how our prescription allows for the calculation of spectral functions of the form $\expval{[\Op(t),\Op(0)]^{n/2}}_{\beta}$. In order to obtain non-trivial contributions to these observables, it is required to study theories beyond the Gaussian regime. Furthermore, we showed that contact Witten diagrams fail to probe this spectral function, so genuine OTO content originates from exchange diagrams that couple distinct Lorentzian segments. We demonstrated this directly for $\expval{[\Op(t),\Op(0)]^{2}}_{\beta}$ in the $\grSKI{4}$ geometry, and proposed a conjecture as to the behavior of the spectral functions $\expval{[\Op(t),\Op(0)]^{n/2}}_{\beta}$.

\textbf{Factorization channels.} 
In addition to obtaining non-trivial spectral functions, we also showed that the calculation of $\expval{[\Op(t),\Op(0)]^{2}}_{\beta}$ has a factorization channel, along the lines explored in \cite{Meltzer:2020qbr}. In principle, the exchange diagram contribution to an arbitrary 4-point function involves two coupled radial integrals, but for the case of the spectral function, these two radial integrals factorize because, in the commutator-squared combination, only the off-diagonal entries of the bulk-to-bulk propagator contribute, as depicted in \eqref{eq:ExcC2}. We conjecture that a similar behavior occurs for higher order spectral functions: a large enough number of exchanges must be considered in order to lead to non-trivial results, but a simple factorization of the radial integrals is present. We support this conjecture by the study of a subset of the required Witten diagrams, but postpone the detailed proof for future work. 

\subsection{Outlook}
Finally, we mention some interesting generalizations of our construction and future research directions.

\textbf{Lorentzian signature and complex geometries.} 
Here, we studied only thermal correlators, but by considering more general complex geometries using a similar piecewise construction, we could also access additional related observables. These include $\Tr[\rho^{\frac{1}{4}}_{\beta}\mathcal{O}\rho^{\frac{1}{4}}_{\beta}\mathcal{O}\rho^{\frac{1}{4}}_{\beta}\mathcal{O}\rho^{\frac{1}{4}}_{\beta}\mathcal{O}]$, correlation functions with respect to the thermo-field double state,  as well as other multi-sheeted geometries associated with the quantum information theory description of black holes \cite{Magan:2025hce}. Relatedly, we describe the $\grSKI{n}$ geometry in a piecewise fashion, but this specific (non-)analytic behavior suggests the possibility of reformulating the geometry as a complexified manifold over an $n$-sheeted Riemann surface, as has been done for the two-fold geometry \cite{Jana:2020vyx}. It would be desirable to connect our analysis to the approach developed in \cite{Loganayagam:2024mnj}, in which Witten diagrams on the complexified $\grSKI{2}$ geometry were studied.

\textbf{Analytic continuations and gauge redundancies.} As noted above, the key nontrivial ingredient in our construction is given by the analytic continuation between Lorentzian segments. While in the main text we chose a specific form of the analytic continuation, which is consistent with unitarity and KMS invariance, other choices are possible (see Appendix \ref{app:grSK2}). For the case of the Schwinger-Keldysh contour, two-fold, the unitarity and KMS constraints have been elegantly formulated in terms of a gauge redundancy \cite{Haehl:2016pec, Haehl:2016uah}. It would be highly desirable to extend this analysis for a large number of time-folds and study a possible relation between the corresponding gauge redundancy and the choice of analytic continuation.

\textbf{The thermal bootstrap program.} 
It will be interesting to study whether the special behavior of the higher-order Witten diagrams for spectral functions we observed here leads to applications to the thermal conformal bootstrap program. The main idea is to use the results from the Witten diagram calculations as data to match with the corresponding OPE expansion, as explored recently in the case of the two-point function \cite{Buric:2025fye}, or the equivalent Lorentzian inversion formula.

\textbf{OTOCs and chaotic dynamics.} 
This project is deeply motivated by the study of quantum chaos, as described by the exponential growth of the out-of-time-order correlator \cite{Shenker:2013pqa}. We have shown that our proposed dual geometry indeed captures a non-trivial spectral function which, while still constructed in terms of the simple data of the ingoing bulk-to-boundary propagator, is distinct from similar four-point functions studied within the context of the two-fold geometry \cite{Loganayagam:2022zmq}. Furthermore, our formalism allows for the calculation of higher-order observables, which allows for a more detailed description of the dynamics of the theory \cite{Vardhan:2025rky}. We do not directly show the emergence of exponential growth, as we expect this to be related to the presence of a long-lived mode, such as is observed in systems with a shift symmetry \cite{Blake:2017ris} or in the case of soft modes due to a specific symmetry-breaking process \cite{Haehl:2018izb}. As noticed earlier, it is clear how to extend our work in order to include these more general fields associated to long-lived modes, and it will be the subject of future research.

\textbf{Complex geometries and the gravitational path integral.}
While this work is deeply rooted within the context of the AdS/CFT correspondence, it can be seen in a wider context as part of the study of the real-time gravitational path integral. The $\grSKI{n}$ geometry we construct here is a saddle-point of this integral, and it shows that multi-sheeted complex geometries play a crucial role in understanding Lorentzian observables. While the fact that our geometry is a saddle of the gravitational path integral is true by construction, we make no claim as to whether it is the dominant saddle. A wider study of the path integral, including more general geometries and topology-changing processes, presents an exciting, albeit highly challenging, avenue of research. In this context, we see our study here as providing some sure initial footing.

\section*{Acknowledgements}
We would like to thank  Mike Blake, Felix Haehl, Jakob Hollweck,  Mukund Rangamani, and Vaios Ziogas for their useful discussion. JV would also like to thank the Centro de Ciencias de Benasque for their hospitality during the course of this project. MA and CS thank the Erwin–Schrödinger Institute (ESI) for the hospitality in April 2024 during the program ``Carrollian physics and holography". CS and JV are supported by the Deutsche Forschungsgemeinschaft (DFG) under
Grant No 406116891 within the Research Training
Group RTG 2522/1. 
\appendix

\section{Gravitational Schwinger-Keldysh two fold solution}\label{app:grSK2}
In this Appendix, we rederive the solution for the scalar field in the gravitational Schwinger-Keldysh geometry as stated in \eqref{eq:grSKsol1}. This solution has already been introduced in \cite{Glorioso:2018mmw,Jana:2020vyx}. Here we want to rederive the $\grSKI{2}$ solution using the technique introduced in section \ref{subsec:gaussian}, in which we do not use the complexified radial coordinate but work with each segment individually. 

As described in section \ref{sec:HoloTimeContours}, the grSK geometry consists of two Lorentzian and one Euclidean segments. A crucial aspect of the derivation of the solution for the $\grSKI{4}$ geometry was that we assumed the non-analytic solution $\Gout(r,k)$ to have two different analytic continuations, captured by two distinct monodromies when passing through the future horizon. We denoted them as $G_{\mathrm{out},+}(r,k)$ and $G_{\mathrm{out},-}(r,k)$. 
The monodomies of these functions at the future horizon are defined as
\begin{align}
	G_{\mathrm{out},\pm}(k,r_{\text{R}}) = e^{\pm \beta \omega}G_{\mathrm{out},\pm}(k,r_{\text{L}}), 
\end{align}
where the index on the radial component denotes the Lorentzian segment. 

In principle, this ambiguity on the analytic continuation is also present for the $\grSKI{2}$ solution, which can be written as
\begin{align}
\label{eq:grSK2generalSolution}
	\phi_{\text{R}}(k, r_{\text{R}}) &= a_1\,  \Gin (k, r_{\text{R}}) + a_2 \, G_{\mathrm{out},+}(k,r_{\text{R}}) + a_3\,  G_{\mathrm{out},-}(k,r_{\text{R}})\,,\\
	\phi_{\text{L}}(k, r_{\text{L}}) &= b_1\,  \Gin (k, r_{\text{L}}) + b_2 \, G_{\mathrm{out},+}(k,r_{\text{L}}) + b_3 \, G_{\mathrm{out},-}(k,r_{\text{L}})\,,
\end{align}
on the two Lorentzian segments.
The six coefficients are yet to be determined, in terms of the sources $J_{\text{R}, \text{L}}$, by the boundary and matching conditions. 

As we defined the functions $\Gin, \, G_{\mathrm{out}, \pm}$ to asymptote to $1$ when going to the conformal boundary, the conditions that the fields $\phi_{{\text{R}}, {\text{L}}}$ go to $J_{\text{R},\text{L}}$ is
\begin{equation}
\begin{split}
	a_1 + a_2 + a_3 &= \JR \\
	b_1 + b_2 + b_3 &= \JL\, . 
\end{split}
\end{equation}
On the patch that covers the horizon, all three functions have to agree with their counterpart. This yields the three conditions
\begin{equation}
\begin{split}
	a_1 &= b_1 \\
	e^{\beta \omega }\, a_2 & = b_2 \\
	e^{-\beta \omega }\, a_3 & = b_3\, .
\end{split}
\end{equation}
As we do not assume any sources on the Euclidean segment and the function $\Gin$ has no poles in the upper half-plane, the matching condition at the point of time-reflection symmetry produces
\begin{align}
	a_2 + a_3 = 
  e^{-\beta \omega} (b_2 + b_3)\, . 
\end{align}
This condition precisely takes care of the KMS property of the correlation functions. 

Solving all these conditions yields
\begin{equation}\label{eq:grSK2coeffs}
\begin{split}
	a_1 &= \nB(\vw) \left(e^{\beta \vw} \, \JR - \JL\right)\,,\\
	a_2 &= \nB(\vw) \left(\JL - \JR\right)\,,\\
	a_3 &= 0\,,\\
	b_1 &= \nB(\vw) \left(e^{\beta \vw} \, \JR - \JL\right)\,,\\
	b_2 &= \nB(\vw) \, e^{\beta \vw}\, \left(\JL - \JR\right)\,,\\
	b_3 &= 0\,, 
\end{split}
\end{equation}
with $\nB(\omega)$ defined in \eqref{eq:BoltzFact}. Plugging the coefficients \eqref{eq:grSK2coeffs} into  \eqref{eq:grSK2generalSolution}, yields exactly the solution \eqref{eq:grSKsol1}. 
The fact that the solution does not contain $G_{\mathrm{out}, -}$ allows one to work only with the two solution branches $\Gin(k, r)$ and $G_{\mathrm{out}, +}(k, r)$, where the latter can be identified with $e^{-\beta \omega\zeta}\Gin(\bar{k}, r)$ as in \eqref{eq:grSKsol2}.

Recall that monodromy at the future horizon was a choice in the analytic continuation of $\Gout$. We could as well have chosen more general monodromies, such that instead of $G_{\mathrm{out}, \pm}$ we could have $G_{\mathrm{out}, \alpha_{1, 2}}$, defined as
\begin{align}
\label{eq:DifferentMonodromies}
	G_{\mathrm{out}, \alpha_{1, 2}}(k, r_\text{R}) = e^{\lambda_{1, 2}\omega \beta} G_{\mathrm{out}, \lambda_{1, 2}}(k, r_\text{L})\, . 
\end{align}

In principle, the corresponding solution does not coincide with \eqref{eq:grSK2coeffs}. But we recall that the functions $G_{\mathrm{out}, \lambda_1}$ and $G_{\mathrm{out}, \lambda_2}$ differ only on their analytic continuation across the horizon. Hence, in a coordinate patch away from the horizon, we stop keeping track of this distinction, just as we did in the main text. 
The field solution for the two Lorentzian segments stemming from the ansatz \eqref{eq:DifferentMonodromies} then coincides with \eqref{eq:grSKsol1}. Therefore, the solution \eqref{eq:grSKsol1} is independent of the choice of monodromies. 

The linear dependence of the fields on the sources allows us to write the solution \eqref{eq:grSKsol1} as a bulk-to-boundary matrix, as is done for the 4-fold in \eqref{eq:solMatrixLR}. For the 2-fold, this reads
\begin{align}
\label{eq:grSK2BBdryMatrix}
	\begin{pmatrix}
\phi_1 \\ 
\phi_2 
\end{pmatrix}
=\nB(\vw)
\begin{pmatrix}
e^{\beta\vw}\Gin -\Gout && -\left(\Gin - \Gout\right) \\ 
e^{\beta\vw}\left(\Gin - \Gout\right) && -e^{\beta\vw}\left(\Gin - \Gout\right) 
\end{pmatrix}
\begin{pmatrix}
\JRI{}\\ 
\JLI{}
\end{pmatrix}\,.
\end{align}

This matrix can also be found as on the $2 \times 2$ block-diagonal of the 4-fold analog \eqref{eq:solMatrixLR}. When turning off the sources $(\JRI{1}, \JLI{1})$, $(\JRI{2}, \JLI{2})$, or $(\JRI{1}, \JLI{2})$, and focus on the remaining two segments, one recovers the grSK$_2$ structure. This corresponds to collapsing the empty segments in the Schwinger-Keldysh contour.

\section{A Conjecture on Higher Commutators}
\label{app:higherCommutators}
\subsection{The Conjecture}
In section \ref{subsec:witten_contact},  we have seen that the commutator-squared ($2$-commutator function) does not receive any contribution from the contact diagrams. We conjecture that this statement generalizes to tree-level diagrams computing higher-order spectral functions as \eqref{eq:TheConjectureMain}, i.e., 
\begin{align}
\label{eq:TheConjecture}
	\braket{\left[\Op(t,\vx),\Op(0)\right]^n}_{\beta}|_{\text{Diagrams with at most $n-2$ internal lines}} &= 0\qquad \text{for all }n\, .  
\end{align}

The conjecture is trivially satisfied for connected diagrams with only cubic interactions. To see this, suppose that there are $I \leq n-2$ internal lines. For connected diagrams, the number of vertices for connected diagrams is $0<V \leq I+1$. For a commutator to the power of $n$ ($n$-commutator function), there are $2n$ external legs required, but in a cubic theory, the minimal number of external legs is 
\begin{align}
	2n =3V - 2I\leq V + 2\leq n+1\, . 
\end{align}

We can then consider theories with at least cubic and quartic interactions. In the following, we will show that the conjecture \eqref{eq:TheConjecture} holds for all tree-level diagrams whose internal lines form a chain, that is, diagrams where each vertex is connected to at most two internal lines. An example of this kind of diagram is showed in fig.\,\ref{fig:Cong:example}. 

\begin{figure}[H]
\centering
\def\tkzscl{0.3}
\begin{tikzpicture}[baseline={([yshift=-.5ex]current bounding box.center)},vertex/.style={anchor=base,
    circle,fill=black!25,minimum size=18pt,inner sep=2pt},scale=\tkzscl]
    \coordinate[label=right:$\text{R}_i$\,] (u_1) at (0,3);
        \coordinate[label=right:$\text{L}_j$\,] (u_2) at (8,3);
         \coordinate[label=right:\,$\text{L}_k$] (d_1) at (1,-3);
          \coordinate[label=left:\,$\text{R}_s$] (d_2) at (-1,-3);
        \coordinate[label=right:\,$\text{R}_m$] (d_3) at (8,-3);
         \coordinate[label=right:\,$\text{L}_q$] (u_3) at (17,3);
         \coordinate[label=left:\,$\text{R}_o$] (u_4) at (15,3);
         \coordinate[label=right:\,$\text{L}_t$] (d_4) at (16,-3);

        \draw[thick] (u_1) -- (0,0);
        \draw[thick] (u_2) -- (8,0);
        \draw[thick] (u_3) -- (16,0);
        \draw[thick] (u_4) -- (16,0);
        \draw[thick] (d_1) -- (0,0);
        \draw[thick] (d_2) -- (0,0); 
        \draw[thick] (d_3) -- (8,0);\draw[thick] (d_4) -- (16,0);  
\draw[thick,  brown] (0,0) -- node[above] {\tiny bulk-bulk} (8, 0);    
\draw[thick,  brown] (8,0) -- node[above] {\tiny bulk-bulk} (16, 0);

        \fill (0, 0) circle (3pt); 
        \fill (8, 0) circle (3pt); 
        \fill (16, 0) circle (3pt); 
        \end{tikzpicture}
        \caption{}
        \label{fig:Cong:example}
\end{figure}
An example of a diagram that we do not consider is given in fig.\,\ref{Fig:cong:NotConsidered}. 

\begin{figure}[H]
\centering
\def\tkzscl{0.3}
\begin{tikzpicture}[baseline={([yshift=-.5ex]current bounding box.center)},vertex/.style={anchor=base,
    circle,fill=black!25,minimum size=18pt,inner sep=2pt},scale=\tkzscl]
        \coordinate[] (cu) at (0,5);
        \coordinate[] (cdl) at (-4,-3);
        \coordinate[] (cdr) at (4,-3);
        \coordinate[label=left:\,$\text{L}_i$] (cu_l) at (-2,7);
         \coordinate[label=right:\,$\text{R}_j$] (cu_r) at (2,7);
        \coordinate[label=left:\,$\text{L}_k$] (cdl_l) at (-7,-1);
         \coordinate[label=left:\,$\text{R}_m$] (cdl_r) at (-7,-5);
        \coordinate[label=right:\,$\text{L}_s$] (cdr_l) at (7,-1);
         \coordinate[label=right:\,$\text{R}_o$] (cdr_r) at (7,-5);
            
        \draw[thick] (cu) -- (cu_l);
        \draw[thick] (cu) -- (cu_r);
        \draw[thick] (cdl) -- (cdl_l);
        \draw[thick] (cdl) -- (cdl_r);
        \draw[thick] (cdr) -- (cdr_l);
        \draw[thick] (cdr) -- (cdr_r);
          \draw[thick,  brown] (cu) -- node[left] {\tiny bulk-bulk} (0,0);
        \draw[thick,  brown] (cdl) --node[above left] {\tiny bulk-bulk} (0,0);
        \draw[thick,  brown] (cdr) --node[above right] {\tiny bulk-bulk} (0,0);

        \fill (cu) circle (3pt);
        \fill (cdl) circle (3pt);
        \fill (cdr) circle (3pt);
        \fill (0, 0) circle (3pt); 
    \end{tikzpicture}\,  
    \caption{}
    \label{Fig:cong:NotConsidered}
\end{figure}

In order to prove the conjecture for this kind of diagrams, we consider a diagram with $n-1$ vertices with $2n$ external lines. 
Recall that the external lines carry, in addition to the momentum label, Schwinger-Keldysh species labels (of the form $\text{R}_i$ or $\text{L}_i$ for $i \leq n$), which specify the segment in the grSK$_{2n}$ geometry. 

In order to make statements about the contribution of Witten diagrams, we employ the matrix language of \eqref{eq:4PointExchange}. This extends straightforwardly to arbitrary diagrams of the form that we consider. For example the diagram depicted in fig.\,\ref{fig:Cong:example}, we have a contribution proportional to
\small
\begin{align}
\label{eq:Cong:example}
\begin{split}
	\sim\int_{\rH}^{\infty} \prod\, dr \sqrt{-g} \,\Tr &\left[ \Jm \Gbdy{\text{R}_s}(r_1,k_1) \Gbdy{\text{R}_i}(r_1,k_2)\Gbdy{\text{L}_k}(r_1,k_3)  \GbbM(r_1, r_2) \,\hat{S} \,\Gbdy{\text{L}_j}(r_2,k_4)\Gbdy{\text{R}_m}(r_2,k_5) \right. \\
	& \:\left.\GbbM(r_2, r_3)\,\hat{S}\,\Gbdy{\text{R}_o}(r_3,k_6)\,\Gbdy{\text{L}_q}(r_3,k_7)\,\Gbdy{\text{L}_t}(r_3,k_8)\right]\,,
	\end{split}
\end{align}
\normalsize
where we recall that $\Jm$ is defined in \eqref{eq:JMatrix} and the sign matrix $\hat{S}$ in \eqref{eq:sMatrix}. 
This is a generalization of the notation used in \eqref{eq:4DiagramM}. Note that for each vertex inside the chain, we have to take one sign matrix $\hat{S}$ to account for contour orientation. However, for the first and last vertex, the orientation is already accounted for as part of the definition of the bulk-bulk propagator.

We are interested in showing that the contribution of certain diagrams vanishes in the commutator function. Hence, it suffices to show that the matrix inside the trace vanishes. Moreover, we do not take into account the actual form of the entries but rather the relations among them. Thus, we can replace the bulk-to-bulk propagator $\GbbM$ in the expression \eqref{eq:Cong:example} with the matrix $\Mbb$ defined in \eqref{eq:BtoB4FoldAux}.

The expression for the $n$-commutator function, $\braket{\left[\Op(t,\vx),\Op(0)\right]^n}_{\beta}$ is given by a linear combination of the diagrams that contribute to the $2n$ point function, just as observed for the commutator-squared.  We can thus consider each possible diagram for the $2n$-point function and commute the Schwinger-Keldysh labels $\text{R}_i \leftrightarrow \text{L}_i$ in order to form the corresponding linear combination\footnote{An equivalent description is the commutation of the momentum labels, as we did in \eqref{eq:Comm2}, but it will be beneficial to consider the change of Schwinger-Keldysh labels. }.

\subsection{Significant Subdiagrams}

\label{sec:diagrams}
Recall that we only consider Witten diagrams with, where each vertex is connected to at maximum two internal lines. To prove the conjecture \eqref{eq:TheConjecture} for these diagrams, we consider diagrams with $n-1$ vertices and $2n$ external legs. 
In this subsection, we want to list significant subdiagrams that necessarily occur in these diagrams. In subsection \ref{sec:conj:WhyVanish}, we show why all the diagrams with these subdiagrams give vanishing contributions. 

We will see that our arguments given in subsection \ref{sec:conj:WhyVanish} also apply if we reverse the order of the matrices involved. Hence we do not need to consider diagrams that are the reverse of other diagram separately. 
Moreover, in the diagrams we consider, more external lines can be attached to the vertices. There are simply not depicted, but also do not spoil the arguments.

Let us now consider diagrams with $n-1$ vertices and $2n$ external legs together with the permutations according to the $n$-commutator function \eqref{eq:commn}. It follows from the assumptions that there has to be at least one vertex with at least three external lines. In order to compute the contribution of this diagram to the higher commutator function, the external legs must be permuted with a corresponding sign (see \eqref{eq:Comm2} for the example of the $2$-commutator function). 
The permutation is depicted by arrows in the Figs\,\ref{fig:Conj:4v2c} - \ref{fig:Conj:3v0c1for1rev}. 
We call the external legs, those Schwinger-Keldysh labels are permuted \textit{partner-legs}.

With these requirements, we can then construct the possibly contributing diagrams, starting from that vertex with at least three external legs, as follows.

Consider first a vertex with at least four external legs that are permuted among themselves. This is depicted in fig.\,\ref{fig:Conj:4v2c}. 

\begin{figure}[H]

\centering
\def\tkzscl{0.5}
\begin{tikzpicture}[baseline={([yshift=-.5ex]current bounding box.center)},vertex/.style={anchor=base,
    circle,fill=black!25,minimum size=18pt,inner sep=2pt},scale=\tkzscl]
        \coordinate[label=left:$\text{L}_i$\,] (u_1) at (-1,3);
        \coordinate[label=right:$\text{R}_i$\,] (u_2) at (1,3);
        \coordinate[label=left:\,$\text{R}_j$] (d_1) at (-1,-3);
        \coordinate[label=right:\,$\text{L}_j$] (d_2) at (1,-3);
        
        \draw[thick] (u_1) -- (0,0);
        \draw[thick] (u_2) -- (0,0);\draw[thick] (d_1) -- (0,0);
        \draw[thick] (d_2) -- (0,0);
        \draw[thick,  brown] (0,0) -- node[above] {\tiny bulk-bulk} (3, 0);
        \draw[thick,  brown] (0,0) -- node[above] {\tiny bulk-bulk} (-3, 0);
        \draw[thick,  brown, dotted] (3.5,0) -- (4.5, 0);

        \draw[thick,  brown, dotted] (-3.5,0) -- (-4.5, 0);
        
        \fill (u_1) circle (1pt);
        \fill (u_2) circle (1pt);
        \fill (d_1) circle (1pt);
        \fill (d_2) circle (1pt);
        \fill (0, 0) circle (3pt);

        \draw[{Latex[round]}-{Latex[round]},solid, shorten >= 5pt,shorten <= 5pt] (u_1) to[out=50,in=130, looseness=1] (u_2);
        \draw[{Latex[round]}-{Latex[round]},solid, shorten >= 5pt,shorten <= 5pt] (d_1) to[out=-50,in=-130, looseness=1] (d_2);
    
    \end{tikzpicture}\: 
    \caption{}
    \label{fig:Conj:4v2c}
\end{figure}

There are another kinds of diagrams where two of the four external legs are permuted with the external legs of another vertex, such as in fig.\,\ref{fig:Conj:4v1c}.

\begin{figure}[H]
\centering
\def\tkzscl{0.5}
\begin{tikzpicture}[baseline={([yshift=-.5ex]current bounding box.center)},vertex/.style={anchor=base,
    circle,fill=black!25,minimum size=18pt,inner sep=2pt},scale=\tkzscl]
        \coordinate[label=above:$\text{R}_i$\,] (u_1) at (-1,3);
        \coordinate[label=above:$\text{R}_j$\,] (u_2) at (1,3);
        \coordinate[label=right:\,$\text{L}_j$] (u_3) at (8,3);
       \coordinate[label=below:\,$\text{L}_i$] (d_1) at (-1,-3);
        \coordinate[label=below:\,$\text{R}_k$] (d_2) at (1,-3);
        \coordinate[label=right:\,$\text{L}_k$] (d_3) at (8,-3);
        \draw[thick] (u_1) -- (0,0);
        \draw[thick] (u_2) -- (0,0);\draw[thick] (d_1) -- (0,0);
        \draw[thick] (d_2) -- (0,0);
        \draw[thick] (u_3) -- (8,0);
        \draw[thick] (d_3) -- (8,0);
        \draw[thick,  brown] (0,0) -- node[above] {\tiny bulk-bulk} (3, 0);
        \draw[thick,  brown] (0,0) -- node[above] {\tiny bulk-bulk} (-3, 0);
        \draw[thick,  brown, dotted] (3.5,0) -- (4.5, 0);
        \draw[thick,  brown, dotted] (-3.5,0) -- (-4.5, 0);
\draw[thick,  brown] (5,0) -- node[above] {\tiny bulk-bulk} (8, 0);
    \draw[thick,  brown] (8,0) -- node[above] {\tiny bulk-bulk} (11, 0);    
    \draw[thick,  brown, dotted] (11.5,0) -- (12.5, 0);

        \fill (u_1) circle (1pt);
        \fill (u_2) circle (1pt);
        \fill (d_1) circle (1pt);
        \fill (d_2) circle (1pt);
        \fill (0, 0) circle (3pt);

        \draw[{Latex[round]}-{Latex[round]},solid, shorten >= 5pt,shorten <= 5pt] (u_1) to[out=180,in=180, looseness=1] (d_1);
        \draw[{Latex[round]}-{Latex[round]},solid, shorten >= 5pt,shorten <= 5pt] (u_2) to[out=30,in=130, looseness=1] (u_3);
        
        \draw[{Latex[round]}-{Latex[round]},solid, shorten >= 5pt,shorten <= 5pt] (d_2) to[out=-30,in=-130, looseness=1] (d_3);
    \end{tikzpicture}\,  
    \caption{}
    \label{fig:Conj:4v1c}
\end{figure}

Next, let us consider a vertex with  at least three external legs, where exactly two of them are permuted in the commutator function. Hence there is at least one external leg whose partner leg is located on a different vertex. This leads to four possibilities that are best discribed by diagrams in figs.\,\ref{fig:Conj:3v1c1c}, \ref{fig:Conj:3v1c1foreward}, \ref{fig:Conj:3v1c1Ultrarev} and \ref{fig:Conj:3v1c1rev}. 
\begin{figure}[H]
\centering
\def\tkzscl{0.5}
\begin{tikzpicture}[baseline={([yshift=-.5ex]current bounding box.center)},vertex/.style={anchor=base,
    circle,fill=black!25,minimum size=18pt,inner sep=2pt},scale=\tkzscl]
           \coordinate[label=right:$\text{R}_j$\,] (u_1) at (0,3);
        \coordinate[label=right:\,$\text{L}_j$] (u_2) at (8,3);
       \coordinate[label=left:\,$\text{L}_i$] (d_1) at (-1,-3);
        \coordinate[label=right:\,$\text{R}_i$] (d_2) at (1,-3);
        \coordinate[label=left:\,$\text{R}_k$] (d_3) at (7,-3);
        \coordinate[label=right:\,$\text{L}_k$] (d_4) at (9,-3);
        \draw[thick] (u_1) -- (0,0);
        \draw[thick] (u_2) -- (8,0);\draw[thick] (d_1) -- (0,0);
        \draw[thick] (d_2) -- (0,0);
        \draw[thick] (d_3) -- (8,0);
        \draw[thick] (d_4) -- (8,0);
        \draw[thick,  brown] (0,0) -- node[above] {\tiny bulk-bulk} (3, 0);
        \draw[thick,  brown] (0,0) -- node[above] {\tiny bulk-bulk} (-3, 0);
        \draw[thick,  brown, dotted] (3.5,0) -- (4.5, 0);
        \draw[thick,  brown, dotted] (-3.5,0) -- (-4.5, 0);
\draw[thick,  brown] (5,0) -- node[above] {\tiny bulk-bulk} (8, 0);
    \draw[thick,  brown] (8,0) -- node[above] {\tiny bulk-bulk} (11, 0);    
    \draw[thick,  brown, dotted] (11.5,0) -- (12.5, 0);

        \fill (0, 0) circle (3pt); 
           \fill (8, 0) circle (3pt); 
         
        \draw[{Latex[round]}-{Latex[round]},solid, shorten >= 5pt,shorten <= 5pt] (d_2) to[out=-110,in=-70, looseness=1] (d_1);
        \draw[{Latex[round]}-{Latex[round]},solid, shorten >= 5pt,shorten <= 5pt] (u_2) to[out=90,in=90, looseness=1] (u_1);
        
        \draw[{Latex[round]}-{Latex[round]},solid, shorten >= 5pt,shorten <= 5pt] (d_3) to[out=-110,in=-70, looseness=1] (d_4);
    \end{tikzpicture}\:
    \caption{}
    \label{fig:Conj:3v1c1c}
\end{figure}

\begin{figure}[H]
\centering
\def\tkzscl{0.5}
\begin{tikzpicture}[baseline={([yshift=-.5ex]current bounding box.center)},vertex/.style={anchor=base,
    circle,fill=black!25,minimum size=18pt,inner sep=2pt},scale=\tkzscl]
           \coordinate[label=right:$\text{R}_j$\,] (u_1) at (0,3);
        \coordinate[label=right:\,$\text{L}_j$] (u_2) at (8,3);
        \coordinate[] (u_3) at (16,3);
       \coordinate[label=left:\,$\text{L}_i$] (d_1) at (-1,-3);
        \coordinate[label=right:\,$\text{R}_i$] (d_2) at (1,-3);
        \coordinate[label=right:\,$\text{L}_k$] (d_3) at (8,-3);
        \coordinate[label=right:\,$\text{R}_k$] (d_4) at (16,-3);

        \draw[thick] (u_1) -- (0,0);
        \draw[thick] (u_2) -- (8,0);
        \draw[thick] (u_3) -- (16,0);\draw[thick] (d_1) -- (0,0);
        \draw[thick] (d_2) -- (0,0);
        \draw[thick] (d_3) -- (8,0);
        \draw[thick] (d_4) -- (16,0);
        \draw[thick,  brown] (0,0) -- node[above] {\tiny bulk-bulk} (3, 0);
        \draw[thick,  brown] (0,0) -- node[above] {\tiny bulk-bulk} (-3, 0);
        \draw[thick,  brown, dotted] (3.5,0) -- (4.5, 0);
        \draw[thick,  brown, dotted] (-3.5,0) -- (-4.5, 0);
\draw[thick,  brown] (5,0) -- node[above] {\tiny bulk-bulk} (8, 0);
    \draw[thick,  brown] (8,0) -- node[above] {\tiny bulk-bulk} (11, 0);    
    \draw[thick,  brown, dotted] (11.5,0) -- (12.5, 0);
\draw[thick,  brown] (13,0) -- node[above] {\tiny bulk-bulk} (16, 0);  
\draw[thick,  brown] (16,0) -- node[above] {\tiny bulk-bulk} (19, 0);
\draw[thick,  brown, dotted] (19.5,0) -- (20.5, 0);

        \fill (0, 0) circle (3pt); 
        \fill (8, 0) circle (3pt); 
        \fill (16, 0) circle (3pt);
         
        \draw[{Latex[round]}-{Latex[round]},solid, shorten >= 5pt,shorten <= 5pt] (d_2) to[out=-110,in=-70, looseness=1] (d_1);
        \draw[{Latex[round]}-{Latex[round]},solid, shorten >= 5pt,shorten <= 5pt] (u_2) to[out=90,in=90, looseness=1] (u_1);
        
        \draw[{Latex[round]}-{Latex[round]},solid, shorten >= 5pt,shorten <= 5pt] (d_3) to[out=-50,in=-130, looseness=1] (d_4);
    \end{tikzpicture}\:
    \caption{}
    \label{fig:Conj:3v1c1foreward}
\end{figure}
\begin{figure}[H]
\centering
\def\tkzscl{0.5}
\begin{tikzpicture}[baseline={([yshift=-.5ex]current bounding box.center)},vertex/.style={anchor=base,
    circle,fill=black!25,minimum size=18pt,inner sep=2pt},scale=\tkzscl]
           \coordinate[label=right:$\text{R}_j$\,] (u_1) at (8,3);
        \coordinate[] (u_2) at (0,3);
        \coordinate[label=right:\,$\text{L}_j$] (u_3) at (16,3);
       \coordinate[label=left:\,$\text{L}_i$] (d_1) at (7,-3);
        \coordinate[label=right:\,$\text{R}_i$] (d_2) at (9,-3);
        \coordinate[label=right:\,$\text{R}_k$] (d_3) at (0,-3);
        \coordinate[label=right:\,$\text{L}_k$] (d_4) at (16,-3);

        \draw[thick] (u_1) -- (8,0);
        \draw[thick] (u_2) -- (0,0);
        \draw[thick] (u_3) -- (16,0);
        \draw[thick] (d_1) -- (8,0);
        \draw[thick] (d_2) -- (8,0);
        \draw[thick] (d_3) -- (0,0);
        \draw[thick] (d_4) -- (16,0);
        \draw[thick,  brown] (0,0) -- node[above] {\tiny bulk-bulk} (3, 0);
        \draw[thick,  brown] (0,0) -- node[above] {\tiny bulk-bulk} (-3, 0);
        \draw[thick,  brown, dotted] (3.5,0) -- (4.5, 0);
        \draw[thick,  brown, dotted] (-3.5,0) -- (-4.5, 0);
\draw[thick,  brown] (5,0) -- node[above] {\tiny bulk-bulk} (8, 0);
    \draw[thick,  brown] (8,0) -- node[above] {\tiny bulk-bulk} (11, 0);    
    \draw[thick,  brown, dotted] (11.5,0) -- (12.5, 0);
\draw[thick,  brown] (13,0) -- node[above] {\tiny bulk-bulk} (16, 0);  
\draw[thick,  brown] (16,0) -- node[above] {\tiny bulk-bulk} (19, 0);
\draw[thick,  brown, dotted] (19.5,0) -- (20.5, 0);

        \fill (0, 0) circle (3pt); 
        \fill (8, 0) circle (3pt); 
        \fill (16, 0) circle (3pt);
         
        \draw[{Latex[round]}-{Latex[round]},solid, shorten >= 5pt,shorten <= 5pt] (d_2) to[out=-110,in=-70, looseness=1] (d_1);
        \draw[{Latex[round]}-{Latex[round]},solid, shorten >= 5pt,shorten <= 5pt] (u_3) to[out=90,in=90, looseness=1] (u_1);      
        \draw[{Latex[round]}-{Latex[round]},solid, shorten >= 5pt,shorten <= 5pt] (d_3) to[out=-50,in=-130, looseness=1] (d_4);
    \end{tikzpicture}\: 
    \caption{}
    \label{fig:Conj:3v1c1Ultrarev}
\end{figure}

\begin{figure}[H]
\centering
\def\tkzscl{0.5}
\begin{tikzpicture}[baseline={([yshift=-.5ex]current bounding box.center)},vertex/.style={anchor=base,
    circle,fill=black!25,minimum size=18pt,inner sep=2pt},scale=\tkzscl]
           \coordinate[label=right:$\text{R}_j$\,] (u_1) at (0,3);
        \coordinate[] (u_2) at (8,3);
        \coordinate[label=right:\,$\text{L}_j$] (u_3) at (16,3);
       \coordinate[label=left:\,$\text{L}_i$] (d_1) at (-1,-3);
        \coordinate[label=right:\,$\text{R}_i$] (d_2) at (1,-3);
        \coordinate[label=right:\,$\text{R}_k$] (d_3) at (8,-3);
        \coordinate[label=right:\,$\text{L}_k$] (d_4) at (16,-3);

        \draw[thick] (u_1) -- (0,0);
        \draw[thick] (u_2) -- (8,0);
        \draw[thick] (u_3) -- (16,0);\draw[thick] (d_1) -- (0,0);
        \draw[thick] (d_2) -- (0,0);
        \draw[thick] (d_3) -- (8,0);
        \draw[thick] (d_4) -- (16,0);
        \draw[thick,  brown] (0,0) -- node[above] {\tiny bulk-bulk} (3, 0);
        \draw[thick,  brown] (0,0) -- node[above] {\tiny bulk-bulk} (-3, 0);
        \draw[thick,  brown, dotted] (3.5,0) -- (4.5, 0);
        \draw[thick,  brown, dotted] (-3.5,0) -- (-4.5, 0);
\draw[thick,  brown] (5,0) -- node[above] {\tiny bulk-bulk} (8, 0);
    \draw[thick,  brown] (8,0) -- node[above] {\tiny bulk-bulk} (11, 0);    
    \draw[thick,  brown, dotted] (11.5,0) -- (12.5, 0);
\draw[thick,  brown] (13,0) -- node[above] {\tiny bulk-bulk} (16, 0);  
\draw[thick,  brown] (16,0) -- node[above] {\tiny bulk-bulk} (19, 0);
\draw[thick,  brown, dotted] (19.5,0) -- (20.5, 0);

        \fill (0, 0) circle (3pt); 
        \fill (8, 0) circle (3pt); 
        \fill (16, 0) circle (3pt);
         
        \draw[{Latex[round]}-{Latex[round]},solid, shorten >= 5pt,shorten <= 5pt] (d_2) to[out=-110,in=-70, looseness=1] (d_1);
        \draw[{Latex[round]}-{Latex[round]},solid, shorten >= 5pt,shorten <= 5pt] (u_3) to[out=90,in=90, looseness=1] (u_1);      
        \draw[{Latex[round]}-{Latex[round]},solid, shorten >= 5pt,shorten <= 5pt] (d_3) to[out=-50,in=-130, looseness=1] (d_4);
    \end{tikzpicture}\:
    \caption{}
    \label{fig:Conj:3v1c1rev}
\end{figure}

Note that if the partner-leg to the external leg on the starting vertex is the only external leg, there has to be another vertex with at least three external lines that does not have this property. 

Next, we turn to diagrams for which there is no permutation among the labels at the same vertex, such as seen in figs.\,\ref{fig:Conj:3v0call} and \ref{fig:Conj:3v0c1for1rev}.

\begin{figure}[H]
\centering
\def\tkzscl{0.4}
\begin{tikzpicture}[baseline={([yshift=-.5ex]current bounding box.center)},vertex/.style={anchor=base,
    circle,fill=black!25,minimum size=18pt,inner sep=2pt},scale=\tkzscl]
           \coordinate[label=left:$\text{R}_k$\,] (u_1) at (0,3);
        \coordinate[] (u_2) at (8,3);
        \coordinate[] (u_3) at (16,3);
        \coordinate[label=right:\,$\text{L}_k$] (u_4) at (24,3);
       \coordinate[label=left:\,$\text{R}_j$] (d_1) at (-1,-3);
        \coordinate[label=right:\,$\text{R}_i$] (d_2) at (1,-3);
        \coordinate[label=right:\,$\text{L}_i$] (d_3) at (8,-3);
        \coordinate[label=right:\,$\text{L}_j$] (d_4) at (16,-3);
         \coordinate[] (d_5) at (24,-3);

        \draw[thick] (u_1) -- (0,0);
        \draw[thick] (u_2) -- (8,0);
        \draw[thick] (u_3) -- (16,0);
        \draw[thick] (u_4) -- (24,0);\draw[thick] (d_1) -- (0,0);
        \draw[thick] (d_2) -- (0,0);
        \draw[thick] (d_3) -- (8,0);
        \draw[thick] (d_4) -- (16,0);
        \draw[thick] (d_5) -- (24,0);
        \draw[thick,  brown] (0,0) -- node[above] {\tiny bulk-bulk} (3, 0);
        \draw[thick,  brown] (0,0) -- node[above] {\tiny bulk-bulk} (-3, 0);
        \draw[thick,  brown, dotted] (3.5,0) -- (4.5, 0);
        \draw[thick,  brown, dotted] (-3.5,0) -- (-4.5, 0);
\draw[thick,  brown] (5,0) -- node[above] {\tiny bulk-bulk} (8, 0);
    \draw[thick,  brown] (8,0) -- node[above] {\tiny bulk-bulk} (11, 0);    
    \draw[thick,  brown, dotted] (11.5,0) -- (12.5, 0);
\draw[thick,  brown] (13,0) -- node[above] {\tiny bulk-bulk} (16, 0);  
\draw[thick,  brown] (16,0) -- node[above] {\tiny bulk-bulk} (19, 0);
\draw[thick,  brown, dotted] (19.5,0) -- (20.5, 0);
\draw[thick,  brown] (21,0) -- node[above] {\tiny bulk-bulk} (24, 0);
\draw[thick,  brown] (24,0) -- node[above] {\tiny bulk-bulk} (27, 0);
\draw[thick,  brown, dotted] (27.5,0) -- (28.5, 0);
    
        \fill (0, 0) circle (3pt); 
        \fill (8, 0) circle (3pt); 
        \fill (16, 0) circle (3pt);
        \fill (24, 0) circle (3pt);
        
        \draw[{Latex[round]}-{Latex[round]},solid, shorten >= 5pt,shorten <= 5pt] (u_1) to[out=50,in=130, looseness=1] (u_4);
        \draw[{Latex[round]}-{Latex[round]},solid, shorten >= 5pt,shorten <= 5pt] (d_1) to[out=-50,in=-130, looseness=1] (d_4);
        \draw[{Latex[round]}-{Latex[round]},solid, shorten >= 5pt,shorten <= 5pt] (d_2) to[out=-50,in=-130, looseness=1] (d_3);
    \end{tikzpicture}\: 
    \caption{}
    \label{fig:Conj:3v0call}
\end{figure}

\begin{figure}[H]
\centering
\def\tkzscl{0.4}
\begin{tikzpicture}[baseline={([yshift=-.5ex]current bounding box.center)},vertex/.style={anchor=base,
    circle,fill=black!25,minimum size=18pt,inner sep=2pt},scale=\tkzscl]
     \coordinate[] (u_m1) at (-8,3);
           \coordinate[label=left:$\text{R}_j$\,] (u_0) at (0,3);
        \coordinate[] (u_1) at (8,3);
        \coordinate[label=right:$\text{L}_j$\,] (u_2) at (16,3);
        \coordinate[] (d_m2) at (-16,-3);
          \coordinate[label=left:\,$\text{R}_k$] (d_m1) at (-8,-3);
       \coordinate[label=left:\,$\text{L}_k$] (d_0) at (-1,-3);
        \coordinate[label=right:\,$\text{R}_i$] (d_1) at (1,-3);
        \coordinate[label=right:\,$\text{L}_i$] (d_2) at (8,-3);
        \coordinate[] (d_3) at (16,-3);

        
        \draw[thick] (u_m1) -- (-8,0);
        \draw[thick] (u_0) -- (0,0);
        \draw[thick] (u_1) -- (8,0);\draw[thick] (u_2) -- (16,0);
        \draw[thick] (d_m1) -- (-8,0);
        \draw[thick] (d_0) -- (0,0);
        \draw[thick] (d_1) -- (0,0);
        \draw[thick] (d_2) -- (8,0);
        \draw[thick] (d_3) -- (16,0);
        
        \draw[thick,  brown] (0,0) -- node[above] {\tiny bulk-bulk} (3, 0);
        \draw[thick,  brown] (0,0) -- node[above] {\tiny bulk-bulk} (-3, 0);
        \draw[thick,  brown, dotted] (3.5,0) -- (4.5, 0);
     
\draw[thick,  brown] (5,0) -- node[above] {\tiny bulk-bulk} (8, 0);
    \draw[thick,  brown] (8,0) -- node[above] {\tiny bulk-bulk} (11, 0);    
    \draw[thick,  brown, dotted] (11.5,0) -- (12.5, 0);
\draw[thick,  brown] (13,0) -- node[above] {\tiny bulk-bulk} (16, 0);  
\draw[thick,  brown] (16,0) -- node[above] {\tiny bulk-bulk} (19, 0);
\draw[thick,  brown, dotted] (19.5,0) -- (20.5, 0);
\draw[thick,  brown, dotted] (-3.5,0) -- (-4.5, 0);
\draw[thick,  brown] (-5,0) -- node[above] {\tiny bulk-bulk} (-8, 0);
\draw[thick,  brown] (-8,0) -- node[above] {\tiny bulk-bulk} (-11, 0);
\draw[thick,  brown, dotted] (-11.5,0) -- (-12.5, 0);


        \fill (-8, 0) circle (3pt); 
        \fill (0, 0) circle (3pt); 
        \fill (8, 0) circle (3pt); 
        \fill (16, 0) circle (3pt);

        \draw[{Latex[round]}-{Latex[round]},solid, shorten >= 5pt,shorten <= 5pt] (d_m1) to[out=-50,in=-130, looseness=1] (d_0);
        \draw[{Latex[round]}-{Latex[round]},solid, shorten >= 5pt,shorten <= 5pt] (u_0) to[out=50,in=130, looseness=1] (u_2);
        \draw[{Latex[round]}-{Latex[round]},solid, shorten >= 5pt,shorten <= 5pt] (d_1) to[out=-50,in=-130, looseness=1] (d_2);
    \end{tikzpicture}\:
    \caption{}
    \label{fig:Conj:3v0c1for1rev}
\end{figure}

With the restrictions above, namely, that we consider connected tree-level diagrams with maximal two internal lines at each vertex, $2n$ external legs and $n-1$ vertices, the $n$-commutator function can be fully decomposed in the diagrams of the form shown in figs.\,\ref{fig:Conj:4v2c}-\ref{fig:Conj:3v0c1for1rev} and diagrams of that form with reversed ordering.

\subsection{Vanishing Contribution of the Significant Subdiagrams}
\label{sec:conj:WhyVanish}

We will now show that all diagrams of the form in figs.\,\ref{fig:Conj:4v2c}-\ref{fig:Conj:3v0c1for1rev}  vanish using the matrix description introduced above. 

In order to compute the $n$-commutator function, we have to consider the grSK$_{2n}$ geometry. As explained in the main text, the bulk-to-boundary propagators and the bulk-to-bulk propagators constructed for $\grSKI{4}$ naturally extend to this higher-folds setting. 
The general bulk-to-boundary propagators, $\Gbdy{\text{R}_i}$, $\Gbdy{\text{L}_i}$, are casted into diagonal matrices, with fully normalizable entries everywhere except for the $i$th entry \eqref{eq:BtoBdyVecLR}. 
Similarly, the bulk-to-bulk propagator can be generalized simply by extending the matrix \eqref{eq:BtoB4FoldAux} in accordance with unitarity and the KMS condition.

It will be useful to decompose all the $2n\times 2n$ matrices into $2 \times 2$ matrices. In the following, lowercase Latin letters refer to the index of the $2 \times 2$ submatrix.

Before we face the diagrams described above in section \ref{sec:diagrams}, we analyse their building blocks. 
For every external leg in a diagram (for example the diagram if fig.\,\ref{fig:Cong:example}), with Schwinger-Keldysh label $\text{R}_i$, the corresponding partner-leg  $\text{L}_i$, which we will permute, may be located either on the same vertex or in a different one.

If the partner-leg is located on the same vertex, this gives rise to a matrix of the form
\begin{align}
\label{eq:commutatorOnSameLeg}
	\Gbdy{[\text{R}_i}(r,k_1) 	\,\Gbdy{\text{L}_i]}(r,k_2) = -\nB(\vw_1)\nB(\vw_2)\mathrm{diag} \begin{pmatrix}
	\vdots \\ 0 \\ \mg_1(r,k_1)\mg_3(r,k_2) - \mg_3(r,k_1)\mg_1(r,k_2)\\
	e^{\beta \vw_1}\mg_3(r,k_1)\mg_2(r,k_2) - e^{\beta \vw_2}\mg_2(r,k_1)\mg_3(r,k_2) \\ 0\\ \vdots
	\end{pmatrix}, 
\end{align}
where $\Gbdy{\text{R}_i}$ and $\Gbdy{\text{L}_i}$ are the generalizations of \eqref{eq:BtoBdyVecLR}. The image of this matrix is on the $i$th subspace. This matrix will enter the expression inside trace (such as in \eqref{eq:Cong:example}). 

From \eqref{eq:commutatorOnSameLeg} we can for instance directly see that the diagram in fig.\,\ref{fig:Conj:4v2c} will thus immediately give rise to a vanishing matrix inside the trace by \eqref{eq:commutatorOnSameLeg}, as the two matrices which are multiplied project on different subspaces. 
The other possibility is that the partner-leg is located at a different vertex. To examine these cases, let us highlight the structure of the matrices involved. 

The bulk-to-bulk propagator in the grSK$_{2n}$ geometry is proportial to a matrix (see \eqref{eq:BtoB4FoldAux}) consisting of diagonal blocks 
\begin{align}
\mathcal{R}
 := \begin{pmatrix}
-\mg_1 & e^{\beta\vw}\mg_3  \\
\mg_3 & -\mg_2
\end{pmatrix}
 \end{align}
  and off-diagonal parts consisting of $2\times 2$ matrices that are all proportional to $\hat{S} \Jm \hat{S}$. 
This specific combination has the property that 
\begin{align}
	\left(\Jm \hat{S}\right)^2 &=0\,,\\
	\left(\hat{S}\Jm \right)^2 &=0\, . 
\end{align}
It will be useful to define the sets
\begin{align}
\label{eq:conj:defkappar}
	\kappa_l &:= \{\mathcal{A} \in \mathds{M}_{2 n \times 2n}\,,\: \text{such that} \,\Jm  \,\mathcal{A} =0 \}\,, \\
	\label{eq:conj:defkappal}
	\kappa_r &:= \{\mathcal{A} \in \mathds{M}_{2 n \times 2n}\,,\:\text{such that} \,\mathcal{A} \,\Jm  =0 \} \, , 
\end{align}
which are the kernel of the left-multiplication  with $\Jm$ and the kernel of the right-multiplication  with $\Jm $. Moreover, we will define
\begin{align}
\label{eq:conj:defkappa}
	\kappa := \kappa_r \cap \kappa_l\, . 
\end{align}
For $2n \times 2n$ matrices, the matrices in $\kappa$ include the ones whose $2 \times 2$ block-sub matrices are proportional to $\hat{S}\Jm \hat{S}$, i.\,e. 
\begin{align}
	\kappa \supseteq \mathds{M}_{n\times n} \otimes \hat{S}\Jm\,\hat{S}. 
\end{align} 
Note also that 
\begin{align}
\label{eq:CongRJ}
\mathcal{R}\, \Jm =
\begin{pmatrix}
 -\mg_1 & e^{\beta\vw}\mg_3  \\
\mg_3 & -\mg_2
\end{pmatrix} \begin{pmatrix}
1& 1 \\ 1&1
\end{pmatrix} &\sim  \hat{S}\Jm\, \\
\label{eq:CongJR}
\Jm\, \mathcal{R} = \begin{pmatrix}
1& 1 \\ 1&1
\end{pmatrix}\begin{pmatrix}
-\mg_1 & e^{\beta\vw}\mg_3  \\
\mg_3 & -\mg_2
\end{pmatrix}  &\sim \Jm  \hat{S}\, , 
\end{align}
which uses the relation $-\mg_1 +e^{\beta\vw}\mg_3+\mg_3-\mg_2 =0$, which follows from \eqref{eq:gCombo}.

The dots in the diagrams on figs.\,\ref{fig:Conj:4v2c}-\ref{fig:Conj:3v0c1for1rev} correspond to matrices of the form 
\begin{align}
\label{eq:DotsMatrixProduct}
\def\tkzscl{0.3}
\begin{tikzpicture}[baseline={([yshift=-.5ex]current bounding box.center)},vertex/.style={anchor=base,
    circle,fill=black!25,minimum size=18pt,inner sep=2pt},scale=\tkzscl]  
\draw[thick,  brown] (-8,0) -- node[above] {\tiny bulk-bulk} (-11, 0);
\draw[thick,  brown, dotted] (-11.5,0) -- (-12.5, 0);
    \draw[thick,  brown] (-13,0) -- node[above] {\tiny bulk-bulk} (-16, 0);
        \fill (-16, 0) circle (3pt); 
        \fill (-8, 0) circle (3pt); 
   \end{tikzpicture}=\hat{S}\,\prod_{p = 1}^{V-2}\,\left[\Mbb(r,k)\,\hat{S}\left(\prod_{f\in a_p}\,\Gbdy{\mathfrak{I}_{f}} \right)\right]\,\Mbb(r,k)\,\hat{S}\,, 
\end{align}
where $\mathfrak{I} = \text{R}, \text{L}$ and $a_p$ is the set of indices for $2 \times 2$ sub-matrices, where the bulk-to-boundary is non-normalizable. Moreover, $V$ is an integer counting the number of vertices that appear in that part of the diagram. The vertex $p$ has $|a_p|$ external legs.

The key to approach the expression \eqref{eq:DotsMatrixProduct} is not to calculate it explicitly, but to decompose $\Mbb$ into a block-diagonal part and a part that lies in $\kappa$
\begin{align}
\label{eq:cong:BulkbulkMatrixDecomposition}
	\Mbb = \Mbb^{\mathrm{diag}} + \Mbb^{\mathrm{offdiag}}\, , 
\end{align}
with $\Mbb^{\mathrm{diag}} = \oplus^{n} \mathcal{R}$ and $\Mbb^{\mathrm{offdiag}}\in \kappa$. 
Plugging \eqref{eq:cong:BulkbulkMatrixDecomposition} into \eqref{eq:DotsMatrixProduct}, yields
\begin{align}
	\def\tkzscl{0.3}
\begin{tikzpicture}[baseline={([yshift=-.5ex]current bounding box.center)},vertex/.style={anchor=base,
    circle,fill=black!25,minimum size=18pt,inner sep=2pt},scale=\tkzscl]  
\draw[thick,  brown] (-8,0) -- node[above] {\tiny bulk-bulk} (-11, 0);
\draw[thick,  brown, dotted] (-11.5,0) -- (-12.5, 0);
    \draw[thick,  brown] (-13,0) -- node[above] {\tiny bulk-bulk} (-16, 0);
        \fill (-16, 0) circle (3pt); 
        \fill (-8, 0) circle (3pt); 
   \end{tikzpicture}&=\hat{S}\prod_{p = 1}^{V-2}\left[(\Mbb^{\mathrm{diag}} + \Mbb^{\mathrm{offdiag}})\hat{S}\left(\prod_{f \in a_p}\Gbdy{\mathfrak{I}_{f}} \right)\right](\Mbb^{\mathrm{diag}} + \Mbb^{\mathrm{offdiag}})\hat{S}. 
\end{align}

Expanding the product and using that $\kappa\, \mathcal{A}\,\kappa \subseteq \kappa$ for any matrix $\mathcal{A}$, we see that the product is a sum of the form 
\begin{align}
\label{eq:conj:diagonalsInBlockDecomposition}
	\tilde{D}_a +\sum_{ b,c}	\, \tilde{D}_b\, \kappa\, \tilde{D}_c\, , 
\end{align}
where $\tilde{D}_{a}, \,\tilde{D}_{b}, \, \tilde{D}_{c}$ are block-diagonal matrices. 
Keeping track on which subspaces these block-diagonal matrices are not proportional to the identity and using \eqref{eq:CongRJ} and \eqref{eq:CongJR}, allows us to decompose the matrix \eqref{eq:DotsMatrixProduct} into matrices
\begin{align}
\label{eq:Conj:DotDecomposition}
\begin{split}
\def\tkzscl{0.3}
\begin{tikzpicture}[baseline={([yshift=-.5ex]current bounding box.center)},vertex/.style={anchor=base,
    circle,fill=black!25,minimum size=18pt,inner sep=2pt},scale=\tkzscl]  
\draw[thick,  brown] (-8,0) -- node[above] {\tiny bulk-bulk} (-11, 0);
\draw[thick,  brown, dotted] (-11.5,0) -- (-12.5, 0);
    \draw[thick,  brown] (-13,0) -- node[above] {\tiny bulk-bulk} (-16, 0);
        \fill (-16, 0) circle (3pt); 
        \fill (-8, 0) circle (3pt); 
   \end{tikzpicture}= D + U + L + Q+ F, 
   \end{split}\, . 
\end{align}
The matrices $D, \, U, \, L, \, Q$ and $F$ have the following properties:
\begin{align}
	D & \text{ is block-diagonal, \, }\\
	 Q &\in \kappa\,,\\
	  U& \in \kappa_l\, ,
	  \\ L &\in \kappa_r \,,\\
   F_{l,k} &=0\text{ if } (l, k)\notin a_p\,\times a_p\, ,  
\end{align}
where the indices on the $2n \times 2n$ matrix $F$ denote the $2\times 2$ submatrix $F_{l, k}$ and $\kappa\, , \kappa_{r, l}$ are defined in \eqref{eq:conj:defkappa}, \eqref{eq:conj:defkappar} and \eqref{eq:conj:defkappal}. 

Moreover, $U$ has only non-zero rows for indices in $a_p$ and $D$ has only non-zero lines for indices in $a_p$. A schematic example for the texture of such a matrix \eqref{eq:DotsMatrixProduct} in grSK$_{12}$ (thus $n=6$)  with $a_p = \{2, 4\}$ is

\begin{align}
\begin{pmatrix}
	d &   &   &   &   &   \vspace{0.0cm}\\
	& d  &   &   &   &   \vspace{0.0cm}\\
	  &   & d  &   &   &   \vspace{0.0cm}\\
	  &   &   & d  &  &  \vspace{0.0cm}\\
	  &   &   &   & d  &  \vspace{0.0cm}\\
	  &   &   &   &   & d \vspace{0.0cm}
\end{pmatrix}
+ 
\begin{pmatrix}
	 & u  &   & u  &   &   \vspace{0.0cm}\\
	 &   &   &   &   &   \vspace{0.0cm}\\
	 & u  &   & u  &   &   \vspace{0.0cm}\\
	  &   &   &   &  &  \vspace{0.0cm}\\
	  & u  &   & u  &   &  \vspace{0.0cm}\\
	  & u  &   & u  &   &  \vspace{0.0cm}
\end{pmatrix} + \begin{pmatrix}
	 &   &   &   &   &   \vspace{0.0cm}\\
	l &   & l  &   & l  & l  \vspace{0.0cm}\\
	 &   &   &   &   &   \vspace{0.0cm}\\
	l  &   & l  &   & l & l \vspace{0.0cm}\\
	  &   &   &   &   &  \vspace{0.0cm}\\
	  &   &   &   &   &  \vspace{0.0cm}
\end{pmatrix}+ \begin{pmatrix}
	 &   &   &   &   &   \vspace{0.0cm}\\
	 &   &   &  f &   &   \vspace{0.0cm}\\
	 &   &   &   &   &   \vspace{0.0cm}\\
	  &  f &   &   &  &  \vspace{0.0cm}\\
	  &   &   &   &   &  \vspace{0.0cm}\\
	  &   &   &   &   &  \vspace{0.0cm}
\end{pmatrix}+ \begin{pmatrix}
	 & q  & q  & q  & q  & q  \vspace{0.0cm}\\
	q &   & q  & q  & q  & q  \vspace{0.0cm}\\
	q & q  &   & q  & q  & q  \vspace{0.0cm}\\
	q  & q  & q  &   & q & q \vspace{0.0cm}\\
	q  & q  & q  & q  &   & q \vspace{0.0cm}\\
	q  & q  & q  &q   & q  &  \vspace{0.0cm}
\end{pmatrix}. 
\end{align}

Here, $d,\, f,\, l,\, q$ and $u$ are $2 \times 2$ matrices with $u \in \kappa_l\, , l \in \kappa_r$ and $q \in \kappa$. 

A general feature in the diagrams in \ref{sec:diagrams} is the sub-diagram of the form of fig.\,\ref{fig:Conj:ComAcross}. 
\begin{figure}[H]
\centering
\def\tkzscl{0.5}
\begin{tikzpicture}[baseline={([yshift=-.5ex]current bounding box.center)},vertex/.style={anchor=base,
    circle,fill=black!25,minimum size=18pt,inner sep=2pt},scale=\tkzscl]
     \coordinate[label=right:$\text{R}_i$\,] (u_m1) at (-8,3);
      \coordinate[label=left:$\text{L}_i$\,] (u_m2) at (-16,3);
         \coordinate[] (u_d1) at (-8,-3);
      \coordinate[] (u_d2) at (-16,-3);
        
        \draw[thick] (u_m2) -- (-16,0);
        \draw[thick] (u_m1) -- (-8,0);
           \draw[thick] (u_d2) -- (-16,0);
        \draw[thick] (u_d1) -- (-8,0);
\draw[thick,  brown] (-8,0) -- node[above] {\tiny bulk-bulk} (-11, 0);
\draw[thick,  brown, dotted] (-11.5,0) -- (-12.5, 0);
    \draw[thick,  brown] (-13,0) -- node[above] {\tiny bulk-bulk} (-16, 0);
 
        \fill (-16, 0) circle (3pt); 
        \fill (-8, 0) circle (3pt);

        \draw[{Latex[round]}-{Latex[round]},solid, shorten >= 5pt,shorten <= 5pt] (u_m2) to[out=50,in=130, looseness=1] (u_m1);
      
    \end{tikzpicture}\,
    \caption{}
    \label{fig:Conj:ComAcross}
\end{figure}

This corresponds to a matrix of the form
\begin{align}
\label{eq:conj:commutatorAcrossgeneralMatrix}
	\Gbdy{[\text{R}_i}(r,k_1) \, \mathcal{A}\,	\Gbdy{\text{L}_i]}(r,k_2), 
\end{align}
where $\mathcal{A}$ is of the form \eqref{eq:DotsMatrixProduct}. Nevertheless,  we may treat it as an arbitrary matrix for the purpose of our calculation. Then we compute

\small
\begin{align}
\label{eq:Conj:ComAcross}\begin{split}
&\Gbdy{[\text{R}_i}(r_1,k_1)  \,\mathcal{A}\,	\Gbdy{\text{L}_i]}(r_2,k_2) =\\
&\: \begin{pmatrix}
	0 & \dots & 0 & \nB(\vw_1)\mg_3(k_1)\mathcal{A}_{1, i}\Gin( k_2)& 0 & \dots & 0 \\
	\vdots &  & \vdots & \vdots & \vdots &  & \vdots \\
		0 & \dots & 0 & \nB(\vw_1\mg_3(k_1))\mathcal{A}_{i-1,n}\Gin( k_2)& 0 & \dots & 0 \\
		\nB(\vw_2)\Gin( k_1)\mathcal{A}_{i, 1}\mg_3(k_2)& \dots & & \tilde{\mathcal{A}}_{ii} &&\dots & e^{\beta \vw_2}\nB(\vw_2)\Gin( k_1)\mathcal{A}_{i, n}\mg_3(k_2)\\
		0 & \dots & 0 &e^{\beta \vw_1} \nB(\vw_1)\mg_3(k_1)\mathcal{A}_{1, i}\Gin( k_2)& 0 & \dots & 0 \\
		0 & & \vdots & \vdots&\vdots & &  \vdots \\
		0 & \dots & 0 & e^{\beta \vw_1}\nB(\vw_1)\mg_3(k_1)\mathcal{A}_{n, i}\Gin( k_2)& 0 & \dots & 0 \\		
\end{pmatrix}\end{split}
\end{align}
\normalsize
where suppressed the $r$ dependences and we used \eqref{eq:BoltzFact}
 and \eqref{eq:gCombo}. Note that all $\mathcal{A}_{m,n}$ are $2\times 2$ matrices and
\begin{align}
\begin{split}
  \tilde{\mathcal{A}}_{ii} = -\nB(\vw_1)\nB(\vw_2)&\left[\text{diag}
\begin{pmatrix}
\mg_1(r_1,k)\\ 
e^{\beta\vw_1}\mg_3(r_1,k_1)
\end{pmatrix} \mathcal{A}_{i,i}
\,\text{diag}
\begin{pmatrix}
\mg_3(r_2,k_2)\\ 
\mg_2(r_2,k_2)
\end{pmatrix} \right. \\
& \phantom{=} \, \left. \, - \text{diag}
\begin{pmatrix}
\mg_3(r_1,k)\\ 
\mg_2(r_1,k_1)
\end{pmatrix} \mathcal{A}_{i,i}
\,\text{diag}
\begin{pmatrix}
\mg_1(r_2,k_2)\\ 
e^{\beta\vw_2}\mg_3(r_2,k_2)
\end{pmatrix} \right]\, . 
\end{split}
\end{align} 
An essential feature of \eqref{eq:Conj:ComAcross} is that the matrix is zero entries those indices do not contain $i$. 
If we now combine  \eqref{eq:Conj:ComAcross} with \eqref{eq:commutatorOnSameLeg}, we can conclude that the diagrams in figs.\,\ref{fig:Conj:4v1c} and \ref{fig:Conj:3v1c1c} both have vanishing contributions.

The diagram in fig.\,\ref{fig:Conj:ComAcross} may be used to construct more general constituents such as in fig.\,\ref{fig:ConjFF}. This corresponds to two copies \eqref{eq:Conj:ComAcross} joined by a contour orientation matrix  $\hat{S}$:
\begin{align}
\label{eq:Conj:forewardforeward}
		\Gbdy{[\text{R}_i}(r,k_1) \,  \mathcal{A}\, 	\Gbdy{\text{L}_i]}(r,k_2) \, \hat{S}\, \Gbdy{[\text{R}_j}(r,k_3)\,   \mathcal{A}'\, \Gbdy{\text{L}_j]}(r,k_4). 
\end{align}

\begin{figure}[H]
\centering
\def\tkzscl{0.5}
\begin{tikzpicture}[baseline={([yshift=-.5ex]current bounding box.center)},vertex/.style={anchor=base,
    circle,fill=black!25,minimum size=18pt,inner sep=2pt},scale=\tkzscl]
     \coordinate[label=right:$\text{R}_i$\,] (u_m1) at (-8,3);
      \coordinate[label=left:$\text{L}_i$\,] (u_m2) at (-16,3);
           \coordinate[] (u_0) at (0,3);

        \coordinate[] (d_m2) at (-16,-3);
          \coordinate[label=left:\,$\text{R}_j$] (d_m1) at (-8,-3);
       \coordinate[label=right:\,$\text{L}_j$] (d_0) at (0,-3);
              
        
        \draw[thick] (u_m2) -- (-16,0);
        \draw[thick] (u_m1) -- (-8,0);
        \draw[thick] (u_0) -- (0,0);
        \draw[thick] (d_m2) -- (-16,0);
        \draw[thick] (d_m1) -- (-8,0);
        \draw[thick] (d_0) -- (0,0);

        \draw[thick,  brown] (0,0) -- node[above] {\tiny bulk-bulk} (3, 0);
        \draw[thick,  brown] (0,0) -- node[above] {\tiny bulk-bulk} (-3, 0);
        \draw[thick,  brown, dotted] (3.5,0) -- (4.5, 0);
     
\draw[thick,  brown] (-5,0) -- node[above] {\tiny bulk-bulk} (-8, 0);
\draw[thick,  brown] (-8,0) -- node[above] {\tiny bulk-bulk} (-11, 0);
\draw[thick,  brown, dotted] (-11.5,0) -- (-12.5, 0);
    \draw[thick,  brown] (-13,0) -- node[above] {\tiny bulk-bulk} (-16, 0);
    \draw[thick,  brown] (-16,0) -- node[above] {\tiny bulk-bulk} (-19, 0);
    \draw[thick,  brown, dotted] (-19.5,0) -- (-20.5, 0);

        \fill (-16, 0) circle (3pt); 
        \fill (-8, 0) circle (3pt); 
        \fill (0, 0) circle (3pt);

        \draw[{Latex[round]}-{Latex[round]},solid, shorten >= 5pt,shorten <= 5pt] (u_m2) to[out=50,in=130, looseness=1] (u_m1);
        \draw[{Latex[round]}-{Latex[round]},solid, shorten >= 5pt,shorten <= 5pt] (d_m1) to[out=-50,in=-130, looseness=1] (d_0);

    \end{tikzpicture}\,  
    \caption{}
    \label{fig:ConjFF}
\end{figure}
Recall that for $\Gbdy{[\text{R}_i}(r,k_1) \, \mathcal{A}\,\Gbdy{\text{L}_i]}(r,k_2) $, for $\mathcal{A}$ of the form \eqref{eq:DotsMatrixProduct}, the off-diagonal entries are all in $\kappa$ unless for $(i, a)$, they are in $\kappa_l$ and for $(a, i)$ they are in $\kappa_r$ for $a \in a_p$. 
Especially, we know that the $(i, j)$ entry of both matrices $\Gbdy{[\text{R}_i}(r,k_1) \,  \mathcal{A}\, 	\Gbdy{\text{L}_i]}(r,k_2)$ and $\Gbdy{[\text{R}_j}(r,k_3)\,\mathcal{A}'\,\Gbdy{\text{L}_j]}(r,k_4)$ lies in $\kappa$.  Consequently, using $\kappa_r \, \hat{S} \, \kappa_l =0$, the matrix \eqref{eq:Conj:forewardforeward} has only $(i,j)$ as non-zero entry. 
Even if we replace $\hat{S}$ by $\hat{S} \prod_{b \in b_{p}}  \Gbdy{\mathfrak{I}_{b}} $, this diagonal matrix would still act on the two-dimensional $i$ and $j$ subspaces as $\hat{S}$ and the statement still holds. 
Combining now 
\eqref{eq:Conj:forewardforeward} and \eqref{eq:commutatorOnSameLeg} shows that the contribution associated to the diagram in fig.\,\ref{fig:Conj:3v1c1foreward} vanishes. 

Let us discuss the diagram in fig.\,\ref{fig:Conj:3v1c1Ultrarev}. Instead of computing it concretely, recall that in the end, we compute a trace of matrices with the matrix $\Jm$ as in \eqref{eq:Cong:example}. This trace is of the form
\begin{align}
\label{eq:Conj:Traceprop}
\begin{split}
&\mathrm{Tr} \left[\Jm \Gbdy{\mathfrak{I}_{a_p}} \Mbb\dots \Mbb\, \Gbdy{\mathfrak{I}_{b_p}}\,\hat{S}\, \Mbb \dots \Mbb \Gbdy{\mathfrak{I}_{c_p}} \right] \\&= \mathrm{Tr} \left[\dots\Mbb\, \Gbdy{\mathfrak{I}_{c_p}} \,\hat{S}\,\left(\hat{S} \,\Jm \,\hat{S}\,\right)\,  \Gbdy{\mathfrak{I}_{a_p}}\, \hat{S}\,\Mbb\, \dots\, \Gbdy{\mathfrak{I}_{b_p}} \hat{S}\Mbb\dots \right], 
\end{split}
\end{align}
with $\Gbdy{\mathfrak{I}_{a_p}} = \prod_{a \in a_p}\Gbdy{\mathfrak{I}_{a}}$ and $\Gbdy{\mathfrak{I}_{b_p}}$ and $\Gbdy{\mathfrak{I}_{c_p}}$ respectively. 
As the combination $\hat{S} \, \Jm \,  \hat{S}$ is also of the form \eqref{eq:cong:BulkbulkMatrixDecomposition}, with $\Mbb^{\mathrm{diag}} =0$, we can treat it as an artificial bulk-to-bulk propagator, which allows us to apply the arguments for fig.\,\ref{fig:Conj:3v1c1foreward} also for fig.\,\ref{fig:Conj:3v1c1Ultrarev}.

The remaining diagrams contain sub-diagrams of the form of fig.\,\ref{fig:cong:forwardbackwardPropotype}. 

\begin{figure}[H]

\centering
\def\tkzscl{0.5}
\begin{tikzpicture}[baseline={([yshift=-.5ex]current bounding box.center)},vertex/.style={anchor=base,
    circle,fill=black!25,minimum size=18pt,inner sep=2pt},scale=\tkzscl]
    \coordinate[label=left:$\text{R}_j$\,] (u_0) at (0,3);
        \coordinate[] (u_1) at (8,3);
        \coordinate[label=right:$\text{L}_j$\,] (u_2) at (16,3);
 
          \coordinate[label=left:\,$\text{R}_i$] (d_1) at (0,-3);
        \coordinate[label=right:\,$\text{L}_i$] (d_2) at (8,-3);
        \coordinate[] (d_3) at (16,-3);

        
        \draw[thick] (u_0) -- (0,0);
        \draw[thick] (u_1) -- (8,0);\draw[thick] (u_2) -- (16,0);

        \draw[thick] (d_1) -- (0,0);
        \draw[thick] (d_2) -- (8,0);
        \draw[thick] (d_3) -- (16,0);
        
        \draw[thick,  brown] (0,0) -- node[above] {\tiny bulk-bulk} (3, 0);
        \draw[thick,  brown] (0,0) -- node[above] {\tiny bulk-bulk} (-3, 0);
        \draw[thick,  brown, dotted] (3.5,0) -- (4.5, 0);
     
\draw[thick,  brown] (5,0) -- node[above] {\tiny bulk-bulk} (8, 0);
    \draw[thick,  brown] (8,0) -- node[above] {\tiny bulk-bulk} (11, 0);    
    \draw[thick,  brown, dotted] (11.5,0) -- (12.5, 0);
\draw[thick,  brown] (13,0) -- node[above] {\tiny bulk-bulk} (16, 0);  
\draw[thick,  brown] (16,0) -- node[above] {\tiny bulk-bulk} (19, 0);
\draw[thick,  brown, dotted] (19.5,0) -- (20.5, 0);

        \fill (0, 0) circle (3pt); 
        \fill (8, 0) circle (3pt); 
        \fill (16, 0) circle (3pt);
        
        \draw[{Latex[round]}-{Latex[round]},solid, shorten >= 5pt,shorten <= 5pt] (u_0) to[out=50,in=130, looseness=1] (u_2);
        \draw[{Latex[round]}-{Latex[round]},solid, shorten >= 5pt,shorten <= 5pt] (d_1) to[out=-50,in=-130, looseness=1] (d_2);
    \end{tikzpicture} \, 
    \caption{}
    \label{fig:cong:forwardbackwardPropotype}
\end{figure}

Therefore, we have to compute the matrix product of \eqref{eq:Conj:ComAcross} with \eqref{eq:Conj:DotDecomposition}
\begin{align}
\label{eq:forwardbackwardMatrixProduct}
	\left(\Gbdy{[\text{R}_i} \, 	(D+U+L+Q+F)\,\Gbdy{\text{L}_i]}\right)\,\hat{S}\, \left(D'+U'+L'+Q'+F'\right)\, . 
\end{align}
Let us denote the indices with offdiagonal entries as $a_p$ and $a_p'$ respectively. 
We expand \eqref{eq:forwardbackwardMatrixProduct} and examine the terms:
\begin{itemize}
\item In general, we have $\Gbdy{[\text{R}_i} \,F\,\Gbdy{\text{L}_i]} =0$, 
\item $\left(\Gbdy{[\text{R}_i}  	(D+U+L+Q+F)\,\Gbdy{\text{L}_i]}\right)\,\hat{S} \left(F'\right)$ is non-zero only on $(i, a)$ for $a \in a_p'$,
\item $\left(\Gbdy{[\text{R}_i}\,D\,\Gbdy{\text{L}_i]}\right)\,\hat{S}\,D'$ is non-zero only on $(i,i)$, 
\item 
$\left(\Gbdy{[\text{R}_i} \, D\,\Gbdy{\text{L}_i]}\right)\,\hat{S}\, L'=0$ as $i \neq a_p'$, 
\item $\left(\Gbdy{[\text{R}_i}  \,	D\,\Gbdy{\text{L}_i]}\right)\,\hat{S} \,U'$ is non-zero only on $(i, a)$ with $a \in a_p'$, 
\item $\left(\Gbdy{[\text{R}_i} \,D\,\Gbdy{\text{L}_i]}\right)\hat{S} \,Q'$ is non-zero only for $(i, a)$, where $a$ is arbitrary, 
\item $\left(\Gbdy{[\text{R}_i}\, U \,\Gbdy{\text{L}_i]}\right)\,\hat{S} \,U'$ is non-zero only for $(i, a)$, $a \in a_p'$, 
\item $\left(\Gbdy{[\text{R}_i} \,U \Gbdy{\text{L}_i]}\right)\,\hat{S} \,D'$ is non-zero only for $(i, a)$ with $a \in a_p$, 
\item $\left(\Gbdy{[\text{R}_i}\,U\, \Gbdy{\text{L}_i]}\right)\hat{S}\, L'=0$ as $a_p \cap a_p' = \emptyset$
\item $\left(\Gbdy{[\text{R}_i} \,U\,  \Gbdy{\text{L}_i]}\right)\hat{S} \,Q'$ is non-zero only for $(i, a)$, $a$ arbitrary, 
\item $\left(\Gbdy{[\text{R}_i}\, L\, \Gbdy{\text{L}_i]}\right)\hat{S} \,D'$ is non-zero only for $(a, i)$ with $a\in a_p$,  
\item $\Gbdy{[\text{R}_i}( L+Q)\Gbdy{\text{L}_i]} \,\hat{S}\,(U'+Q') =0$ due to the properties of the matrices \eqref{eq:Conj:DotDecomposition}, 
\item $\left(\Gbdy{[\text{R}_i}\, L\, \Gbdy{\text{L}_i]}\right)\hat{S} \,L'=0$ as $a_p \cap a_p' = \emptyset$,  
\item $\left(\Gbdy{[\text{R}_i} \,Q\, \Gbdy{\text{L}_i]}\right)\,\hat{S} \,D'$ is non-zero only for $(a,b)$ with either $a =i$ or $b=i$,  
\item $\left(\Gbdy{[\text{R}_i}\,Q\, \Gbdy{\text{L}_i]}\right)\,\hat{S}\, L'$ is non-zero only for $(i, a)$ with $a$ arbitrary\, .  
\end{itemize}
Note that, due the commutation in the $j$ segment in the diagram in fig.\,\ref{fig:cong:forwardbackwardPropotype}, we are only interested in the $(a, j)$ and $(j, a)$ components for arbitrary $a$, as these are the only entries that survive the commutation in the $j$ labels (see \eqref{eq:Conj:ComAcross}). 
Thus, the only terms that appear in the matrix for the diagram in fig.\,\ref{fig:cong:forwardbackwardPropotype} are
\begin{align}
\label{eq:conj:forwardBackwardResult}
	\Gbdy{[\text{R}_j}\left[\left(\Gbdy{[\text{R}_i} \left(D + U \right) \Gbdy{\text{L}_i]}\right)\hat{S} \,Q'\,+ \left(\Gbdy{[\text{R}_i} \,Q \,\Gbdy{\text{L}_i]}\right)\hat{S} \left(D' + L'\right)\right]\Gbdy{\text{L}_j]}\, , 
\end{align}
which have only non-zero entries at indices $(j, i)$ and $(i, j)$.

Analog considerations hold for the diagram in fig.\,\ref{fig:conj:forwardbackwardreverse}. 

\begin{figure}[H]
\centering
\def\tkzscl{0.5}
\begin{tikzpicture}[baseline={([yshift=-.5ex]current bounding box.center)},vertex/.style={anchor=base,
    circle,fill=black!25,minimum size=18pt,inner sep=2pt},scale=\tkzscl]
    \coordinate[label=left:$\text{R}_j$\,] (u_0) at (0,3);
        \coordinate[] (u_1) at (8,3);
        \coordinate[label=right:$\text{L}_j$\,] (u_2) at (16,3);
 
          \coordinate[] (d_1) at (0,-3);
        \coordinate[label=left:\,$\text{L}_i$] (d_2) at (8,-3);
        \coordinate[label=right:\,$\text{R}_i$] (d_3) at (16,-3);

        
        \draw[thick] (u_0) -- (0,0);
        \draw[thick] (u_1) -- (8,0);\draw[thick] (u_2) -- (16,0);

        \draw[thick] (d_1) -- (0,0);
        \draw[thick] (d_2) -- (8,0);
        \draw[thick] (d_3) -- (16,0);
        
        \draw[thick,  brown] (0,0) -- node[above] {\tiny bulk-bulk} (3, 0);
        \draw[thick,  brown] (0,0) -- node[above] {\tiny bulk-bulk} (-3, 0);
        \draw[thick,  brown, dotted] (3.5,0) -- (4.5, 0);
     
\draw[thick,  brown] (5,0) -- node[above] {\tiny bulk-bulk} (8, 0);
    \draw[thick,  brown] (8,0) -- node[above] {\tiny bulk-bulk} (11, 0);    
    \draw[thick,  brown, dotted] (11.5,0) -- (12.5, 0);
\draw[thick,  brown] (13,0) -- node[above] {\tiny bulk-bulk} (16, 0);  
\draw[thick,  brown] (16,0) -- node[above] {\tiny bulk-bulk} (19, 0);
\draw[thick,  brown, dotted] (19.5,0) -- (20.5, 0);

        \fill (0, 0) circle (3pt); 
        \fill (8, 0) circle (3pt); 
        \fill (16, 0) circle (3pt);
        
        \draw[{Latex[round]}-{Latex[round]},solid, shorten >= 5pt,shorten <= 5pt] (u_0) to[out=50,in=130, looseness=1] (u_2);
        \draw[{Latex[round]}-{Latex[round]},solid, shorten >= 5pt,shorten <= 5pt] (d_2) to[out=-50,in=-130, looseness=1] (d_3);
    \end{tikzpicture} \, 
    \caption{}
    \label{fig:conj:forwardbackwardreverse}
\end{figure}

From the shape of matrix \eqref{eq:conj:forwardBackwardResult}, we can apply the matrix \eqref{eq:commutatorOnSameLeg} to see that the matrix for the diagram in fig.\,\ref{fig:Conj:3v1c1rev} vanishes. 

The only diagrams that remain are diagrams in fig.\,\ref{fig:Conj:3v0call} and fig.\,\ref{fig:Conj:3v0c1for1rev}. 
For the first one, we multiply \eqref{eq:conj:forwardBackwardResult} with the vertex sign matrix $\hat{S}$ and \eqref{eq:Conj:DotDecomposition}. Looking at the diagram fig.\,\ref{fig:Conj:3v0call} and having \eqref{eq:Conj:ComAcross} in mind, we are only interested in $2 \times 2$ entries involving $k$ as index. Thus only 
\begin{align}
\label{eq:cong:intermediatefinialdiagram}
\begin{split}
		&\Gbdy{[\text{R}_j}\left[\left(\Gbdy{[\text{R}_i} \left(D + U \right) \Gbdy{\text{L}_i]}\right)\hat{S} \left(Q'\right)  + \left(\Gbdy{[\text{R}_i}\, Q \,\Gbdy{\text{L}_i]}\right)\hat{S} \left(D' + L'\right)\right]\Gbdy{\text{L}_j]}\, \hat{S}\,Q''  \\&= \Gbdy{[\text{R}_j}\left(\Gbdy{[\text{R}_i}\, Q \,\Gbdy{\text{L}_i]}\right)\hat{S}\,D'\,\Gbdy{\text{L}_j]}\, \hat{S}\,Q''
		\end{split}
\end{align}
matters. 
We see that from the block-diagonal matrix $D'$, only the $(i, i)$ and $(j, j)$ components enters. 
Let us look closer on how this block-diagonal matrix appears in \eqref{eq:Conj:DotDecomposition}. 
From the discussion above, we know that the block-diagonal of \eqref{eq:Conj:DotDecomposition} takes the form of the block-diagonal of \eqref{eq:conj:diagonalsInBlockDecomposition}. 
Moreover, the block-diagonal matrices appearing in \eqref{eq:conj:diagonalsInBlockDecomposition} satisfy
\begin{align}
	\left(\Tilde{D}_a\right)_{i, i} \sim 	\left(\Tilde{D}_b\right)_{i, i} \sim \left(\Tilde{D}_c\right)_{i, i} \sim \mathcal{R}
\end{align}
and analog for $(j,j)$ as all other possible entries would stem from external legs and must not appear in the $(i, i)$ nor the $(j, j)$ entry. 

We conclude that contributions of $D'$ in \eqref{eq:cong:intermediatefinialdiagram} are multiplies of $\mathcal{R}$. Therefore, we get, using \eqref{eq:CongJR} or \eqref{eq:CongRJ}, 
\begin{align}
\Gbdy{[\text{R}_k}	\, \left[\Gbdy{[\text{R}_j}\, \left[\left(\Gbdy{[\text{R}_i}\, Q\, \Gbdy{\text{L}_i]}\right)\,\hat{S}\, \left(\,D'\,\right)\right]\,\Gbdy{\text{L}_j]}\, \hat{S}\,Q\right]\,\Gbdy{\text{L}_k]}=0\, . 
\end{align}
This concludes the proof why the diagram in fig.\,\ref{fig:Conj:3v0call} also does not contribute. 

Analog considerations hold for the diagram in fig.\,\ref{fig:Conj:3v0c1for1rev}. There we us again \eqref{eq:conj:forwardBackwardResult} to compute
\small
\begin{align}
\begin{split}
		&\Gbdy{[\text{R}_k}\left(D'' + U'' + L'' + Q'' + F''\right)\Gbdy{\text{L}_k]}\hat{S}\Gbdy{[\text{R}_j}\left[\left(\Gbdy{[\text{R}_i} \left(D + U \right) \Gbdy{\text{L}_i]}\right)\hat{S} Q'  + \left(\Gbdy{[\text{R}_i} Q \Gbdy{\text{L}_i]}\right)\hat{S} \left(D' + L'\right)\right]\Gbdy{\text{L}_j]}\\&= \Gbdy{[\text{R}_k}\, Q''\, \Gbdy{\text{L}_k]}\,\hat{S}\,\Gbdy{[\text{R}_j}\left[\left(\Gbdy{[\text{R}_i}\,\left( D+U \right) \,\Gbdy{\text{L}_i]}\right)\hat{S}\,Q'\right]\,\Gbdy{\text{L}_j]} \\ & =0\, . 
		\end{split}
\end{align}
\normalsize
A different way to see that the diagram in fig.\,\ref{fig:Conj:3v0c1for1rev} does not contribute is along the lines of the vanishing of the contribution of fig.\,\ref{fig:Conj:3v1c1Ultrarev} above, where we used the trace property as in \eqref{eq:Conj:Traceprop}. Using the trace property of the expressions for the Witten diagrams, relates the contributions of fig.\,\ref{fig:Conj:3v0call} with the contributions of fig.\,\ref{fig:Conj:3v0c1for1rev}. 

Moreover, these diagrams \,\ref{fig:Conj:3v0call}  and \ref{fig:Conj:3v0c1for1rev} also contain the case of diagram fig.\,\ref{fig:Conj:3v0c3for}, where all partner-legs are on the same vertex, 
\begin{figure}[H]
\centering
\def\tkzscl{0.5}
\begin{tikzpicture}[baseline={([yshift=-.5ex]current bounding box.center)},vertex/.style={anchor=base,
    circle,fill=black!25,minimum size=18pt,inner sep=2pt},scale=\tkzscl]
           \coordinate[label=right:$\text{R}_j$\,] (u_1) at (0,3);
        \coordinate[label=right:\,$\text{L}_j$] (u_2) at (8,3);
       \coordinate[label=left:\,$\text{R}_k$] (d_1) at (-1,-3);
        \coordinate[label=right:\,$\text{R}_i$] (d_2) at (1,-3);
        \coordinate[label=right:\,$\text{L}_i$] (d_3) at (7,-3);
        \coordinate[label=right:\,$\text{L}_k$] (d_4) at (9,-3);
        \draw[thick] (u_1) -- (0,0);
        \draw[thick] (u_2) -- (8,0);\draw[thick] (d_1) -- (0,0);
        \draw[thick] (d_2) -- (0,0);
        \draw[thick] (d_3) -- (8,0);
        \draw[thick] (d_4) -- (8,0);
        \draw[thick,  brown] (0,0) -- node[above] {\tiny bulk-bulk} (3, 0);
        \draw[thick,  brown] (0,0) -- node[above] {\tiny bulk-bulk} (-3, 0);
        \draw[thick,  brown, dotted] (3.5,0) -- (4.5, 0);
        \draw[thick,  brown, dotted] (-3.5,0) -- (-4.5, 0);
\draw[thick,  brown] (5,0) -- node[above] {\tiny bulk-bulk} (8, 0);
    \draw[thick,  brown] (8,0) -- node[above] {\tiny bulk-bulk} (11, 0);    
    \draw[thick,  brown, dotted] (11.5,0) -- (12.5, 0);
    
        \fill (0, 0) circle (3pt); 
           \fill (8, 0) circle (3pt); 
         
        \draw[{Latex[round]}-{Latex[round]},solid, shorten >= 5pt,shorten <= 5pt] (d_2) to[out=-50,in=-130, looseness=1] (d_3);
        \draw[{Latex[round]}-{Latex[round]},solid, shorten >= 5pt,shorten <= 5pt] (u_2) to[out=90,in=90, looseness=1] (u_1);
        
        \draw[{Latex[round]}-{Latex[round]},solid, shorten >= 5pt,shorten <= 5pt] (d_1) to[out=-50,in=-130, looseness=1] (d_4);
    \end{tikzpicture}\:
    \caption{}
    \label{fig:Conj:3v0c3for}
\end{figure}
or even only two of the partner-legs are located on the same vertex, as in fig.\,\ref{fig:Conj:3v0c3forfor}. 

\begin{figure}[H]
\centering
\def\tkzscl{0.5}
\begin{tikzpicture}[baseline={([yshift=-.5ex]current bounding box.center)},vertex/.style={anchor=base,
    circle,fill=black!25,minimum size=18pt,inner sep=2pt},scale=\tkzscl]
           \coordinate[] (u_1) at (0,3);
        \coordinate[label=right:\,$\text{L}_j$] (u_2) at (8,3);
       \coordinate[label=left:\,$\text{R}_k$] (d_1) at (-1,-3);
        \coordinate[label=right:\,$\text{R}_i$] (d_2) at (1,-3);
        \coordinate[label=right:\,$\text{L}_i$] (d_3) at (7,-3);
        \coordinate[label=right:\,$\text{L}_k$] (d_4) at (9,-3);
        \coordinate[] (d_5) at (16,-3);
        \coordinate[label=right:$\text{R}_j$\,] (u_4) at (16,3);
        
        \draw[thick] (u_1) -- (0,0);
        \draw[thick] (u_2) -- (8,0);\draw[thick] (d_1) -- (0,0);
        \draw[thick] (d_2) -- (0,0);
        \draw[thick] (d_3) -- (8,0);
        \draw[thick] (d_4) -- (8,0);
        \draw[thick] (u_4) -- (16,0);
        \draw[thick] (d_5) -- (16,0);
        \draw[thick,  brown] (0,0) -- node[above] {\tiny bulk-bulk} (3, 0);
        \draw[thick,  brown] (0,0) -- node[above] {\tiny bulk-bulk} (-3, 0);
        \draw[thick,  brown, dotted] (3.5,0) -- (4.5, 0);
        \draw[thick,  brown, dotted] (-3.5,0) -- (-4.5, 0);
\draw[thick,  brown] (5,0) -- node[above] {\tiny bulk-bulk} (8, 0);
    \draw[thick,  brown] (8,0) -- node[above] {\tiny bulk-bulk} (11, 0);    
    \draw[thick,  brown, dotted] (11.5,0) -- (12.5, 0);
    \draw[thick,  brown] (13,0) -- node[above] {\tiny bulk-bulk} (16, 0);  
    \draw[thick,  brown] (16,0) -- node[above] {\tiny bulk-bulk} (19, 0);  
    \draw[thick,  brown, dotted] (19.5,0) -- (20.5, 0);
        \fill (0, 0) circle (3pt); 
           \fill (8, 0) circle (3pt); 
           \fill (16, 0) circle (3pt); 
        \draw[{Latex[round]}-{Latex[round]},solid, shorten >= 5pt,shorten <= 5pt] (d_2) to[out=-50,in=-130, looseness=1] (d_3);
        \draw[{Latex[round]}-{Latex[round]},solid, shorten >= 5pt,shorten <= 5pt] (u_2) to[out=90,in=90, looseness=1] (u_4);
        
        \draw[{Latex[round]}-{Latex[round]},solid, shorten >= 5pt,shorten <= 5pt] (d_1) to[out=-50,in=-130, looseness=1] (d_4);
    \end{tikzpicture}\:
    \caption{}
    \label{fig:Conj:3v0c3forfor}
\end{figure}

The vanishing of of the contribution of fig.\,\ref{fig:Conj:3v0c3for} also follows from applying  \eqref{eq:Conj:ComAcross} three times. 

We conclude that no connected diagram with at most two internal lines at each vertex and $n-1$ vertices and $2n$ external lines can contribute to the $n$th power of the commutator.  Of course, to definitely establish the conjecture \eqref{eq:TheConjecture}, one should consider tree-level diagrams with more than two internal lines between vertices, such as in fig.\,\ref{Fig:cong:NotConsidered}, which would require extending our matrix formalism to include more general tensor structures. We shall leave this to future investigations. 

In this Appendix, we discussed many diagrams that do not give any contribution to the $n$-commutator function. But what kind of diagram actually gives rise to a non-vanishing contribution?
Let us consider a diagram of the form

\begin{figure}[H]

\centering
\def\tkzscl{0.4}
\begin{tikzpicture}[baseline={([yshift=-.5ex]current bounding box.center)},vertex/.style={anchor=base,
    circle,fill=black!25,minimum size=18pt,inner sep=2pt},scale=\tkzscl]
    \coordinate[label=above:$\text{L}_j$\,] (u_m2) at (-20,3);
     \coordinate[label=above:$\text{R}_i$\,] (u_m1) at (-15,3);
           \coordinate[label=above:$\text{R}_k$\,] (u_0) at (-10,3);
        \coordinate[label=above:$\text{L}_m$\,] (u_1) at (-5,3);
        \coordinate[label=above:$\text{L}_q$\,] (u_2) at (0,3);
        \coordinate[label=below:\,$\text{R}_j$] (d_m2) at (-20,-3);
          \coordinate[label=below:\,$\text{L}_i$] (d_m1) at (-15,-3);
       \coordinate[label=below:\,$\text{L}_k$] (d_0) at (-10,-3);
        \coordinate[label=below:\,$\text{R}_m$] (d_1) at (-5,-3);
        \coordinate[label=below:\,$\text{R}_q$] (d_2) at (0,-3);

        \draw[thick] (u_m2) -- (-20,0);
        \draw[thick] (u_m1) -- (-15,0);
        \draw[thick] (u_0) -- (-10,0);
        \draw[thick] (u_1) -- (-5,0);
        \draw[thick] (u_2) -- (0,0);
        \draw[thick] (d_m2) -- (-20,0);
        \draw[thick] (d_m1) -- (-15,0);
        \draw[thick] (d_0) -- (-10,0);
        \draw[thick] (d_1) -- (-5,0);
        \draw[thick] (d_2) -- (0,0);
        
        \draw[thick,  brown] (-17,0) -- node[above] {\tiny bulk-bulk} (-20, 0);
        \draw[thick,  brown] (-12,0) -- node[above] {\tiny bulk-bulk} (-15, 0);
        \draw[thick,  brown] (-7,0) -- node[above] {\tiny bulk-bulk} (-10, 0);
        \draw[thick,  brown] (-2,0) -- node[above] {\tiny bulk-bulk} (-5, 0);
\draw[thick,  brown, dotted] (0,0) -- (1.5, 0);
 \draw[thick,  brown] (-17,0) --   (-15, 0);
 \draw[thick,  brown] (-12,0) --  (-10, 0);
 \draw[thick,  brown] (-7,0) --   (-5, 0);
 \draw[thick,  brown] (-2,0) --   (-0, 0);        
        \fill (-20, 0) circle (3pt); 
        \fill (-15, 0) circle (3pt); 
        \fill (-10, 0) circle (3pt); 
        \fill (-5, 0) circle (3pt); 
        \fill (0, 0) circle (3pt);
        
        \draw[{Latex[round]}-{Latex[round]},solid, shorten >= 5pt,shorten <= 5pt] (u_m2) to[out=180,in=180, looseness=1] (d_m2);
        \draw[{Latex[round]}-{Latex[round]},solid, shorten >= 5pt,shorten <= 5pt] (u_m1) to[out=180,in=180, looseness=1] (d_m1);
        \draw[{Latex[round]}-{Latex[round]},solid, shorten >= 5pt,shorten <= 5pt] (u_0) to[out=180,in=180, looseness=1] (d_0);
        \draw[{Latex[round]}-{Latex[round]},solid, shorten >= 5pt,shorten <= 5pt] (u_1) to[out=180,in=180, looseness=1] (d_1);
        \draw[{Latex[round]}-{Latex[round]},solid, shorten >= 5pt,shorten <= 5pt] (u_2) to[out=180,in=180, looseness=1] (d_2);
    \end{tikzpicture}\:
    \caption{}
    \label{fig:Conj:nonVanishing}
\end{figure}

We use identity \eqref{eq:commutatorOnSameLeg} to evaluate the contribution of this diagram to be
\begin{align}
\begin{split}
&	\int_{\rH}^{\infty} \left( \prod_{p=1}^{n}\, dr_p\right) \sqrt{-g} \,\Tr \left[ \Jm \Gbdy{[\text{R}_j}(r_1,k_1) 	\,\Gbdy{\text{L}_j]}(r_1,k_2) \Mbb\Gbdy{[\text{R}_i}(r_2,k_3) 	\,\Gbdy{\text{L}_i]}(r_2,k_4) \Mbb \, \dots\right] \\ 
&\:\sim \int_{\rH}^{\infty} \left( \prod_{p=1}^{n}\, dr_p \right)\,  \sqrt{-g} \,\Tr \left[  \prod_{p=1}^{n} \left( \mathcal{P}(r_p) \, \hat{S}\, \Jm \right)\right]\, .\label{eq:Conj:nonVanishing}\end{split}
\end{align}
with
\begin{align}
	\mathcal{P} = -\nB(\vw_1)\nB(\vw_2)\mathrm{diag} \begin{pmatrix}
	\mg_1(r,k_1)\mg_3(r,k_2) - \mg_3(r,k_1)\mg_1(r,k_2)\\
	e^{\beta \vw_1}\mg_3(r,k_1)\mg_2(r,k_2) - e^{\beta \vw_2}\mg_2(r,k_1)\mg_3(r,k_2) 
	\end{pmatrix}, 
\end{align}

We see that in \eqref{eq:Conj:nonVanishing} only the off-diagonal entries of the bulk-bulk propagator matrix $\Mbb$ enter. This kind of diagram gives a non-vanishing contribution. 
As the Haeviside functions of the radial coordinate in the bulk-bulk propagator \eqref{eq:BtoBMatrix} only appear in the block-diagonal parts, the contribution of the diagram in fig.\,\ref{fig:Conj:nonVanishing} factorizes into $n$ radial integrals.

\section{A toy model: scalars in BTZ background}\label{app:toy_model}
In order to be even more concrete in our calculations, it will be useful to consider the simpler case where the background geometry is the three dimensional BTZ black hole. We consider then a minimally coupled scalar with a cubic interaction
\begin{equation}\label{eq:BTZScalar}
\begin{split}
S[\vphi] &= \int d^{3}x \sqrt{-g}\left[-\frac{1}{2}\left(\nabla_{A}\vphi\nabla^A\vphi + m^2 \vphi^2\right) - \lambda \vphi^3\right]
+S_{\text{c.t.}}\,.
\end{split}
\end{equation}
The gaussian dynamics and contact interactions of this theory were studied in \cite{Jana:2020vyx} within the context of the $\grSKI{2}$ geometry. A more general model was studied in \cite{Loganayagam:2022zmq}, which allowed for non-minimal scalars whose dynamics are modulated by a dilaton coupling. Here we will consider the minimally coupled scalar for clarity of the notation, but remark that many of the calculations can be easily generalized to include scalars with a non-trivial dilaton.

As we have seen, the key ingredient for the evaluation of Witten diagrams in the $\grSKI{n}$ geometry is the ingoing bulk-to-boundary propagator. For the case of $d=2$, we can solve the equation of motion \eqref{eq:EoMIn} and boundary conditions \eqref{eq:NearHorizon} in terms of a regularized hypergeometric function, the result is
\begin{equation}\label{eq:BtoBdyBTZ}
\Gin^{\Delta}(z,k) = \frac{\Gamma\left(\mP_+ +\frac{\Delta}{2}\right)\Gamma\left(\mP_{-} + \frac{\Delta}{2}\right)}{\Gamma\left(\Delta - 1\right)}
z^{\Delta}(1+z)^{-i\mw}\,\pFqReg{2}{1}{\mP_{+}+\frac{\Delta}{2}, \mP_{-} + \frac{\Delta}{2}}{1-i\,\mw}{1-z^2}\,,
\end{equation}
where $z=\frac{\rH}{r}$, and we introduce dimensionless light-cone variables
\begin{equation}
\label{eq:deffrakw}
\mP_{\pm} = \frac{i}{2}\left(-\mw\pm \mq\right)\,,\qquad \mw = \frac{\vw}{\rH}\,, \qquad \mq = \frac{k}{\rH}\,.
\end{equation}
The scaling dimension $\Delta$ is given by the usual relation: $m^2 = \Delta(\Delta-2)$. As indicated in section \ref{subsec:gaussian}, the outgoing bulk-to-boundary propagator then follows from simple time-reversal symmetry
\begin{equation}
\Gout^{\Delta}(z,k) = \left(\frac{1-z}{1+z}\right)^{i\mw}\Gin^{\Delta}(z,\bar{k})\,,
\end{equation}
and we introduce $\tilde{\mP}_{\pm}$, which corresponds to the light-cone momenta with $\mw\rightarrow - \mw$.

The retarded two-point function is then easily evaluated as
\begin{equation}
K_{\Delta}(k) \equiv \lim_{z\rightarrow 0}\left(z^{2-\Delta}\pi + \text{c.t.}\right)  = 2\left(\Delta - 1\right)\frac{\mG(k,\Delta)}{\mG(k,\tilde{\Delta})}\,,
\end{equation}
where
\begin{equation}
\mG(k,\Delta) = \Gamma\left(\mP_{+} + \frac{\Delta}{2}\right)\Gamma\left(\mP_{-} + \frac{\Delta}{2}\right)\Gamma\left(1-\Delta\right)\,,
\end{equation}
and $\tilde{\Delta} =2-\Delta$ is the shadow dimension. We observe that the analytic structure of the two-point function follows simply from that of the $\Gamma$-function and, as observed in \cite{Loganayagam:2022zmq}, this allows us to easily keep track of the pole structure of the higher order correlators.

For the purpose of evaluating the Witten diagrams, it will be useful to consider the Mellin representation of the ingoing bulk-to-boundary propagator
\begin{equation}
\begin{split}
\Gin^{\Delta}(z,k) &=
z^{\Delta}\left(1+z\right)^{-i\mw}\int_{\mathcal{C}}\frac{ds}{2\pi i}\,\frac{\Gamma(s)
\Gamma\left(1-\Delta+s\right)}{\Gamma\left(1-\Delta+2s\right)}\frac{\mG(k,\Delta-2s)}{\mG(k,\tilde{\Delta})}z^{-2s}\,,
\end{split}
\end{equation}
which follows from the Mellin representation of the hypergeometric function. The integration contour runs parallel to the imaginary axis, separating the poles at $s=-n$ and $s=1-\tDelta -n$ on the left, and $s=\mP_{\pm} +n$ on the right.

Let us now proceed to the evaluation of \eqref{eq:WInt}:
\begin{equation}\label{eq:WIntBTZ}
\begin{split}
\mW(k_1,k_2) &= \int_{r}\Gin^{\Delta}(r,k_1)\Gout^{\Delta}(r,k_2)
\times \left(\Gin^{\Delta'}(k_1+k_2,r)-\Gout^{\Delta'}(k_1+k_2,r)\right) - (k_1\leftrightarrow k_2)\,,
\end{split}
\end{equation}
where for simplicity, we assume the external fields are all identical, with scaling dimension $\Delta$, while the exchanged field is distinct, with scaling dimension $\Delta'$. Using the Mellin representation:
\small
\begin{equation}
\begin{split}
\mW(k_1,k_2) &= \left(\prod_{i=1}^{3}\int_{\mathcal{C}}\frac{ds_i}{2\pi i}\Gamma(s_i)\right)
\frac{\Gamma\left(1-\Delta+s_1\right)}{\Gamma\left(1-\Delta+2s_1\right)}
\frac{\Gamma\left(1-\Delta+s_2\right)}{\Gamma\left(1-\Delta+2s_2\right)}
\frac{\Gamma\left(1-\Delta'+s_3\right)}{\Gamma\left(1-\Delta'+2s_3\right)}\\ 
&\quad 
\times
\frac{\mG(k_1,\Delta-2s_1)}{\mG(k_1,\tilde{\Delta})}
\frac{\mG(\bar{k}_2,\Delta-2s_2)}{\mG(\bar{k}_2,\tilde{\Delta})}
\int_{0}^{1}\frac{dz}{z^{d+1}}z^{2\Delta+\Delta'-2(s_1+s_2+s_3)}\left(1+z\right)^{-i\mw_1}\left(1-z\right)^{i\mw_2}\\ 
&\quad 
\times
\left[\left(1+z\right)^{-i\mw_1-i\mw_2}\frac{\mG(k_1+k_2,\Delta'-2s_3)}{\mG(k_1+k_2,\tilde{\Delta}')}-\left(1-z\right)^{i\mw_1+i\mw_2}\frac{\mG(\bar{k}_1+\bar{k}_2,\Delta'-2s_3)}{\mG(\bar{k}_1+\bar{k}_2,\tilde{\Delta}')}\right] \\ 
&- (k_1\leftrightarrow k_2)\,.
\end{split}
\end{equation}
\normalsize
In principle, it seems somewhat counterintuitive that we have so many contributions. The expression above has four distinct terms, and then we have a similar expression as a function of $k_3$ and $k_4$, for a total of sixteen terms; this reflects the fact that the commutator we are computing is better expressed in the left/right basis but the solutions are given in terms of the F/P basis objects $\Gin$ and $\Gout$.

We can perform the radial integral to obtain
\small
\begin{equation}
\begin{split}
\mW(k_1,k_2) &= \left(\prod_{i=1}^{3}\int_{\mathcal{C}}\frac{ds_i}{2\pi i}\Gamma(s_i)\right)
\frac{\Gamma\left(1-\Delta+s_1\right)}{\Gamma\left(1-\Delta+2s_1\right)}
\frac{\Gamma\left(1-\Delta+s_2\right)}{\Gamma\left(1-\Delta+2s_2\right)}
\frac{\Gamma\left(1-\Delta'+s_3\right)}{\Gamma\left(1-\Delta'+2s_3\right)}\\ 
&\quad 
\times
\frac{\mG(k_1,\Delta-2s_1)}{\mG(k_1,\tilde{\Delta})}
\frac{\mG(\bar{k}_2,\Delta-2s_2)}{\mG(\bar{k}_2,\tilde{\Delta})}\\ 
&\quad 
\times
\left[\Gamma\left(1+i\mw_2\right)\pFqReg{2}{1}{-2i\mw_1-i\mw_2,2\Delta+\Delta'-2 (s_1+s_2+s_3)}{1+2\Delta+\Delta'+i\mw_2-2(s_1+s_2+s_3)}{-1}\frac{\mG(k_1+k_2,\Delta'-2s_3)}{\mG(k_1+k_2,\tilde{\Delta}')} \right. \\ 
&\quad 
\left. 
-\Gamma\left(1+2i\mw_2+i\mw_1\right)\pFqReg{2}{1}{-i\mw_1,2\Delta+\Delta'-2 (s_1+s_2+s_3)}{1+2\Delta+\Delta'+2i\mw_2+i\mw_1-2(s_1+s_2+s_3)}{-1}\frac{\mG(\bar{k}_1+\bar{k}_2,\Delta'-2s_3)}{\mG(\bar{k}_1+\bar{k}_2,\tilde{\Delta}')}\right] \\ 
&- (k_1\leftrightarrow k_2)\,.
\end{split}
\end{equation}
\normalsize
Next, we evaluate the integrals over the Mellin parameters by deforming the contour so as to enclose the poles at $s_i = -n_i$ and $s_i = \Delta/\Delta' - 1 -n_i $:
\small
\begin{equation}
\begin{split}
\mW(k_1,k_2) &= \sum_{\delta_1,\delta_2 = \{\Delta,\tDelta\}}\sum_{\delta_3 = \{\Delta',\tDelta'\}}\left(\prod_{i=1}^{3}\sum_{n_i=0}^{\infty}\frac{(-1)^{n_i}}{\Gamma(1+n_i)}
\frac{\Gamma\left(1-\delta_i-n_i\right)}{\Gamma\left(1-\delta_i-2n_i\right)}\right)
\frac{\mG(k_1,\delta_1+2n_1)}{\mG(k_1,\tilde{\Delta})}
\frac{\mG(\bar{k}_2,\delta_2+2n_2)}{\mG(\bar{k}_2,\tilde{\Delta})}\\ 
&\quad 
\times
\left[\Gamma\left(1+i\mw_2\right)\pFqReg{2}{1}{-2i\mw_1-i\mw_2,\sum_i(\delta_i+2 n_i)}{1+i\mw_2+\sum_i(\delta_i+2 n_i)}{-1}\frac{\mG(k_1+k_2,\delta_3+2n_3)}{\mG(k_1+k_2,\tilde{\Delta}')} \right. \\ 
&\quad 
\left. 
-\Gamma\left(1+2i\mw_2+i\mw_1\right)\pFqReg{2}{1}{-i\mw_1,\sum_i(\delta_i+2 n_i)}{1+2i\mw_2+i\mw_1+\sum_i(\delta_i+2 n_i)}{-1}\frac{\mG(\bar{k}_1+\bar{k}_2,\delta_3+2n_3)}{\mG(\bar{k}_1+\bar{k}_2,\tilde{\Delta}')}\right] \\ 
&- (k_1\leftrightarrow k_2)\,.
\end{split}
\end{equation}
\normalsize
While this expression is not very illuminating, we remark that it allows an easy distinction of the analytic structure of the correlator, as described in \cite{Loganayagam:2022zmq}. The possible poles are all contained within the poles of the functions $\mG(k,\delta)$, i.e., the two-point functions.

\section{Connection to the Skenderis-van Rees prescription}\label{app:VRees}

The goal of this Appendix is to relate the prescription of the main text to the proposal by Skenderis-van Rees \cite{Skenderis:2008dg, Skenderis:2008dh}. Here we follow the convention of \cite{Botta-Cantcheff:2019apr}, and then relate with the results in the main text. To exemplify the procedure, we first rework the gravitational Schwinger-Keldysh (grSK$_2$) solution in this framework. 

We consider multiple copies of maximally extended Schwarzschild black holes whose dual is the thermofield double state. 
The full bulk geometry is depicted in fig.\,\ref{fig:appendix2Fold}. 

As before, we study the dynamics of a probe scalar field in this full geometry. 
The dual contour on the field theory side consists of a Schwinger-Keldysh contour with times $t \in [0, T]$, connected to an imaginary time segment of length $\beta /2$ with another Schwinger-Keldysh contour in the time domain $t \in [-T, 0]$. The latter is again connected to an imaginary time segment of length $\beta /2$ and is then identified with the beginning of the contour (see fig. \ref{fig:appendix2Fold}). 
As we are only interested in the first Schwinger-Keldysh contour with a thermal circle of length $\beta$, 
we require the solution to be normalizable in all segments except those belonging to the first Schwinger-Keldysh contour. From the field theory perspective, this corresponds to collapsing the second real-time contour that has the time domain $t \in [-T, 0]$. 

For simplicity, we consider the Lorentzian, planar BTZ black hole with the metric
\begin{align}
    ds^2=-(r^2-1) dt^2+\frac{dr^2}{(r^2-1)}+r^2dx^2
    \,.
\end{align}
This is in contrast to the metric in \cite{Botta-Cantcheff:2019apr}, where the authors use global coordinates. The analysis does not depend on this choice. 
The Euclidean version of the metric reads
\begin{align}
    d s^2=(r^2-1)\,d \tau^2+\frac{d r^2}{(r^2-1)}+r^2 dx^2
    \,.
\end{align}
All quantities are written in units of $r_+$, where $r_+$ is the Schwarzschild radius. 
The solution of the Klein-Gordon equation $ (\Box-m^2)\varphi(t,r,x)=0$ in these spacetimes can be expanded into plane waves 
\begin{align}
\varphi(t,r,x) = \int_{k} \, e^{-i \mw t + i \mq x}\, f(\pm\mw,\mq,r)\, , 
\end{align}
where we defined the integral in \eqref{eq:MomSpace}, and use the dimensionless frequency and momentum $\mw$ and momenta $\mq$, which were defined in \eqref{eq:deffrakw}. 

The two linearly independent solutions $f(\pm\mw,\mq,r)$ are given in terms of the regularized hypergeometric function 
\begin{align}
\label{eq:functionfforBTZ}
    f(\mw,\mq,r)
    =\mathscr{C}_{\mw \mq\Delta}\,
    r^{-\Delta}
    \left(1-\frac{1}{r^2}\right)^{i\frac{\mw}{2}}
    \pFqReg{2}{1}{i\frac{\mw - \mq}{2}+\frac{\Delta}{2}, i\frac{\mw + \mq}{2} + \frac{\Delta}{2}}{1+i\,\mw}{1-\frac{1}{r^2}}\,,
\end{align}
with the normalization factor
\begin{equation}
    \mathscr{C}_{\mw \mq\Delta}=
    \frac{
    \Gamma\left(
    \frac{\Delta}{2}
    +\frac{i}{2}(\mw-\mq)
    \right)
    \Gamma\left(
    \frac{\Delta}{2}
    +\frac{i}{2}(\mw+\mq)
    \right)
    }{
    \Gamma(\Delta-1)
        }
\end{equation}
chosen such that the asymptotic expansion of $f$ close to the conformal boundary located at $r \rightarrow  \infty$ reads 
\begin{align}
f(\mw,\mq,r) \sim 1 \cdot r^{\Delta -2}.
\end{align}
The asymptotics of $f$ allow for Dirichlet boundary conditions \eqref{eq:DirichletBoundaryConditions}. 

Note the different sign in the frequency compared to \eqref{eq:BtoBdyBTZ}. 
The poles of $f(\mw, \mq, r)$ lie in the upper half plane. 
Moreover, we are using $t$-$r$ coordinates here in contrast to ingoing coordinates. 
In particular, the relation between $f$ and the bulk-boundary propagator $\Gin^{\Delta}(z,k)$ defined in \eqref{eq:BtoBdyBTZ} is given by \eqref{eq:Appendix:RelationtoGin}. 


\subsection{Two-fold}

\begin{figure}[H]
    \centering
    \includegraphics[width=0.75\linewidth]{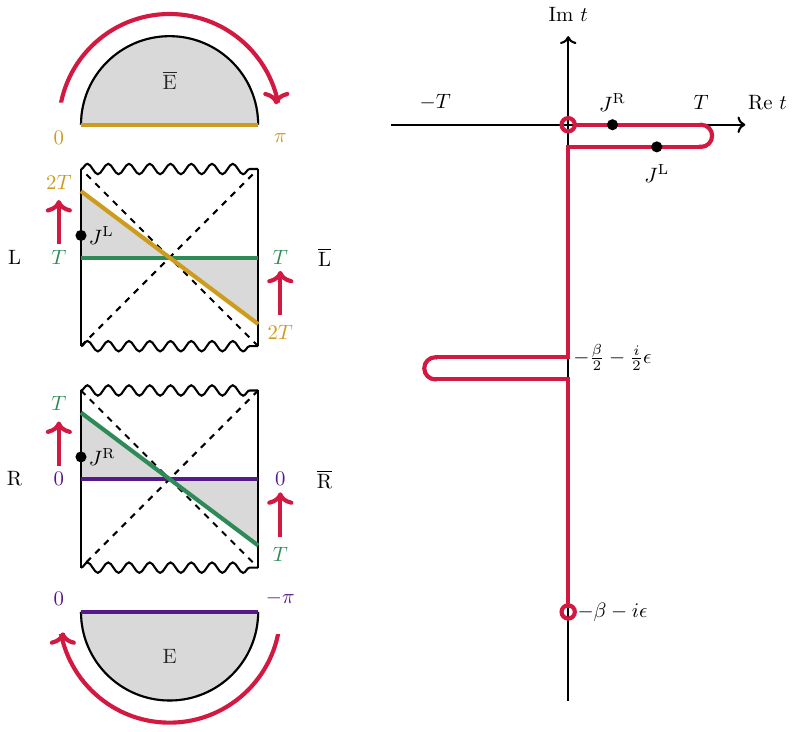}
    \caption{Left: Geometry of the two-fold with equal-time slices. Right: Contour in the field theory. }
    \label{fig:appendix2Fold}
\end{figure}

In contrast to the main text, we want to match the field in segments along a constant $t$ slice. Therefore, we consider a geometry of the form depicted in fig.\,\ref{fig:appendix2Fold}. 

The matching conditions are\footnote{For simplicity, we dropped the the additional arguments $r$ and $x$ of $\varphi(s,r,x).$}
\begin{subequations}
\begin{align}
    \varphi^\text{E}(\tau=0)
    &=\varphi^\text{R}(s=0)
    \quad
    &-\partial_\tau\varphi^\text{E}(\tau=0)
    &=i\partial_s\varphi^\text{R}(s=0)
    \label{eq:App2Fold_matching_a}
    \,,\\
    \varphi^\text{R}(s=T)
    &=\varphi^\text{L}(s=T)
    \quad
    &-i\partial_s\varphi^\text{R}(s=T)
    &=i\partial_s\varphi^\text{L}(s=T)
    \label{eq:App2Fold_matching_b}
    \,,\\
    \varphi^{\overline{\text{E}}}(\tau=0)
    &=\varphi^\text{L}(s=2T)
    \quad
    &\partial_\tau\varphi^{\overline{\text{E}}}(\tau=0)
    &=i\partial_s\varphi^\text{L}(s=2T)
    \label{eq:App2Fold_matching_c}
    \,,\\
    \varphi^{\overline{\text{E}}}(\tau=\pi)
    &=\varphi^{\overline{\text{L}}}(s=2T)
    \quad
    &\partial_\tau\varphi^{\overline{\text{E}}}(\tau=\pi)
    &=i\partial_s\varphi^{\overline{\text{L}}}(s=2T)
    \label{eq:App2Fold_matching_d}
    \,,\\
    \varphi^{\overline{\text{R}}}(s=T)
    &=\varphi^{\overline{\text{L}}}(s=T)
    \quad
    &-i\partial_s\varphi^{\overline{\text{R}}}(s=T)
    &=i\partial_s\varphi^{\overline{\text{L}}}(s=T)
    \label{eq:App2Fold_matching_e}
    \,,\\
    \varphi^\text{E}(\tau=-\pi)
    &=\varphi^{\overline{\text{R}}}(s=0)\,,
    \quad
    &-\partial_\tau\varphi^\text{E}(\tau=-\pi)
    &=i\partial_s\varphi^{\overline{\text{R}}}(s=0)
    \label{eq:App2Fold_matching_f}
    \,,
\end{align}
\end{subequations}

There are various signs appearing in the derivative conditions that warrant discussion. The matching of two Lorentzian segments to one another always includes one segment with a reversed causal structure. Since both time directions are anti-parallel this yields a sign difference in the matching conditions \eqref{eq:App2Fold_matching_b} and \eqref{eq:App2Fold_matching_c}. The matching of the first Euclidean segment E to R \eqref{eq:App2Fold_matching_a} or $\overline{\text{R}}$ \eqref{eq:App2Fold_matching_f} is chosen to respect the Wick rotation that defines the Euclidean time. Because of $t=-i\tau$, both time directions are related by a clockwise $\frac{\pi}{2}$ rotation in the complex time plane. For the matching of the other Euclidean segment $\overline{\text{E}}$ to L \eqref{eq:App2Fold_matching_c} or $\overline{\text{L}}$ \eqref{eq:App2Fold_matching_d} however, the Lorentzian segments time direction runs backwards while the Euclidean time direction remains unchanged, generating an additional sign difference in the condition.

This prescription deviates from the Skenderis-van Rees \cite{Skenderis:2008dg} proposal, as the matching surface lies entirely outside the black hole horizon. As such, our prescription does not consider any phase-term associated with an analytic continuation across the black hole horizon. Instead, the phases emerge solely from the matching with the euclidean geometries. 

Starting with the first Lorentzian segment, the solution reads
\begin{align}
\label{eq:App_2Fold_Solution}
\begin{split}
    \varphi^\text{R}(s,r,x)=
    \int_k \, e^{-i\mw s+i\mq x}\, 
    &\left(\frac{1}{2}
    \Tilde{\phi}^\text{R}(\mw,\mq)
    \left[f(\mw,\mq,r)+f(-\mw,\mq,r)\right]\right. \\
    &
    \left.\phantom{\frac{1}{2}}+A^\text{R}_{\mw \mq}
    \left[f(\mw,\mq,r)-f(-\mw,\mq,r)\right]
    \right)
    \,,
\end{split}
\end{align}
where $s \in [0, T]$. In addition,  $\Tilde{\phi}^\text{R}(\mw,\mq)$ and $A^\text{R}_{\mw \mq}$ are related to the Fourier transform of this segments source and 1-point function respectively.

On the second Lorentzian segment, we choose the following ansatz
\begin{align}
    \label{eq:App_2Fold_SolutionFR}
    \begin{split}
    \varphi^\text{L}(s,r,x)=
    \int_{k}
    \,e^{i\mw s+i\mq x}\,
    \bigg(
    &\frac{1}{2}
    \Tilde{\phi}^\text{L}(-\mw,\mq)
    \left[f(\mw,\mq,r)+f(-\mw,\mq,r)\right]
    \\
    &
    -A^\text{L}_{-\mw \mq}
    \left[f(\mw,\mq,r)-f(-\mw,\mq,r)\right]
    \bigg)
    \,,
    \end{split}
\end{align}
where the domain of the parameter $s$ for this segment is now $s \in [T, 2\,T]$. Note that the sign of the frequency is reversed in \eqref{eq:App_2Fold_SolutionFR} as compared to \eqref{eq:App_2Fold_Solution}; this turns out to be beneficial for the evaluation of the matching conditions above. We will discuss later how $\Tilde{\phi}^\text{L}(-\mw,\mq)$ is related to the Fourier transform of the source on this segment.

We do not consider any sources on the Euclidean segments, hence $ \varphi^\text{E}$ and $ \varphi^{\overline{\text{E}}}$ contain only normalizable modes of the form
\begin{subequations}
\begin{align}
    \varphi^\text{E}(\tau,r,x)&=
    \int_k \, e^{-\mw\tau+i\mq x}\, 
    I_{\mw \mq}[f(\mw,\mq,r)-f(-\mw,\mq,r)]
    \,,\\
    \varphi^{\overline{\text{E}}}(\tau,r, x)&=
    \int_k\,
    e^{-\mw\tau+i\mq x}\,
    F_{\mw \mq}[f(\mw, \mq, r)-f(-\mw,\mq,r)]
    \,,
\end{align}
\end{subequations}
where $\tau$ takes values in the interval $[-\pi, 0]$ on the Euclidean segment E, and $[0,\pi]$ on the Euclidean segment ${\overline{\text{E}}}$. For simplicity, we also consider no sources on the segments $\overline{\text{R}}$ and $\overline{\text{L}}$; the respective normalizable modes on those segments 
read
\begin{subequations}
\begin{align}
    \varphi^{\overline{\text{R}}}(s,r, x)=&
    \int_k \,e^{-i\mw s+i\mq x} 
A^{\overline{\text{R}}}_{\mw \mq}
    \left[f(\mw,\mq,r)-f(-\mw,\mq,r)\right]
    \,,
    \\
    \varphi^{\overline{\text{L}}}(s,r, x)=&
    \int_k \,e^{i\mw s+i\mq x}\,
    \left(-A^{\overline{\text{L}}}_{-\mw \mq}\right)
    \left[f(\mw,\mq,r)-f(-\mw,\mq,r)\right]
    \,.
\end{align}
\end{subequations}
Close to the matching surfaces we require a distinction of cases. At late times far in the future of a source somewhere at $\Hat{s}\in(0,T)$ we close the contour of the $\mw$ integral using an infinite semi-circle in the lower half plane and hence only $f(-\mw,\mq,r)$ contributes poles. At early times far to the past of the source we close the contour above where only $f(\mw,\mq,r)$ contributes poles. We can thus rewrite the solutions near the matching surfaces to
\begin{align}
\begin{split}
    \varphi^\text{R}(s\sim0,r, x)=&
    \int_{k}\,
    e^{-i\mw s+i\mq x} \,
    \left(\frac{1}{2}\Tilde{\phi}^\text{R}(\mw,\mq)+A^{R}_{\mw \mq}\right)
    [f(\mw,\mq,r)-f(-\mw,\mq,r)]
    \,,\\
    \varphi^\text{R}(s\sim T,r, x)=&
    \int_{k}\,
    e^{-i\mw s+i\mq x}\,
    \left(-\frac{1}{2}\Tilde{\phi}^\text{R}(\mw,\mq)+A^{R}_{\mw \mq}\right)
    [f(\mw,\mq,r)-f(-\mw,\mq,r)]
    \,.
    \end{split}
\end{align}
Similarly, on the second segment L where $s\in[T,2T]$ we get the distinction
\begin{subequations}
\begin{align}
\begin{split}
    \varphi^\text{L}(s\sim T,r, x)=&
    \int_{k}\,
    e^{i\mw s+i\mq x}\,
    \left(
    -\frac{1}{2}\Tilde{\phi}^\text{L}(-\mw,\mq)
    -A^{R}_{-\mw \mq}\right)
    [f(\mw,\mq,r)-f(-\mw,\mq,r)]
    \,,\\
    \varphi^\text{L}(s\sim 2T,r, x)=&
    \int_k\,
    e^{i\mw s+i\mq x}\,
    \left(\frac{1}{2}\Tilde{\phi}^\text{L}(-\mw,\mq)
    -A^{R}_{-\mw \mq}\right)
    [f(\mw,\mq,r)-f(-\mw,l\mq,r)]
    \,.
    \end{split}
\end{align}
\end{subequations}
The matching conditions for the continuity of the field and its derivative both yield the same equations at each matching surface. Subsequently we are left with six independent equations to determine our six coefficients 
\begin{subequations}
\begin{align}
    &I_{\mw \mq}=
    \frac{1}{2}
    \Tilde{\phi}^\text{R}(\mw,\mq)
    +A^\text{R}_{\mw \mq}
    \,,\\
    &\left(-\frac{1}{2}
    \Tilde{\phi}^\text{R}(\mw,\mq)
    +A^\text{R}_{\mw \mq}
    \right)
    e^{-i\mw T}
    =
    \left(-\frac{1}{2}
    \Tilde{\phi}^\text{L}(-\mw,\mq)
    -A^\text{L}_{-\mw \mq}
    \right)
    e^{i\mw T}
    \,,\\
    &F_{\mw \mq}=
    \left(\frac{1}{2}
    \Tilde{\phi}^\text{L}(-\mw,\mq)
    -A^\text{L}_{-\mw \mq}
    \right)
    e^{2i\mw T}
    \,,\\
    &F_{\mw \mq}
    e^{-\mw \pi}=
    -A^{\overline{\text{L}}}_{-\mw \mq}
    e^{2i\mw T}
    \,,\\
    &
    A^{\overline{\text{R}}}_{\mw \mq}
    e^{-i\mw T}=
    -A^{\overline{\text{L}}}_{-\mw \mq}
    e^{i\mw T}
    \,,\\
    &A^{\overline{\text{R}}}_{\mw \mq}=
    I_{\mw \mq}
    e^{\mw \pi}
    \,,
\end{align}
\end{subequations}
Taken together, these conditions fully determine all coefficients in terms of the sources
\begin{subequations}
\begin{align}
    I_{\mw \mq}=&
    \frac{1}{e^{2\pi\mw}-1}
    \left(
    -\Tilde{\phi}^\text{R}(\mw,\mq)
    +e^{2i\mw T}
    \Tilde{\phi}^\text{L}(-\mw,\mq)
    \right)
    \,,\\
    F_{\mw \mq}=&
    \frac{e^{2\pi\mw}}{e^{2\pi\mw}-1}
    \left(
    -\Tilde{\phi}^\text{R}(\mw,\mq)
    +e^{2i\mw T}
    \Tilde{\phi}^\text{L}(-\mw,\mq)
    \right)
    \,,\\
    A^\text{R}_{\mw \mq}=&
    \frac{1}{e^{2\pi\mw}-1}
    \Big(
    -\frac{1}{2}
    \left(1+e^{2\pi\mw}\right)
    \Tilde{\phi}^\text{R}(\mw,\mq)
    \phantom{\Big(}+e^{2i\mw T}
    \Tilde{\phi}^\text{L}(-\mw,\mq)
    \Big)
    \,,\\
    A^\text{L}_{-\mw \mq}=&
    \frac{1}{e^{2\pi\mw}-1}
    \Big(
    e^{2\pi\mw}e^{-2i\mw T}
    \Tilde{\phi}^\text{R}(\mw,\mq)
    \phantom{\Big(}
    -\frac{1}{2}
    \left(1+e^{2\pi\mw}\right)
    \Tilde{\phi}^\text{L}(-\mw,\mq)
    \Big)
    \,,\\
    A^{\overline{\text{R}}}_{\mw \mq}=&
    \frac{e^{\pi\mw}}{e^{2\pi\mw}-1}
    \left(
    -\Tilde{\phi}^\text{R}(\mw,\mq)
    +e^{2i\mw T}\Tilde{\phi}^\text{L}(-\mw,\mq)
    \right)
    \,,\\
    A^{\overline{\text{L}}}_{-\mw \mq}=&
    \frac{e^{\pi\mw}}{e^{2\pi\mw}-1}
    \left(\Tilde{\phi}^\text{R}(\mw,\mq)
    -\Tilde{\phi}^\text{L}(-\mw,\mq)
    \right)
    \,.
\end{align}
\end{subequations}
Before giving the full solution, we want to switch back from our contour parametrization $s$ to the time coordinate $t$ of the complex time plane. On the first Lorentzian segment $R$, the parametrization agrees with the real time. On the second Lorentzian segment $L$, the relation is $t=2T-s$, as $t \in [0, T]$ and the minus sign reflects the anti-time ordering. 

Using how $\tilde{\phi}(\mw,\mq)$ is related to the fourier transform of the sources $\JRI{}, \, \JLI{}$, 
\begin{align}
\begin{split}
    \label{eq:Appendix:ConnectionPhiTildeToSource}
    \JRI{}(\mw,\mq)&=\frac{1}{4\pi^2}
    \Tilde{\phi}^\text{R}(\mw,\mq)
    \,. \\
    \JLI{}(\mw,\mq)&=\frac{1}{4\pi^2}
    \Tilde{\phi}^\text{L}(-\mw,\mq)
    e^{2i\mw T}
    \,,
\end{split}
\end{align}
and additionally applying the transformation from $s$ to $t$ to the other exponential functions in the fields
allows us to give the bulk field solutions as
\begin{subequations}
\label{eq:AppendixgrSK2Solution}
\begin{align}
\begin{split}
    \varphi^\text{R}(t,r, x)
    =&
    \int_k\,
    \frac{1}{e^{2\pi\mw }-1}
    e^{-i\mw t+i\mq x}
    \\
    \Big[
    &\left(
    -\JRI{}
    +\JLI{}
    \right)
    f(\mw,\mq,r)
    +
    \left(
    e^{2\pi\mw }
    \JRI{}
    -\JLI{}
    \right)
    f(-\mw,\mq,r)
    \Big]
    \,,
\end{split}\\
\begin{split}
    \varphi^\text{L}(t,r, x)
    =&
    \int_k\,
    \frac{1}{e^{2\pi\mw }-1}
    e^{-i\mw t+i\mq x}
    \\
    \Big[
    &e^{2\pi\mw}
    \left(
    -\JRI{}
    +\JLI{}
    \right)
    f(\mw,\mq,r)
    +
    \left(
    e^{2\pi\mw}
    \JRI{}
    -\JLI{}
    \right)
    f(-\mw,\mq,r)
    \Big]
    \,.
    \end{split}
\end{align}
\end{subequations}
In order to make contact with the gravitational Schwinger-Keldysh solution \eqref{eq:grSKsol1}, we first restore the units of $r_+$ which leads to $2 \pi \mw = \beta \omega$, and identify
\begin{align}
\label{eq:Appendix:RelationtoGin}
\begin{split}
    e^{-i \omega t+ i k x}\, f(\omega , k, r) & = e^{-i \omega (t+r_*)}\Gout(\omega, k)\, \\
    e^{-i \omega t+ i k  x}\, f(-\omega , k, r) & = e^{-i \omega (t+r_*)}\Gin(\omega, k)\,. 
    \end{split}
\end{align}
With this identification, the solution \eqref{eq:AppendixgrSK2Solution} agrees with the well-known gravitational Schwinger-Keldysh solution \eqref{eq:grSKsol1}.

\subsection{Four-fold}
\begin{figure}[H]
    \centering
    \includegraphics[width=0.75\linewidth]{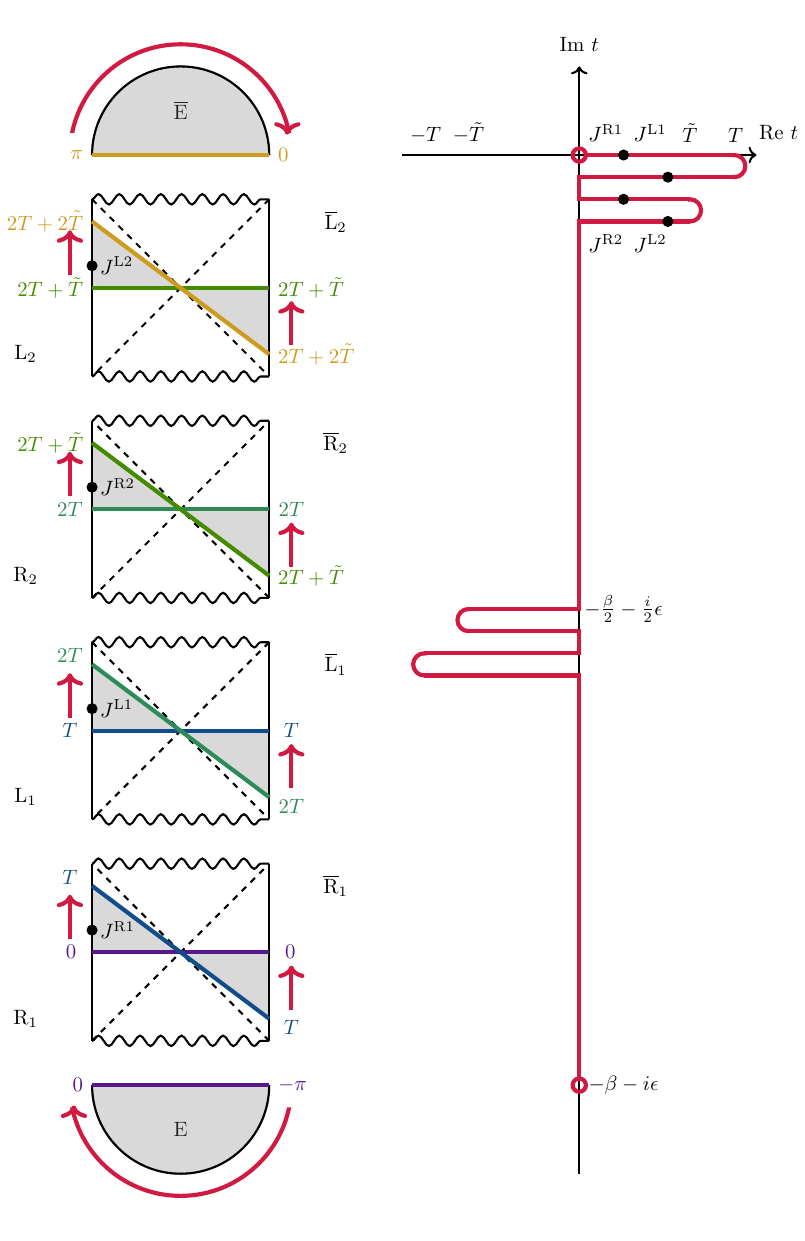}
    \caption{Left: Geometry of the four-fold with equal-time slices. Right: Contour in the field theory. }
    \label{fig:appendix4Fold}
\end{figure}

The solution for the four-fold geometry can be obtained in a similar fashion. We glue together four copies of two-sided black holes along constant time sliced as depicted in fig.\,\ref{fig:appendix4Fold}. 

Again, we consider also a non-vanishing field on the segments $\overline{R}_{1, 2}$, $\overline{L}_{1, 2}$ in order to solve the matching conditions. Later we turn off the sources on these segments and focus only on the field in the segments $R_{1, 2}$, $L_{1, 2}$. 

The matching conditions follow the same logic as for the two-fold
\small
\begin{subequations}
\begin{align}
    \varphi^\text{E}(\tau=0)
    &=\varphi^\text{R1}(s=0)
    \quad
    &-\partial_\tau\varphi^\text{E}(\tau=0)
    &=i\partial_s\varphi^\text{R1}(s=0)
    \,,\\
    \varphi^\text{R1}(s=T)
    &=\varphi^\text{L1}(s=T)
    \quad
    &-i\partial_s\varphi^\text{R1}(s=T)
    &=i\partial_s\varphi^\text{L1}(s=T)
    \,,\\
    \varphi^\text{L1}(s=2T)
    &=\varphi^\text{R2}(s=2T)
    \quad
    &-i\partial_s\varphi^\text{L1}(s=2T)
    &=i\partial_s\varphi^\text{R2}(s=2T)
    \,,\\
    \varphi^\text{R2}(s=2T+\Tilde{T})
    &=\varphi^\text{L2}(s=2T+\Tilde{T})
    \quad
    &-i\partial_s\varphi^\text{R2}(s=2T+\Tilde{T})
    &=i\partial_s\varphi^\text{L2}(s=2T+\Tilde{T})
    \,,\\
    \varphi^{\overline{\text{E}}}(\tau=0)
    &=\varphi^\text{L2}(s=2T+2\Tilde{T})
    \quad
    &\partial_\tau\varphi^{\overline{\text{E}}}(\tau=0)
    &=i\partial_s\varphi^\text{L2}(s=2T+2\Tilde{T})
    \,,\\
    \varphi^{\overline{\text{E}}}(\tau=\pi)
    &=\varphi^{\overline{\text{L}2}}(s=2T+2\Tilde{T})
    \quad
    &\partial_\tau\varphi^{\overline{\text{E}}}(\tau=\pi)
    &=i\partial_s\varphi^{\overline{\text{L}2}}(s=2T+2\Tilde{T})
    \,,\\
    \varphi^{\overline{\text{R}2}}(s=2T+\Tilde{T})
    &=\varphi^{\overline{\text{L}2}}(s=2T+\Tilde{T})
    \quad
    &-i\partial_s\varphi^{\overline{\text{R}2}}(s=2T+\Tilde{T})
    &=i\partial_s\varphi^{\overline{\text{L}2}}(s=2T+\Tilde{T})
    \,,\\
    \varphi^{\overline{\text{L}1}}(s=2T)
    &=\varphi^{\overline{\text{R}2}}(s=2T)
    \quad
    &-i\partial_s\varphi^{\overline{\text{L}1}}(s=2T)
    &=i\partial_s\varphi^{\overline{\text{R}2}}(s=2T)
    \,,\\
    \varphi^{\overline{\text{R}1}}(s=T)
    &=\varphi^{\overline{\text{L}1}}(s=T)
    \quad
    &-i\partial_s\varphi^{\overline{\text{R}1}}(s=T)
    &=i\partial_s\varphi^{\overline{\text{L}1}}(s=T)
    \,,\\
    \varphi^\text{E}(\tau=-\pi)
    &=\varphi^{\overline{\text{R}1}}(s=0)
    &-\partial_\tau\varphi^\text{E}(\tau=-\pi)
    &=i\partial_s\varphi^{\overline{\text{R}1}}(s=0)
    \,.
\end{align}
\end{subequations}
\normalsize
These conditions ensure the continuity of the scalar field and its first derivative across the entire spacetime manifolds. Solving these equations determines all ten coefficients with respect to the $\Tilde{\phi}(\mw,\mq)$, which is connected to the sources (see equation \eqref{eq:Appendix:ConnectionPhiTildeToSource}). We give the four coefficients of the relevant segments $R_{1, 2}, \, L_{1, 2}$
\begin{subequations}
\begin{align}
    A^\text{R1}_{\mw \mq}= 
    \frac{1}{e^{2 \pi  \mw }-1}
    \Big(&
    -\frac{1}{2}(1+e^{2 \pi  \mw })
    \Tilde{\phi}^\text{R1}(\mw,\mq)
    +\Tilde{\phi}^\text{L1}(-\mw,\mq) 
    e^{2 i \mw T}
    \nonumber\\
    &
    -\Tilde{\phi}^\text{R2}(\mw,\mq) 
    e^{-2 i \mw T}
    +\Tilde{\phi}^\text{L2}(-\mw,\mq) 
    e^{i \mw  (2T+2\Tilde{T})}
    \Big)
    \,,\\
    A^\text{L1}_{-\mw \mq}=
    \frac{1}{e^{2 \pi  \mw }-1}
    \Big(&
    \Tilde{\phi}^\text{R1}(\mw,\mq) 
    e^{2\pi\mw }
    e^{-2i \mw T}
    -\frac{1}{2}(e^{2 \pi  \mw }+1)
    \Tilde{\phi}^\text{L1}(-\mw,\mq)
    \nonumber\\
    &
    -\Tilde{\phi}^\text{L2}(-\mw,\mq) 
    e^{2 i \mw \Tilde{T} }
    +\Tilde{\phi}^\text{R2}(\mw,\mq) 
    e^{-4 i \mw T }
    \Big)
    \,,\\
    A^\text{R2}_{\mw \mq}=
    \frac{1}{e^{2 \pi  \mw }-1}
    \Big(&
    -\Tilde{\phi}^\text{R1}(\mw,\mq) 
    e^{2\pi\mw }
    e^{2i \mw T}
    +\Tilde{\phi}^\text{L1} 
    e^{2\pi\mw }
    e^{2i\mw T}
    \nonumber\\
    &
    -\frac{1}{2}(e^{2 \pi  \mw }+1)
    \Tilde{\phi}^\text{R2}(\mw,\mq)
    +\Tilde{\phi}^\text{L2}(-\mw,\mq)
    e^{i \mw  (4 T+2\Tilde{T})}
    \Big)
    \,,\\
    A^\text{L2}_{-\mw \mq}=
    \frac{1}{e^{2 \pi  \mw }-1}
    \Big(&
    \Tilde{\phi}^\text{R1}(\mw,\mq) 
    e^{2 \mw \pi }
    e^{- i\mw (2T+2\Tilde{T})}
    -\Tilde{\phi}^\text{L1}(-\mw,\mq) 
    e^{2 \pi \mw }
    e^{-2i\mw\Tilde{T}) }
    \nonumber\\
    &
    +\Tilde{\phi}^\text{R2}(\mw,\mq) 
    e^{2 \pi\mw}
    e^{- i\mw  (4 T+2\Tilde{T})}
    -\frac{1}{2}(e^{2 \pi  \mw } +1)
    \Tilde{\phi}^\text{L2}(-\mw,\mq)
    \Big)
    \,.
\end{align}
\end{subequations}
Reinserted into the bulk field solutions, and after transforming from the contour parameter $s$ to the real time $t$, we get the four fields in terms of the sources. These can be compactly written as a matrix

\small
\begin{align}
\label{eq:Appendix:4FoldFinalMatrix}
\begin{split}
    &\begin{pmatrix}
        \varphi_{1}\\
        \varphi_{2}\\
        \varphi_{3}\\
        \varphi_{4}\\
    \end{pmatrix}  = 
    \int_{k}  \, e^{-i \mw t + i \mq x}\,  \\ &\begin{pmatrix}
        -f(\mw)+e^{2 \pi \mw}f(-\mw) &f(\mw)-f(-\mw)&f(-\mw)-f(\mw)&f(\mw)-f(-\mw)\\
         e^{2 \pi \mw}(f(-\mw)-f(\mw)) &e^{2 \pi \mw}f(\mw) - f(-\mw)&f(-\mw)-f(\mw)&f(\mw) - f(-\mw)\\
         e^{2 \pi \mw}\left(f(\mw)-f(-\mw)\right)&e^{2 \pi \mw}\left(f(\mw)-f(-\mw)\right)&f(-\mw)-f(\mw)&f(\mw) - f(-\mw)\\
         e^{2 \pi \mw}\left(f(-\mw)-f(\mw)\right)&e^{2 \pi \mw}\left(f(\mw)-f(-\mw)\right)&e^{2 \pi \mw}\left(f(-\mw)-f(\mw)\right)&e^{2 \pi \mw} f(\mw)-f(-\mw)\\
    \end{pmatrix} \begin{pmatrix}
\JRI{1}\\ 
\JLI{1}\\ 
\JRI{2}\\ 
\JLI{2}
\end{pmatrix}\,,
\end{split}
\end{align}
\normalsize
where we suppressed the momentum dependence in $f$ and the sources. 
Using the identification \eqref{eq:Appendix:RelationtoGin}, as well as changing the units, the solution \eqref{eq:Appendix:4FoldFinalMatrix} indeed coincides with the corresponding solution in the main text \eqref{eq:solMatrixLR}. 
While this analysis is done for the three-dimensional black hole, we never explicitly use the form of $f$ in \eqref{eq:functionfforBTZ}. Thus the analysis straighforewardly generalizes to arbitrary dimensions. 
Moreover, this construction allows us to also consider real-time contours for other states. For example, for correlation functions the thermofield double state, we simply have to turn on sources on the regions $\overline{R}_{1,2}, \, \overline{L}_{1,2}$, or for considering a perturbed thermal state, one can turn on sources on the Euclidean segments.

\bibliographystyle{JHEP}
\bibliography{References}
\end{document}